\documentclass[11pt,aps,prc,tightenlines,showpacs,
showkeys,superscriptaddress]{revtex4}
\usepackage{graphicx,color}
\def\thalf{{\textstyle{\frac{1}{2}}}}
\def\tquar{{\textstyle{\frac{1}{4}}}}
\def\teight{{\textstyle{\frac{1}{8}}}}
\newcommand{\feyn}[1]{{#1}\!\!\!{\slash}}	
\def\ni{\noindent}

\def\simge{%
    \mathrel{\rlap{\raise 0.511ex
        \hbox{$>$}}{\lower 0.511ex \hbox{$\sim$}}}}
\def\simle{%
    \mathrel{\rlap{\raise 0.511ex
        \hbox{$<$}}{\lower 0.511ex \hbox{$\sim$}}}}
\begin{document}
\title{Isospin Asymmetry in Nuclei and Neutron Stars}

\author{A.W. Steiner} 
\affiliation{School of Physics and Astronomy, University of Minnesota,
Minneapolis, MN 55455-0112}
\affiliation{Theoretical Division, Los Alamos National Laboratory,
Los Alamos, NM 87545}

\author{M. Prakash} 

\author{J.M. Lattimer}
\affiliation{Department of Physics and Astronomy, SUNY at Stony Brook,
Stony Brook, New York 11794-3800}

\author{P.J. Ellis}
\affiliation{School of Physics and Astronomy, University of Minnesota,
Minneapolis, MN 55455-0112}


\begin{abstract}   
The roles of isospin asymmetry in nuclei and neutron stars are 
investigated using a range of potential and field-theoretical models
of nucleonic matter. The parameters of these models are fixed by fitting
the properties of homogeneous bulk matter and closed-shell nuclei.  We
discuss and unravel the causes of correlations among the neutron skin
thickness in heavy nuclei, the pressure of beta-equilibrated matter at
a density of 0.1 fm$^{-3}$, the derivative of the nuclear symmetry
energy at the same density and the radii of moderate mass neutron
stars.  Constraints on the symmetry properties of nuclear matter from
the binding energies of nuclei are examined.  The extent to which
forthcoming neutron skin measurements will further delimit the symmetry
properties is investigated.  The impact of symmetry energy constraints
for the mass and moment of inertia contained within neutron star
crusts and the threshold density for the nucleon direct Urca process,
all of which are potentially measurable, is explored.  We also comment
on the minimum neutron star radius, assuming that only nucleonic
matter exists within the star.
\end{abstract}
\maketitle
%\preprint{LA-UR-04-6745} 
\newpage
\tableofcontents
\newpage
\section{INTRODUCTION}

Strongly interacting matter in which more neutrons than protons exist
is encountered in both heavy nuclei and neutron stars.  In stable
nuclei the net asymmetry $I=(N-Z)/(N+Z)$ ranges up to about 0.24, but
the neutron-to-proton asymmetry, $\delta = (n_n-n_p)/n$ where $n_n$
and $n_p$ are the number densities of neutrons and protons and
$n=n_n+n_p$, approaches unity in the nuclear surface. In the future,
rare-isotope accelerator experiments will extend the range of $I$ to
values well in excess of 0.2. In contrast, $I$ could be as large as
0.95 in the interiors of neutron stars. The physical properties of
nuclei, such as their masses, neutron and proton density distributions
(including their mean radii), collective excitations, fission
properties, matter and momentum flows in high energy heavy-ion
collisions, etc. all depend on the isospin structure of the strong
interactions between nucleons (i.e., $nn$ and $pp$ interactions versus
$np$ interactions).  The energetics associated with the $n-p$
asymmetry can be characterized by the so-called symmetry energy,
$E_{sym}$, which is the leading coefficient of an expansion of the
total energy with respect to asymmetry: $E(n,\delta)\approx
E_0(n)+E_{sym}\delta^2\cdots$.  The energy $\hat{\mu} = \mu_n - \mu_p
\cong 4 E_{sym}\delta$, where $\mu_n$ and $\mu_p$ are the neutron and
proton chemical potentials, respectively, is crucial in determining
reaction rates involving electrons and neutrinos, particle abundances,
etc., in astrophysical contexts such as supernova dynamics,
proto-neutron star evolution, the $r-$process, the long-term cooling
of neutron stars, and the structure of cold-catalyzed neutron stars
(i.e. their masses, radii and crustal extent), etc.  The pervasive
role of the isospin dependence of strong interactions in nuclear
processes in the laboratory and the cosmos is sketched in Fig. 1. In
this work some of these connections will be discussed.

\begin{figure}[htb]
\begin{center}
\includegraphics[scale=0.65,angle=0]{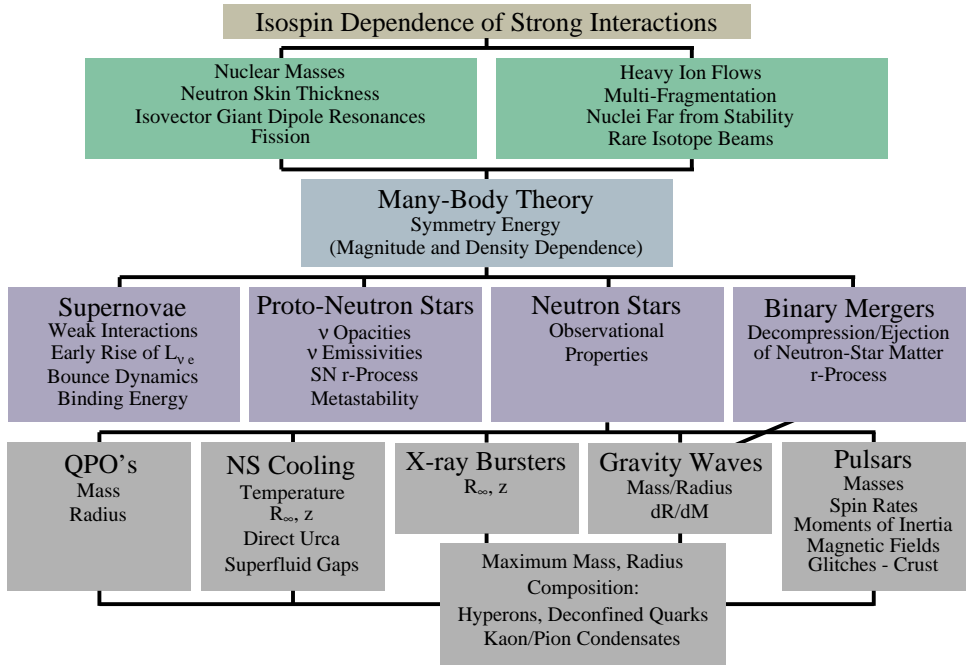}
\end{center}
\caption{The multifaceted influence of the nuclear 
symmetry energy.}
\label{figinfluence}
\end{figure}

Recently, several empirical relationships have been discovered that
underscore the role of isospin interactions in nuclei and neutron
stars. These include correlations between:

\begin{enumerate}

\item {\em The neutron star radius $R$ and the
pressure $P$ of neutron-star matter}: Lattimer and Prakash
\cite{Lattimer00,Lattimer01} found that the quantity $RP^{-1/4}$ is
approximately constant, for a given neutron star mass, for a wide
variety of equations of state when the pressure $P$ of
beta-equilibrated neutron-star matter is evaluated at a density in the
range $n_0$ to $2n_0$, where $n_0$ denotes equilibrium nuclear matter
density. Since the pressure of nearly pure neutron matter (a good
approximation to neutron star matter) near $n_0$ is approximately given 
by $n^2\partial E_{sym}/\partial n$, the density dependence of the
symmetry energy just above $n_0$ will be a critical factor in
determining the neutron star radius.  

\item {\em The neutron skin thickness in nuclei and the pressure of
pure neutron matter at sub-nuclear density:} Typel and Brown
\cite{Brown00,Typel01} have noted that model calculations of the
difference between neutron and proton rms radii $\delta R ={\langle
r_n^2\rangle}^{1/2} - {\langle r_p^2\rangle}^{1/2}$ are linearly
correlated with the pressure of pure neutron matter at a density below
$n_0$ characteristic of the mean density in the nuclear surface (e.g.,
0.1 fm$^{-3})$.  The density dependence of the symmetry energy
controls $\delta R$ (we will call this the   
neutron skin thickness) in a
heavy, neutron-rich nucleus.  Explicitly, $\delta R$ is proportional
to a specific average of $[1-E_{sym}(n_0)/E_{sym}(n)]$ in the nuclear
surface, see Refs.~\cite{Krivine84,Lattimer96} 
and Eq. (\ref{eq:newt}) below. 

\item {\em The neutron skin thickness in nuclei
and the neutron star radius:} Horowitz and Piekarewicz
\cite{Horowitz01} have pointed out that models that yield smaller
neutron skins in heavy nuclei tend to yield smaller neutron star
radii. These authors, along with others \cite{Horowitz01b,Michaels00},
have also pointed out the need for an accurate measurement of the
neutron skin.
\end{enumerate}

Unlike proton distributions, neutron distributions in nuclei have
remained uncertain to this date.  Recent studies of neutron densities
from a global analysis of medium-energy proton scattering on
$^{208}$Pb indicate that $0.07 < \delta R < 0.16$ fm
\cite{Clark03}. Related information is also available from an analysis
of antiprotonic atom data that gives $\delta R = 0.15\pm 0.02$
fm~\cite{Trzcinska01}. 
In the latter work, nucleon density distributions are
parameterized by Fermi functions and it is found that the half-density 
radii for neutrons and protons in heavy nuclei are the same, but the
diffuseness parameter for the neutrons is larger than that for the 
protons. Skin thicknesses as large as 0.2 fm were obtained in earlier 
analyses; see the discussion of Karataglidis et al. \cite{Karataglidis02}.
Since these studies involve strongly interacting
probes, even to this date the value of $\delta R$ for a nucleus such
as $^{208}$Pb is not accurately known.
This situation should improve
as it is expected that the neutron rms radius will be determined to
about 1\% accuracy by measuring the parity-violating electroweak
asymmetry in the elastic scattering of polarized electrons from
$^{208}$Pb \cite{Horowitz01b}, an experiment planned at the Jefferson
Laboratory \cite{Michaels00}.

A common theme in the evaluation of the neutron skin thickness and the
pressure of neutron star matter below and above the saturation density
$n_0$ is the density dependence of the isospin asymmetric part of the
nuclear interaction.  To highlight this dependence, we examine
theoretical predictions of $\delta R$ and the neutron star radius for
an extensive set of potential and field-theoretical models.  For the
former we employ Skyrme-like potential model interactions, including a
parameterization~\cite{Akmal98} of the microscopic calculations of  
Akmal and Pandharipande~\cite{Akmal97}. For the latter
field-theoretical models we use Walecka-type models~\cite{Serot86} in
which nucleons interact through the exchange of $\sigma$, $\omega$ and
$\rho$ mesons augmented with additional mesonic couplings (see, for
example, Horowitz and Piekarewicz~\cite{Horowitz01}) to better
describe the properties of nuclei and explore the properties of
neutron stars.  In order to provide baseline calculations, our
investigations will be restricted to the case in which the high
density phase of neutron stars contains nucleons (and enough electrons
and muons to neutralize the matter) only.

Our goals in this work are to identify the causes of the correlations
mentioned above through a semi-analytic analysis of detailed
calculations and to uncover other possible correlations.  Toward these
goals, we investigate an assortment of potential and field-theoretical
models that cover a range of symmetry properties and
incompressibilites, but which are constrained by the
empirical properties of isospin symmetric and asymmetric nuclear
matter, and the binding energies and radii of closed shell nuclei. The
properties of nuclei are investigated through Hartree-Fock-Bogoliubov
calculations for potential models and Hartree calculations for field
theoretical models.  In both categories, our calculations include
models that match the Akmal and Pandharipande~\cite{Akmal97} equation
of state (EOS) as results for nuclei with this model are not available
in the literature.  Additional insights are provided through
analytical and numerical analyses of isospin asymmetric semi-infinite
matter in the potential and field theoretical approaches.  The results
of these calculations are utilized in understanding several
correlations in conjunction with inferences from nuclear masses.  The
neutron skin-pressure and neutron star radius-pressure correlations
are reexamined to provide improved relations.  Our calculations also
provide an estimate of the smallest neutron star radii that ensue from
these models.

The presentation is organized as follows.  In Sec. II, the Hamiltonian
(Lagrangian) densities of the potential (field-theoretical) models
used in this work are described. The variational approach to determine
the bulk and surface properties of isospin asymmetric semi-infinite
nucleonic matter in these models is developed in Sec. III. This
section also contains a discussion of how these properties enter in
the liquid droplet model description of nuclei. Our results and
discussion of the various correlations, and their origins are
presented in Sec. IV. Section V contains a brief description of other
related correlations.  Discussion and conclusions are contained in
Sec. VI. Appendix A lists the properties of the models used in this
work. The coupling strengths of some newly-constructed models are
given in Appendix B.

\section{EFFECTIVE THEORIES}

In both potential and field-theoretical model descriptions of
many-body systems, the Hohenberg-Kohn-Sham theorem
\cite{Hohenberg64,Kohn65} allows the total Hamiltonian density to be
expressed in terms of the local particle number and kinetic energy
densities of the various species.  This simplification is of great
utility in the study of both heavy nuclei and infinite matter since
microscopic calculations, such as variational/Monte Carlo studies
starting from a Hamiltonian constructed on the basis of
nucleon-nucleon and many-nucleon interactions, are computer intensive
and are only beginning to be undertaken for heavy nuclei (for a review
of such studies on light nuclei, see Ref. \cite{Pieper01}).  In practice,
the strengths of the various interactions in the  density
functional approach are determined by appealing to available data. In
the following, we consider both potential and field theoretical
approaches so that distinguishing traits between these approaches can
be identified.

\subsection{Potential Models }

A commonly employed non-relativistic effective Hamiltonian density
stems from Skyrme's work \cite {Skyrme59} in which finite-range forces
between nucleons were approximated by effective zero-range forces.
For low relative momenta (i.e., with $s-$ and $p-$wave interactions
only), the resulting Hamiltonian density takes the form
\cite{Vautherin72}
\begin{eqnarray}
{\cal H} = {\cal H}_B + \thalf\left[Q_{nn}(\vec{\nabla} n_n)^2
+2Q_{np}\vec{\nabla} n_n \cdot
\vec{\nabla} n_p+Q_{pp}(\vec\nabla n_p)^2\right]
+ {\cal H}_C + {\cal H}_J
\,, \label{eq:basicham}
\end{eqnarray}
where the term associated with spatially homogeneous bulk matter is of the form
\begin{eqnarray}
{\cal H}_B &=& \frac{\hbar^2}{2 m_n} \tau_n + 
\frac{\hbar^2}{2 m_p} \tau_p + 
n \left(\tau_n + \tau_p \right) \left[ \frac{t_1}{4} 
\left( 1 + \frac{x_1}{2} \right)
+ \frac{t_2}{4} \left( 1 + \frac{x_2}{2} \right) \right] \nonumber \\
&& + \left( \tau_n n_n + \tau_p n_p \right) \left[ \frac{t_2}{4} 
\left( \frac{1}{2} + x_2 \right)
- \frac{t_1}{4} \left( \frac{1}{2} + x_1 \right) \right] \nonumber \\
&& + \frac{t_0}{2} 
\left[ \left( 1 + \frac{x_0}{2} \right) n^2 - 
\left( {\textstyle \frac{1}{2}} + x_0 \right) 
\left( n_n^2 + n_p^2 \right) \right]  \nonumber \\ 
&& + \frac{t_3}{12} \left[ \left( 1 + \frac{x_3}{2} \right) n^2 
- \left(\frac 12 + x_3\right)
\left(n_n^2 + n_p^2\right) \right]~n^\epsilon \,. 
\label{SkyH}
\end{eqnarray}
Above, $m_n$ and $m_p$ are the masses of the neutron and the proton
(these will often be taken to be equal and then denoted by $M$), $n_n$
and $n_p$ are the number densities of neutrons and protons, and $t_0$,
$t_1$, $t_2$, $t_3$, $x_0$, $x_1$, $x_2$, $x_3$, and $\epsilon$ are
parameters that give the strengths of the various potential
interactions.  We recall that total baryon density is $n=n_n+n_p$. 
The first two
terms represent the kinetic energy densities for the neutron and the
proton respectively, the sixth term is the density-dependent form of
the zero-range multi-body force, and the remaining terms constitute
the $s-$ and $p-$wave parts of the two-body interaction.

The coefficients $Q$ associated with the spatially varying part of the 
Hamiltonian density in Eq.~(\ref{eq:basicham}) are given by 
\begin{eqnarray}
Q_{nn}=Q_{pp} &=& {3\over16} 
\left[ t_1 \left(1 - x_1 \right) - t_2
\left( 1 + x_2 \right) \right]\,, \\
Q_{np}=Q_{pn} &=& {1\over8}
 \left[ 3 t_1 \left(1 + \frac{x_1}{2} 
\right) - t_2 \left( 1 + \frac{x_2}{2} \right) \right] \,.
\end{eqnarray}
Note that for a Skyrme-type force, the $Q$'s are constants.  However,
in general, it is possible for the $Q$'s to be density-dependent
functions.  Also note that, for Skyrme-type forces, $Q_{nn}=Q_{pp}$,
but this does not have to be true in general.  Generally, we have
$Q_{np}=Q_{pn}$ however.

The term ${\cal H}_C$ in Eq. (\ref{eq:basicham}) arising from the 
Coulomb interaction is
\begin{equation}
{\cal H}_C(\vec{r}\,)=\frac{e^2n_p(\vec{r}\,)}{2}\int d^3 r' 
\frac{n_p(\vec{r}^{\,\prime})}{|\vec{r}-\vec{r}^{\,\prime}|}
-\frac{3e^2}{4}\left(\frac{3}{\pi}\right)^{\!1/3}n_p(\vec{r}\,)^{4/3}\;,
\end{equation}
where $e$ denotes the proton's electric charge and 
the second term is an exchange correction \cite{Reinhard99}. 
The term ${\cal H}_J$ arising from the spin orbit interaction is
\begin{eqnarray}
{\cal H}_J&=&-\frac{W_0}{2}\left(n_n\vec{\nabla}\cdot\vec{J}_n
+n_p\vec{\nabla}\cdot\vec{J}_p+n\vec{\nabla}\cdot\vec{J}\,\right)\nonumber\\
&&\qquad\qquad +\frac{t_1}{16}\left(\vec{J}_n^{\,2}
+\vec{J}_p^{\,2}-x_1\vec{J}^{\,2}\right)
-\frac{t_2}{16}\left(\vec{J}_n^{\,2}+\vec{J}_p^{\,2}+x_2\vec{J}^{\,2}\right)\;,
\end{eqnarray}
where the neutron spin-orbit density 
$\vec{J}_n=\sum_i^{neutron}\psi_i^{\dagger}
\vec{\sigma}\times\vec{\nabla}\psi_i$, similarly for protons and 
$\vec{J}=\vec{J}_n+\vec{J}_p$.

The use of Skyrme's effective interactions has successfully reproduced
many of the global properties of nuclei including binding energies,
charge radii, etc. \cite{Blaizot80}.  The properties of bulk symmetric nuclear 
matter at
the saturation density $n_0$, can be expressed directly in terms of
$n_0$, and the model parameters. Explicitly, the energy per nucleon
$E/A$, the effective mass $M^{*}$, the pressure per particle $P/n$ and the
compressibility $K$ are given by
\def\thalf{{\textstyle{\frac{1}{2}}}}
\begin{eqnarray}
&&\frac{E}{A} = C n_0^{2/3} \left( 1 + \beta n_0 \right) + 
\frac{3 t_0}{8} n_0 + \frac{t_3}{16} n_0^{\epsilon+1}\;, \\
&&C = \frac{3 \hbar^2}{10 M} \left( \frac{3 \pi^2 }{2} \right)^{2/3}  
\quad;\quad\beta = \frac{M}{2\hbar^2} \left[ \frac{1}{4} 
\left( 3 t_1 + 5 t_2 \right)+ t_2 x_2 \right]\;, \\
&&M^{*}/M = \left(1+ \beta n_0 \right)^{-1} \;,\\
&& \frac{P(n_0)}{n_0} = 0 = \frac 23 C n_0^{2/3} 
\left(1 + \frac 52 \beta n_0\right) 
+ \frac 38 t_0 n_0 + \frac {t_3}{16} (\epsilon+1) n_0^{\epsilon+1}\;, \\
&&K =9n^2\frac{\partial^2E/A}{\partial n^2}\Bigg|_{n=n_0}
= - 2 C n_0^{2/3} + 10 C \beta n_0^{5/3} + \frac{9 t_3}{16} 
\epsilon \left( \epsilon+1 \right) n_0^{\epsilon+1} \;.
\end{eqnarray}
The symmetry energy of bulk matter, $E_{sym}$, is
\begin{eqnarray}
E_{sym}(n) = \frac{n^2}{2}\frac{\partial^2E/A}{\partial\alpha^2}
\Bigg|_{\alpha=0} &=&\frac{5}{9} C n^{2/3} + \frac{10 C M}{3 \hbar^2}
\left[ \frac{t_2}{6} \left(1 + \frac{5}{4} x_2 \right) - 
\frac{1}{8} t_1 x_1 \right] n^{5/3} \nonumber \\ 
&&- \frac{t_3}{24} 
\left({\textstyle \frac{1}{2}} + x_3 \right) n^{\epsilon+1} - 
\frac{t_0}{4} \left( {\textstyle \frac{1}{2}} + x_0 \right) n\;, 
\end{eqnarray}
where the isospin asymmetry density is denoted by $\alpha=n_n-n_p$.
The volume symmetry energy of equilibrium nuclear matter $E_{sym}(n_0)$
will be denoted $S_v$. We note that the symmetry energy asymptotically 
decreases at large densities if $\epsilon>2/3$ or 
$t_1x_1>t_2(4+5x_2)/3$. (This poor
high-density behavior can be remedied \cite{Prakash88}.) For 
models in which the symmetry energy monotonically increases with density,
the negative sign of the multi-body term involving $t_3$ generally
results in neutron star radii which are smaller than in field-theoretical 
models (see Sec. \ref{section:radii}).

\subsection{Field Theoretical Models}

The structure of nuclei and the properties of high-density nucleonic 
matter have also been studied utilizing the mean field
approximation to a Walecka-type Lagrangian that couples nucleons to
scalar ($\sigma$), vector ($\omega_\mu$), and vector-isovector
($\vec{\rho}_\mu$) mesons and photons ($A_{\mu}$) \cite{Muller96}. The
Lagrangian, supplemented by non-linear interactions, is 
\begin{eqnarray}
{\cal L} &=& \bar{\Psi} \left[ i \feyn{\partial} - 
g_{\omega} \feyn{\omega} - \thalf g_{\rho} 
\feyn{\vec{\rho}} \cdot 
\vec{\tau} - M + g_{\sigma} \sigma - \thalf e 
\left( 1 + \tau_3 \right) \feyn{A} \right] \Psi  
+ {\textstyle \frac{1}{2}} \left( \partial_{\mu} \sigma \right)^2 
\nonumber \\ &&
- V(\sigma) - {\textstyle \frac{1}{4}} f_{\mu \nu} f^{\mu \nu} 
+ {\textstyle \frac{1}{2}} m^2_{\omega}\omega^{\mu}\omega_{\mu} 
- {\textstyle \frac{1}{4}} \vec{B}_{\mu \nu} \cdot \vec{B}^{\mu \nu}
+ {\textstyle \frac{1}{2}} m^2_{\rho} \vec{\rho}^{\,\mu} \cdot 
\vec{\rho}_{\mu} 
- {\textstyle \frac{1}{4}} F_{\mu \nu} F^{\mu \nu} 
\nonumber \\ &&
+ \frac{\zeta}{24} g_{\omega}^4 \left(\omega^\mu \omega_\mu\right)^2
+ \frac{\xi}{24} g_{\rho}^4 \left(\vec{\rho}^{\,\mu} \cdot 
\vec{\rho}_{\mu}\right)^2 + g_{\rho}^2 f (\sigma, \omega_\mu\omega^\mu) 
\vec{\rho}^{\,\mu} \cdot \vec{\rho}_{\mu}\;, 
\label{FLTL}
\end{eqnarray}
where $\vec{\tau}$ are the SU(2) 
isospin matrices, $f_{\mu \nu} = \partial_{\mu} {\omega}_{\nu} -
\partial_{\nu} {\omega}_{\mu}$, $\vec{B}_{\mu \nu} = \partial_{\mu}
\vec{\rho}_{\nu} - \partial_{\nu} \vec{\rho}_{\mu}$, $F_{\mu \nu}
= \partial_{\mu} {A}_{\nu} - \partial_{\nu} {A}_{\mu}$ and the scalar 
meson potential
\begin{equation}
V(\sigma)=\thalf m_\sigma^2\sigma^2+\frac{\kappa}{6}(g_\sigma\sigma)^3
+\frac{\lambda}{24}(g_\sigma\sigma)^4\;. 
\end{equation}
In Sec. IV, we discuss the choice of the couplings $g_{\sigma}$,
$g_{\omega}$, $g_{\rho}$, $\kappa$ (dimensionful) and $\lambda$, as
well as $\zeta$ and $\xi$ that allow the high-density behavior of the
equation of state to be varied. The first five of these are
constrained to reproduce the properties of equilibrium nuclear matter
and the empirical symmetry energy. Horowitz and Piekarewicz
\cite{Horowitz01} have extended the non-linear Walecka model by adding
$\sigma^2\vec{\rho}^{\,\mu} \cdot \vec{\rho}_{\mu}$ and
$\omega_\mu\omega^\mu\vec{\rho}^{\,\mu} \cdot \vec{\rho}_{\mu}$ terms
to the Lagrangian so as to modify the density dependence of the
symmetry energy at supranuclear densities (in the absence of this
coupling the dependence is linear at large densities). In order to
provide additional freedom in varying the symmetry energy we have
adopted a general function $f$, namely
\begin{equation}
f (\sigma, \omega_\mu\omega^\mu) = \sum_{i=1}^{6} a_i \sigma^i 
+ \sum_{j=1}^{3} b_j \left(\omega_{\mu} \omega^{\mu}\right)^{j}\;.
\label{eq:ffun}
\end{equation}

\def\thalf{{\textstyle{\frac{1}{2}}}}
Utilizing the mean field approximation with the Lagrangian~(\ref{FLTL}), 
the binding energy per particle in symmetric matter at equilibrium
is given by
\begin{equation}
B\equiv M-\frac{E}{A} = M-\frac{1}{n_0} \left[V(\sigma_0)
+\thalf m_\omega\omega_0^2
+\frac{\zeta}{8}(g_\omega\omega_0)^4+\frac{2}{\pi^2}
\int\limits_0^{k_F} dk k^2\sqrt{k^2+M^{*2}} \right]\;,
\end{equation}
where the Dirac effective mass is  $M^{*} = M - g_{\sigma}\sigma_0$, and
the Fermi momentum and energy of symmetric nuclear matter are,
respectively,
\begin{equation}
k_F = \left( \frac{3 \pi^2 n}{2} \right)^{1/3} \,, \qquad
E_F^{*}=\sqrt{k_F^2+M^{* 2}} .
\end{equation}
The equilibrium pressure per particle is
\begin{equation}
\frac{P(n_0)}{n_0} = 0 = \frac{1}{n_0} \left[- V(\sigma_0)
+\thalf m_\omega\omega_0^2
+\frac{\zeta}{24}(g_\omega\omega_0)^4+\frac{2}{3\pi^2}
\int\limits_0^{k_F} dk \frac{k^4}{\sqrt{k^2+M^{*2}}} \right]\;,
\end{equation}
and the compressibility at equilibrium is given by
\begin{equation}
K=9n_0\left[\frac{m_\omega^2}{g_\omega^2}+\frac{\zeta}{2}
(g_\omega\omega_0)^2\right]^{-1}\!\!\! +3\frac{k_F^2}{E_F^*}
-9n_0\left(\frac{M^{*}}{E_F^{*}}\right)^{\!2}\left[\left(\frac{1}{g_\sigma^2}
\frac{\partial^2}{\partial\sigma_0^2}+\frac{3}{g_\sigma M^*}\frac{\partial}
{\partial\sigma_0}\right)V(\sigma_0)-3\frac{n_0}{E_F^*}\right]^{-1}\!\!\!.
\end{equation}
The symmetry energy of bulk matter is
\begin{eqnarray}
E_{sym}(n) = \frac{k_F^2}{6 E_F^{*}} + \frac{ n }
{8 \left(m_{\rho}^2/g_{\rho}^2 + 2 f (\sigma_0, \omega_0^2)  \right)}\;.
\label{rsym}
\end{eqnarray}
In the above equations, the subscript ``0'' denotes mean field values
of $\sigma$ and $\omega$.  For the case $f=0$, the symmetry energy 
in Eq.~(\ref{rsym}) varies linearly with the density at large densities. The 
presence of the function $f$ permits variations in this density dependence. 

\subsection{The EOS of Akmal-Pandharipande-Ravenhall}

The results of the variational microscopic calculations of Akmal and
Pandharipande \cite{Akmal97}, in which many-body and special
relativistic corrections are progressively incorporated into prior
models, have been parameterized by Akmal, Pandharipande and Ravenhall
(hereafter APR) \cite{Akmal98}.  The bulk effective Hamiltonian
density is
\begin{eqnarray}
{\cal H}_{B,APR} &=& \left( \frac{\hbar^2}{2 m} + 
\left( p_3 + \left( 1 - x \right) 
p_5 \right) n e^{-p_4 n} \right) \tau_n + 
\left( \frac{\hbar^2}{2 m} + \left( p_3 + x 
p_5 \right) n e^{-p_4 n} \right) \tau_p \nonumber \\ && + 
g_1 \left( 1 - \left( 1 - 2 x \right)^2 \right) +
g_2 \left( 1 - 2 x \right)^2 \,,
\label{APRH}
\end{eqnarray}
where $\tau_n$ and $\tau_p$ are the kinetic energy densities of the
neutrons and protons, respectively and $x=n_p/n$.  Contributions from 
potential interactions 
are contained in the functions $g_1$ and $g_2$ that have two
different forms corresponding to the low density phase (LDP) and the high
density phase (HDP) that contains a 
neutral pion condensed phase. Explicitly, 
\begin{eqnarray}
g_{1 LDP} &=& -n^2 \left( p_1 + p_2 n + p_6 n^2 + 
\left( p_{10} + p_{11} n \right) e^{-p_9^2 n^2} \right) \,,\\
g_{2 LDP} &=& -n^2 \left( p_{12}/n + p_7 + p_8 n + 
p_{13} e^{-p_9^2 n^2} \right) \,,\\
g_{1 HDP} &=& g_{1 LDP} -n^2 \left( p_{17} \left( n - p_{19} \right)
+ p_{21} \left( n - p_{19} \right)^2 e^{p_{18} \left( n - p_{19} \right) } 
\right)\,,\\
g_{2 HDP} &=& g_{2 LDP} -n^2 \left( p_{15} \left( n - p_{20} \right)
+ p_{14} \left( n - p_{20} \right)^2 e^{p_{16} \left( n - p_{20} \right)} 
\right) \,.
\end{eqnarray}
\begin{table}[t]
\begin{ruledtabular}
\caption{Coupling strengths for the Hamiltonian 
density of APR in Eq.~(\ref{APRH}). 
The dimensions are such that the Hamiltonian density 
is in MeV fm$^{-3}$.} 
\begin{tabular}{|rrrrrrrrrrr|}
$p_1$ &$p_2$ &$p_3$ &$p_4$ &$p_5$ &$p_6$ &$p_7$ &$p_8$ &$p_9$ &$p_{10}$
&$p_{11}$\\
 337.2 & $-382$& 89.8&0.457&$-59.0$&$-19.1$&214.6&$-384$&6.4&69&$-33$\\
$p_{12}$ &$p_{13}$ &$p_{14}$ &$p_{15}$ &$p_{16}$ &$p_{17}$ &$p_{18}$ &
$p_{19}$ &$p_{20}$ &$p_{21}$& \\
0.35&0&0&287&$-1.54$&175.0&$-1.45$&0.32&0.195&0&
\end{tabular}
\label{APRps}
\end{ruledtabular}
\end{table}
The numerical values of the various parameters $p_i$ for the recommended
``A18+$\delta v$+UIX$^{*}$'' EOS, that
incorporates three-nucleon interactions and relativistic boost
corrections, are collected in Table~\ref{APRps}. For this EOS, 
$n_0=0.16~{\rm fm}^{-3}$, $E/A=-16~{\rm MeV}$, $M^*/M=0.7$,
$K=266~{\rm MeV}$, and $E_{sym}(n_0)=32.6~{\rm MeV}$.

For later use the Q's for the EOS of APR, calculated
using the procedure detailed in Ref. \cite{Pethick95}, are
\begin{eqnarray}
Q_{nn} & = & \tquar e^{-p_4 n}
\left[ -6 p_5 - p_4 (p_3 - 2 p_5) (n_n + 2 n_p) \right]\;,
\nonumber \\
Q_{pp} & = & \tquar e^{-p_4 n}
\left[ -6 p_5 - p_4 (p_3 - 2 p_5) (n_p + 2 n_n) \right]\;,
\nonumber \\
Q_{np}=Q_{pn} & = & \teight e^{-p_4 n}
\left[ 4 (p_3 - 4 p_5) - 3 p_4 (p_3 - 2 p_5) (n_n + n_p)\right]\;.
\end{eqnarray}
Note that $Q_{nn}\ne Q_{pp}$ except in the case that $n_n=n_p$.

\begin{figure}
\begin{center}
\includegraphics[scale=0.40,angle=0]{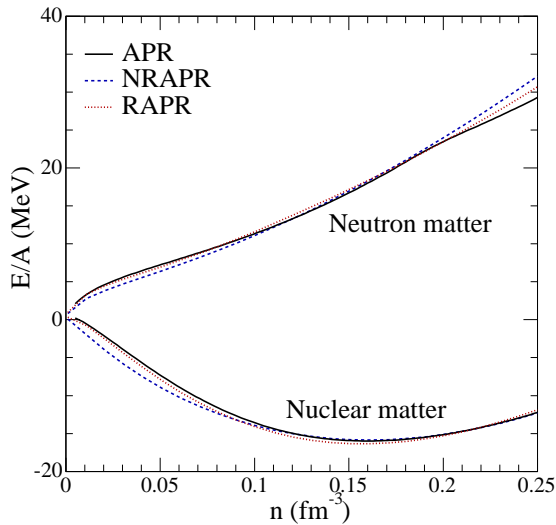}
\end{center}
\caption{Binding energy per particle versus density
in nuclear matter and neutron matter
for the EOS's of APR (solid line), NRAPR (dashed line), and RAPR
(dotted line).} 
\label{figapr}
\end{figure}

A Hartree-Fock calculation of finite nuclei using the bulk Hamiltonian
density in Eq.~(\ref{APRH}) augmented by gradient, Coulomb  
and spin-orbit terms
is more complicated than that using the Skyrme-like Hamiltonian in
Eq.~(\ref{SkyH}) because of terms that vary exponentially with density
in Eq.~(\ref{APRH}).  Therefore, we have fit the APR equation of state
for bulk nuclear and neutron matter up to 3/2 nuclear matter density 
to that of a Skyrme-like Hamiltonian.  
For a realistic description of nuclei, the
density dependence of the effective masses is important, so we have
constrained the fit so that the effective masses match as closely as
possible those of the APR equation of state. We have also adjusted the
coefficient of the spin-orbit interaction so as to closely match the
charge radii and binding energies of $^{208}$Pb, $^{90}$Zr, and
$^{40}$Ca. This procedure leads to the parameter set given in Table
\ref{Skyts} and is referred to as NRAPR below. Figure \ref{figapr}
shows that in matter a good fit to the exact APR binding 
energy/particle is obtained. A fit was also made using 
the field-theoretical approach, referred to as RAPR, and the parameters 
are tabulated in Table III. As shown in Fig. \ref{figapr} the fit  
is even a little better than in the non-relativistic case. 

\begin{table}[hbt]
\begin{ruledtabular}
\caption{Parameters for the Skyrme 
Hamiltonian obtained from a fit to the EOS given by the APR  
Hamiltonian, Eq.~(\ref{APRH}); the dimensions are such that ${\cal H}$
is in MeV fm$^{-3}$. This model is referred to as NRAPR in the
text.}
\begin{tabular}{|ccccc|}
$t_0$ & $t_1$ & $t_2$ & $t_3$ & $x_0$\\
$-2719.7$  & 
417.64  & 
$-66.687$  & 
15042  &
0.16154 \\
$x_1$ & $x_2$ & $x_3$ & $\epsilon$ & $W_0$ \\
$-$0.047986 & 
0.027170 & 
0.13611 & 
0.14416 & 
41.958  \\ 
\end{tabular}
\label{Skyts}
\end{ruledtabular}
\end{table}

\begin{table}[t]
\begin{ruledtabular}
\caption{Coupling strengths for the Lagrangian in Eq. (\ref{FLTL})
that fit the APR Hamiltonian density in Eq.~(\ref{APRH}) up to 3/2
nuclear matter density; the dimensions of $a_i$ and $b_i$ are such
that ${\cal L}$ is in MeV$^4$. This model is referred to as RAPR in
the text.}
\begin{tabular}{|cccccc|}
$m_{\sigma}$ & $g_{\sigma}$ & $g_{\omega}$ &
$g_{\rho}$ & $\kappa$ & $\lambda$ \\
494.28 MeV &
7.8144 &
9.2629 &
10.288 &
4.3789 MeV &
0.078205 \\
$\zeta$ & $\xi$ & $a_1$ & $a_2$ & $a_3$ & $a_4$ \\
0.065381 &
0.30524 &
0.18566 &
0.73588 &
7.5995 $\times 10^{-3}$ &
3.4620 $\times 10^{-4}$ \\
$a_5$ & $a_6$ & $b_1$ & $b_2$ & $b_3$ & \\
6.7880 $\times 10^{-6}$ &
5.4824 $\times 10^{-8}$ &
0.039169 &
3.1323 $\times 10^{-9}$ &
9.8907 $\times 10^{-10}$ & \\
\end{tabular}
\label{FLTs}
\end{ruledtabular}
\end{table}

It is
difficult to utilize these low-density fits to APR at densities of
relevance to neutron stars, largely because of the phase transition
that is present in APR. There is not enough freedom in either the
present potential model approach or the field-theoretical approach to
simultaneously match the equation of state at low densities and
satisfy the constraint on the maximum mass neutron star.
We have also attempted fitting the low-density  and high-density
phases of APR separately. Although this is somewhat difficult to accomplish
with the field-theoretical model above, it is easily done with the
potential model. However, this patchwork approach falls outside 
our goal of studying correlations that are generic to equations of
state without softening phase transitions. In what follows, therefore,
results for nuclei are calculated using the NRAPR and RAPR models, whereas
those for neutron stars are from calculations using the original EOS
of APR.

\section{SEMI-INFINITE NUCLEONIC MATTER}

\begin{figure} \begin{center}
\includegraphics[scale=0.5,angle=90]{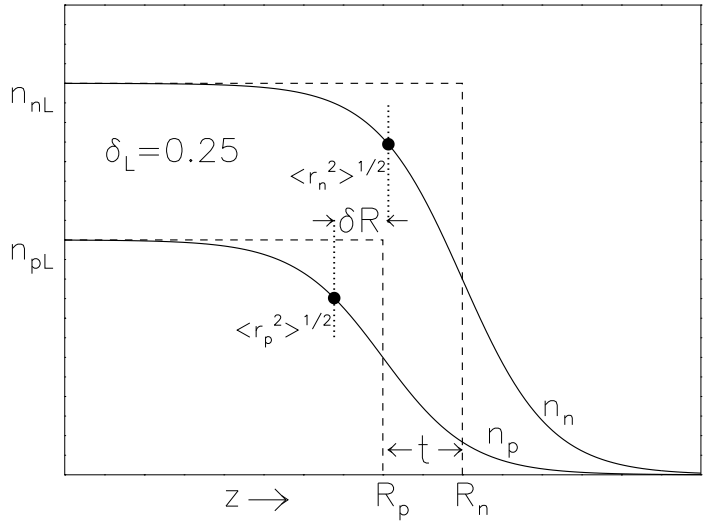}
\end{center} 
\caption{Schematic diagram of the surface regions of semi-infinite
matter indicating the relation between the neutron skin thickness,
$\delta R$, and the corresponding quantity $t=R_n-R_p$ determined from
the squared-off neutron and proton density distributions.}
\label{figt}
\end{figure}

The isospin dependence of the nuclear force has its most important
ramifications in the bulk and surface energies of nuclei.  In finite
nuclei, these effects are coupled, but certain properties like the
neutron skin thickness are affected more by the surface effects.  The
role of the surface is most easily ascertained by an investigation of
semi-infinite matter \cite{Boguta77,Stocker91}. In this case,
matter is assumed to vary only along one axis (the $z$-axis) and is
assumed to be uniform in the two remaining spatial directions.  Matter
at the $z \rightarrow -\infty$ limit is saturated matter at a
specified proton fraction, whereas the $z \rightarrow +\infty$ limit is
the vacuum (see Fig. \ref{figt}).  Matter in these two limits are in
chemical and pressure equilibrium with each other.  Thus, matter in
the $z\rightarrow-\infty$ limit is at the saturation density for the
reference proton fraction, with vanishing pressure.

We note that in sufficiently neutron-rich matter, such as neutron star
matter, the (non-relativistic) chemical potential for neutrons at the 
saturation density is positive.  In this situation, which 
occurs when the reference proton fraction is less than about 0.35, the 
pressure at the saturation density will also be positive and there must 
be a finite density of neutrons in the $z\rightarrow+\infty$ region with 
the same pressure and neutron chemical potential.

We will study how the surface energy and the neutron skin thickness
vary in the semi-infinite approximation
due to modifications of nuclear model parameters (including those
that delineate isospin effects).  These features can be connected to
experimental properties of laboratory nuclei.

\subsection{Variational Approach} 

The density profiles of semi-infinite  matter (schematically
illustrated in Fig.~\ref{figt})
are obtained by minimizing the Hamiltonian (energy) density subject to
the constraints of baryon number and charge conservation. The baryon
and isospin asymmetry densities, 
\begin{equation}\label{eq:den}
n=n_n+n_p\,, \qquad\alpha=n_n-n_p\,,  
\end{equation}
respectively (and fields and functions of them), are  $z$-dependent
even though we do not explicitly indicate them as such. The neutron
excess is defined by $\delta=\alpha/n$.  We will denote quantities at
$z=-\infty$ by the use of the subscript ``$L$''. Primes will generally
indicate derivatives with respect to the coordinate $z$.  Pressure
equilibrium between the dense nuclear phase and the vacuum requires that
the pressure $P_L=0$.  Therefore, the density $n_L$ is the saturation
density for matter with a neutron excess $\delta_L$.  The value of
$n_L$, the surface profiles $n_n$ and $n_p$, and the surface energy
(to be defined below), are all determined for a given nuclear force
model by the single quantity $\delta_L$.

We denote the saturation density of symmetric matter by
$n_0\equiv n_L(\delta_L=0)$.  The symmetry energy function $E_{sym}(n)$ is
generally a function of density and temperature $T$, but we consider only
$T=0$ matter in this paper. For isospin symmetric matter
$E_{sym}(n_0)\equiv S_v$ is the standard volume symmetry energy
parameter whose value lies in the range 25--35 MeV.  We also define the
squared-off neutron and proton radii (see Fig.~\ref{figt}), $R_n$ and $R_p$, by
\begin{eqnarray} n_{nL}(R_n-L) = \int\limits_{-L}^{\infty} n_n~d z \quad; 
\quad n_{pL}(R_p-L) = \int\limits_{-L}^{\infty} n_p~d z \,, 
\end{eqnarray}
where $L\rightarrow \infty$.  Although these integrals
formally diverge, the quantities $R_n$ and $R_p$ are finite.

The variation of the Hamiltonian density
\begin{eqnarray}
\delta~\int\limits_{-\infty}^{\infty} \left[ {\cal H} - \mu n - \mu_\alpha 
\alpha \right] d z =0
\label{eq:var}
\end{eqnarray}
results in the two Euler-Lagrange equations
\begin{eqnarray}
\frac{\partial {\cal H}}{ \partial n } - \mu & = & \frac{d}{d z}
\frac{\partial {\cal H}}{ \partial n^{\prime} } \,, \label{eq:elden} \\
\frac{\partial {\cal H}}{ \partial \alpha } - \mu_\alpha & = & 
\frac{d}{d z} \frac{\partial {\cal H}}
{ \partial \alpha^{\prime} } \,, \label{eq:elalpha}
\end{eqnarray}
where $2\mu=\mu_n+\mu_p$ and $2\mu_\alpha=\mu_n-\mu_p$ are the Lagrange
multipliers (equivalently, chemical potentials) associated with the
constraints of baryon number and charge conservation, respectively.
These chemical potentials are related to the individual neutron and
proton chemical potentials through $\mu n + \mu_\alpha
\alpha = \mu_n n_n + \mu_p n_p$.

For Hamiltonians of the form of Eq. (\ref{eq:basicham}),
without ${\cal H}_C$ and ${\cal H}_J$, we find
\begin{eqnarray}
\frac{\partial {\cal H_B}}{ \partial n } - \mu & = &
\frac{n^{\prime \prime}}{4} \left( Q_{nn}+2Q_{np}+Q_{pp} \right) +
\frac{{\alpha}^{\prime \prime}}{4}
\left(Q_{nn}-Q_{pp}\right) \nonumber \\
& & + {1\over8} \left[
2{\partial Q_{np}\over\partial n}
\left((n^\prime)^2+(\alpha^\prime)^2\right)
+ \left({\partial Q_{nn}\over\partial n}
+ {\partial Q_{pp}\over\partial n}\right)
\left((n^\prime)^2-(\alpha^\prime)^2\right) \right. \nonumber \\
& & \left. + 2 \left( {\partial Q_{nn}\over\partial \alpha}
+ 2 {\partial Q_{np}\over\partial \alpha}
+ {\partial Q_{pp}\over\partial \alpha} \right)
n^\prime \alpha^\prime
+ 2 \left( {\partial Q_{nn}\over\partial \alpha}
- {\partial Q_{pp}\over\partial \alpha} \right)
\left(\alpha^{\prime}\right)^2 \right] \,, \label{eq:elden2} \\
\frac{\partial {\cal H_B}}{ \partial \alpha } - \mu_{\alpha} & = &
\frac{{\alpha}^{\prime \prime}}{4}
\left( Q_{nn} - 2Q_{np} + Q_{pp} \right)
+ \frac{n^{\prime \prime}}{4}
\left(Q_{nn}-Q_{pp}\right) \nonumber \\
& & - {1\over8} \left[
2 {\partial Q_{np}\over\partial\alpha}
\left( (n^\prime)^2+(\alpha^\prime)^2\right)
+ \left( {\partial Q_{nn}\over\partial\alpha}
+ {\partial Q_{pp}\over\partial\alpha} \right)
\left((n^\prime)^2-(\alpha^\prime)^2\right) \right. \nonumber \\
& & \left. - 2 \left( {\partial Q_{nn}\over\partial n}
-2 {\partial Q_{np}\over\partial n}
+ {\partial Q_{pp}\over\partial n} \right)
n^{\prime} \alpha^{\prime}
- 2 \left( {\partial Q_{nn}\over\partial n}
- {\partial Q_{pp}\over\partial n} \right)
\left(n^{\prime}\right)^2 \right] \,. \label{eq:elalpha2}
\end{eqnarray}
Both sides of these
equations vanish for $z\rightarrow \pm \infty$.

The surface energy is the difference between the total energy and the
bulk energy of an equivalent number of neutrons and protons at the
saturation density.  Equivalently, the surface thermodynamical
potential per unit area is expressed \cite{Lattimer85} 
by the same integral that is
minimized in Eq. (\ref{eq:var}):
\begin{eqnarray}
\sigma = \int\limits_{-\infty}^{\infty} \left[ 
{\cal H} - \mu n - \mu_\alpha\alpha \right] d z \,.
\label{sigma}
\end{eqnarray}
The quantity $\sigma$ is also known as the surface tension.  The
integrand vanishes at $z\rightarrow \pm\infty$,  expressing the fact that
the pressure vanishes there.  
Expanding $\sigma(\delta_L)$ in terms of the neutron excess $\delta_L$  
\begin{equation}
\sigma(\delta_L) = \sigma(\delta_L=0)+\frac{\partial\sigma}
{\partial\delta_L^2}\Biggm\vert_{\delta_L=0}\delta_L^2 +\ldots
\equiv \sigma_0-\sigma_\delta\delta_L^2 +\ldots 
\label{eq:sigma_delta}
\end{equation}
allows us to identify $\sigma_\delta$ as the symmetry parameter of the
surface tension.  We note that the surface {\it energy} (which is the
product of surface tension and a characteristic surface area),
as distinct from the surface thermodynamic potential, has a surface
symmetry parameter identical to this, but of opposite sign
\cite{Lattimer85}.

The neutron skin thickness is defined by
\begin{eqnarray}
t = R_n-R_p = \int\limits_{-\infty}^{\infty} \left( \frac{ n_n }{n_{nL}} - 
\frac{ n_p }{n_{pL} }\right) d z \,. 
\label{eq:tdef}
\end{eqnarray}
Using the relations
\begin{eqnarray}
n_{n L} = {\textstyle \frac{1}{2}} n_L \left( 1 + \delta_L \right)
\quad;  \quad 
n_{p L} = {\textstyle \frac{1}{2}} n_L \left( 1 - \delta_L \right) 
\end{eqnarray}
the neutron skin thickness can be expressed in the equivalent form
\begin{eqnarray}
t = 
\frac{2\delta_L}{n_L(1-\delta_L^2)}\int\limits_{-\infty}^{\infty} n \left(
\frac{\delta}{\delta_L}-1 \right) dz \,. 
\label{eq:thick2}
\end{eqnarray}
Were the neutron and proton density distributions describable by Fermi
functions defined by half-density radii $z_{n,p}$ and diffuseness
parameters $a_{n,p}$:
\begin{eqnarray}
n_{n,p} = \frac{n_{(n,p)L}}{1 + e^{(z - z_{n,p})/a_{n,p}}}\,,
\end{eqnarray}
the skin thickness would be $t \cong z_n - z_p$, to lowest order in the
diffuseness parameters $a_{n,p}$. We note that the definition of skin
thickness might be problematic in finite nuclei where there are often
oscillations of the density profiles near the center. In this case,
the skin thickness is better expressed as the difference between the
neutron and proton rms radii ${\langle r_{n,p}^2\rangle}^{1/2}$.  To
lowest order in diffuseness corrections, one finds that $\delta
R={\langle r_n^2\rangle}^{1/2} - {\langle r_p^2\rangle }^{1/2} \cong
\sqrt{3\over 5}~t$.

\subsubsection{Potential Models}
The Euler-Lagrange equations (\ref{eq:elden2}) and (\ref{eq:elalpha2}),
multiplied respectively by $n^\prime$ and
$\alpha^\prime$, then added and integrated, yield
\begin{equation}
(Q_{nn}+2Q_{np}+Q_{pp}){(n^{\prime})^2\over8}+(Q_{nn}-2Q_{np}+Q_{pp})
{(\alpha^{\prime})^2\over8}
+(Q_{nn}-Q_{pp}){n^\prime\alpha^\prime\over4} = 
{\cal H}_B - \mu n - \mu_\alpha \alpha\,.
\label{eq:surfbulk}
\end{equation}
This equation holds even in the case in which the $Q$'s are
density-dependent.  The right-hand side of this equation is the bulk
thermodynamic potential density, $\Omega_B/V = -P_B$, where
$P_B=n^2[\partial({\cal H}_B/n)/\partial n]$ is the bulk pressure.
Note from Eq. (\ref{eq:surfbulk}) that the bulk and gradient terms
contribute equally to the surface tension in Eq. (\ref{sigma}), which
can therefore be written as
\begin{eqnarray}
\sigma (\delta_L) &=& 2\int\limits_{-\infty}^{\infty} \left[ 
{\cal H}_B - \mu n - \mu_\alpha\alpha\right] d z\,. \label{eq:fac2}
\label{eq:nrend}
\end{eqnarray}

To lowest order the bulk Hamiltonian should be quadratic in the neutron
excess $\delta$:
\begin{eqnarray}
{\cal H}_B = {\cal H}_B (n,\delta=0) + E_{sym}(n) n \delta^2\,.
\label{eq:quadham}
\end{eqnarray}
Since the bulk pressure vanishes for $z\rightarrow-\infty$, the chemical
potentials, to second order in $\delta_L$, are 
\begin{eqnarray}
\mu_n = -B + S_v \delta_L
\left( 2 - \delta_L \right)\,, \label{eq:chempot1} \\
\mu_p = -B - S_v
\delta_L \left( 2 + \delta_L \right)\,, \label{eq:chempot2}
\end{eqnarray}
where $B\equiv-{\cal H}_B(n_0,\delta=0)/n_0$ is the symmetric matter bulk
binding energy per particle. Therefore the surface tension becomes
\begin{eqnarray}
\sigma (\delta_L) = 2 \int\limits_{-\infty}^{\infty} 
\left[ {\cal H}_B(n,\delta=0) + n \left(B+ \delta^2 E_{sym}(n) +
\delta_L^2 S_v -2 \delta \delta_L S_v 
\right)\right] d z \,. \label{eq:surften}
\end{eqnarray}
We are interested in evaluating this in the case that $\delta_L$ is
small so that we can take $Q_{nn}=Q_{pp}$; this is exactly true for
standard Skyrme Hamiltonians and approximately so for the APR
Hamiltonian.  It is also reasonable in this case to use
Eq. (\ref{eq:surfbulk}) with $(\alpha^{\prime})^2$ set to zero which
results in a quadrature formula for the surface density profile:
\begin{equation}
\sqrt{Q_{nn}+Q_{np}}~dn = 2 \sqrt{{\cal H}_B - \mu n - \mu_\alpha
\alpha}~dz\,.
\label{eq:quad_profile}
\end{equation}
Utilizing this result, Eq. (\ref{eq:surften}) can now be written as 
\begin{equation}
\sigma (\delta_L) = \sqrt{Q_{nn}+Q_{np}}\int\limits_{0}^{n_L}
\left[ {\cal H}_B(n,\delta=0) + n\left(B +\delta^2 E_{sym}(n) +
\delta_L^2 S_v -2 \delta \delta_L S_v \right)\right]^{1/2} dn
\;.\label{eq:surften1}
\end{equation}
Again setting  $Q_{nn}=Q_{pp}$, neglecting derivatives of $\alpha$
and further assuming that the dependence of the $Q$'s on $\alpha$ can
be neglected, the right hand side of Eq. (\ref{eq:elalpha2}) is zero
so that
\begin{equation}
4 E_{sym}(n) \delta \simeq \mu_n - \mu_p\,. \label{eq:th2}
\end{equation}
Using Eqs. (\ref{eq:chempot1}) and  (\ref{eq:chempot2}) this gives
the useful result
\begin{equation}
\frac{\delta}{\delta_L} = \frac{S_v}{E_{sym}(n)} \;. \label{eq:deltas}
\end{equation}
(Note that the use of Eq.~(\ref{eq:th2}), which is strictly valid 
to order $\delta$ in bulk matter, requires that the variation of
$\delta$ in the surface tracks its behavior in the bulk.)  
Substitution of these relations in Eq. (\ref{eq:surften1}) yields the
quadrature
\begin{equation}
\sigma (\delta_L) = \sqrt{Q_{nn}+Q_{np}}\int\limits_{0}^{n_L}
\left[ {\cal H}_B(n,\delta=0) + nB -
nS_v\delta_L^2\left({S_v\over E_{sym}(n)}-1\right)\right]^{1/2} dn
\;.\label{eq:surften2}
\end{equation}
Then the surface tension of symmetric semi-infinite matter is
\begin{equation}
\sigma_0\equiv\sigma(\delta_L=0) = \sqrt{ Q_{nn} + Q_{np} }
\int\limits_0^{n_L} \left[ {\cal H}_B(n,\delta=0) + nB\right]^{1/2} dn\,,
\label{eq:w0}
\end{equation}
and expansion of the argument of the integral in Eq. (\ref{eq:surften2})
allows us to write the surface symmetry parameter in the convenient form
\begin{equation}
\sigma_{\delta} = {S_v\over2} \sqrt{ Q_{nn} + Q_{np} }
\int\limits_0^{n_L} n \left[ \frac{S_v}{E_{sym}(n)}-1 \right]
\left[ {\cal H}_B(n,\delta=0) + nB \right]^{-1/2} d n \,.
\label{eq:wd2}
\end{equation}
Our results will be given in terms of the surface energy 
$E_s=4\pi r_{0}^2\sigma_0$ and the surface symmetry energy
$S_s=4\pi r_{0}^2\sigma_\delta$, where 
$(4\pi r_0^3/3) (0.16~{\rm fm}^{-3}) =1$.
It is often useful to eliminate the $Q$'s by forming the ratio of Eqs. 
(\ref{eq:wd2}) and (\ref{eq:w0}). This yields
\begin{equation}
\frac{S_s}{S_v} = {E_s\over2}~~{ 
\int\limits_0^{n_L} n \left[S_v/E_{sym}(n)-1 \right]
\left[ {\cal H}_B(n,\delta=0) +nB \right]^{-1/2} d n\over 
\int\limits_0^{n_L} \left[ {\cal H}_B(n,\delta=0) +nB \right]^{1/2} dn
} \;.\label{eq:sssv1}
\end{equation}
This equation and Eqs. (\ref{eq:w0}) and (\ref{eq:wd2})
are central to the results to be presented in Sec. IV below.  
Finally, the profile equation (\ref{eq:surfbulk}) and the 
relation for $\delta$, Eq. (\ref{eq:deltas}), can be
employed in the expression (\ref{eq:thick2}) to yield the relation
\begin{equation}
t = \frac{1}{n_L}~\frac {2\delta_L}{(1-\delta_L^2)} ~\frac{\sigma_\delta}{S_v}
\label{eq:surften3}
\end{equation}
between the skin thickness and the surface tension that was conceived
in the droplet model \cite{Myers69,Krivine83,Lipparini82} and later
shown \cite{Krivine84} to be quite general.

The analytical description of the nuclear surface based on the
leptodermous expansion was pioneered by Myers and
Swiatecki~\cite{Myers69} and developed further in several 
works~\cite{Lipparini82,Krivine83,Krivine84,Lattimer85,Treiner86}.
Our analysis here offers a convenient expression for $S_s/S_v$ 
(Eq. (\ref{eq:sssv1}))
in terms of the energy density functional for isospin symmetric
infinite nucleonic matter.

\subsubsection{Field Theoretical Models}

For spatially non-uniform semi-infinite isospin asymmetric matter, 
the Euler-Lagrange equations corresponding to the Lagrangian (\ref{FLTL}) 
yield the meson and baryon field equations
\begin{eqnarray}
\sigma_0'' &=& m_{\sigma}^2 \sigma_0 - g_{\sigma} n_s +
\frac{\kappa}{2} g_{\sigma}^3 \sigma_0^2 + 
\frac{\lambda}{6} g_{\sigma}^4 \sigma_0^3 
-g_\rho^2\rho_0^2\frac{\partial f}{\partial\sigma_0}\;,\label{eq:field1} \\
\omega_0'' &=& m_{\omega}^2 \omega_0 - g_{\omega} n
+\frac{\zeta}{6}g_\omega^4\omega_0^3+g_\rho^2\rho_0^2
\frac{\partial f}{\partial\omega_0}\;,\label{eq:field2} \\ 
\rho_0'' &=& m_{\rho}^2 \rho_0 - \thalf g_{\rho} \alpha 
+2g_\rho^2\rho_0f+\frac{\xi}{6}g_\rho^4\rho_0^3\;,\label{eq:field3} \\
0 &=& \left( i \feyn{\partial} - g_{\omega} \omega_0 \gamma_0 + 
{\textstyle \frac{1}{2}} g_{\rho} \rho_0 \gamma_0 - M + g_{\sigma}
\sigma_0 \right) \psi_{n} \;,\\ 
0 &=& \left( i \feyn{\partial} - g_{\omega} \omega_0 \gamma_0 - 
{\textstyle \frac{1}{2}} g_{\rho} \rho_0 \gamma_0 - M + g_{\sigma}
\sigma_0 \right) \psi_{p} \;,
\end{eqnarray}
where primes indicate derivatives with respect to $z$, 
the subscript ``$0$'' indicates that the  mean field
approximation has been used and $\rho_0$ refers to the mean field of the 
neutral rho meson. We shall adopt the Thomas-Fermi approximation 
since it provides a reasonably accurate simplification of the Hartree 
approach \cite{Boguta77}. Here the meson fields 
are assumed to vary sufficiently slowly that the nucleons can be regarded 
as moving in locally constant fields at every point in space. Thus the 
Fermi momenta and the effective mass depend upon $z$, i.e., $k_{F_n}(z)$, 
$k_{F_p}(z)$ and $M^*(z)$. The local neutron, proton and scalar densities 
at a given $z$ are then
\begin{eqnarray}
n_n &=& \frac{k_{F_n}^3}{3\pi^2}\ \ ;\ \  n_p=\frac{k_{F_p}^3}{3\pi^2}\;,
 \nonumber \\ n_s &=& 2  \int\limits_0^{k_{F_n}} \frac {d^3k}{(2\pi)^3}\, 
\frac {M^*} { \sqrt{k^2 + M^{* 2}} }  
+ 2  \int\limits_0^{k_{F_p}} \frac {d^3k}{(2\pi)^3}\, 
\frac {M^*} { \sqrt{k^2 + M^{* 2}} } \;, 
\end{eqnarray}
respectively. The energy eigenvalues of the Dirac equations for the 
neutron and proton fields are
\begin{equation}
e_n \left(k\right) = \sqrt{k^2 + M^{* 2}} + g_{\omega} \omega_0 -
{\textstyle \frac{1}{2}} g_{\rho} \rho_0\quad;\quad
e_p \left(k\right) = \sqrt{k^2 + M^{* 2}} + g_{\omega} \omega_0 +
{\textstyle \frac{1}{2}} g_{\rho} \rho_0 \;,
\end{equation}
respectively. The local Thomas-Fermi Hamiltonian density is then 
\begin{eqnarray}
{\cal H} (z) &=& \frac{1}{2} \left( \sigma_0^{\prime 2} + 
m_{\sigma}^2 \sigma_0^2 \right)  - \frac{1}{2} \left( \omega_0^{\prime 2} 
+ m_{\omega}^2 \omega_0^2 \right) - \frac{1}{2}
\left( \rho_0^{\prime 2} + m_{\rho}^2 \rho_0^2 \right) 
+ \frac{\kappa}{6} \left( g_{\sigma} \sigma_0\right)^3 
+ \frac{\lambda}{24} \left( g_{\sigma} \sigma_0 \right)^4\nonumber \\
&& -\frac{\zeta}{24}(g_\omega\omega_0)^4-\frac{\xi}{24}(g_\rho\rho_0)^4
-g_\rho^2\rho_0^2f + 2 \int\limits_0^{k_{F_n}}
\frac{d^3 k}{\left( 2 \pi \right)^3}~e_n + 2 \int\limits_0^{k_{F_p}} 
\frac{d^3k}{\left( 2 \pi \right)^3}~e_p\;.
\label{eq:rhamz}
\end{eqnarray}
Note that, in contrast to the nonrelativistic case, the energy density 
is not simply expressible in terms of just the baryon densities and their 
gradients. The ground state of the system is obtained by minimizing 
${\cal H} (z)-\mu_nn_n-\mu_pn_p$ with respect to the local Fermi 
wavenumbers $k_{F_n}(z)$ and $k_{F_p}(z)$. The Lagrange multipliers 
(chemical potentials) that fix the number of neutrons and protons
are then 
\begin{equation}
\mu_n=e_n(k_{F_n}) \qquad;\qquad  \mu_p=e_p(k_{F_p})\;. \label{eq:field4}
\end{equation}
The algebraic nature of these two coupled equations relating the neutron
and proton density profiles to the meson fields stems from the Thomas-Fermi
approximation. 

In order to obtain the density profiles $n_n(z)$ and $n_p(z)$,
Eqs.~(\ref{eq:field1})--(\ref{eq:field3}) and (\ref{eq:field4}) are 
supplemented by the boundary conditions
\begin{eqnarray}
 \left. \begin{array}{ccc}
\sigma_0(-L) =\sigma_{0L}\,, \quad 
&\omega_0(-L)=\omega_{0L}\,, \quad
&\rho_0(-L)=\rho_{0L}\,,  
\nonumber \\
\sigma_0(L)=0\,, \quad
&\omega_0(L)=0\,, \quad
&\rho_0(L)=0\,, 
\end{array}
\right\} L \rightarrow \infty
\end{eqnarray}
where the subscript ``$0L$'' indicates values of the fields in
equilibrium at the specified neutron excess $\delta_L$.  
(Note that the meson fields are finite, but
exponentially small, in regions of space where the baryon density
vanishes as prescribed by the Thomas-Fermi approximation.)
These conditions guarantee that the derivatives of the fields vanish at
the boundaries. In practice, we impose these conditions at points
$(-L,L)$ that are sufficiently far from the surface to ensure that the
surface energy is unchanged by small variations in $L$. 

The numerical solution of the differential equations 
requires some care because of their behavior at the boundaries. 
We find it best to use a relaxation method. Starting with small values 
for the field derivatives, we integrate from $-L$ outwards. The
integration has to be stopped when one of the fields becomes negative,
say at $L_1>-L$.  This solution can then be used as an initial guess 
for relaxation in the region $(-L_2,L_2)$, where $L_2=(L_1+L)/2$. 
After convergence is obtained, $L_2$ can be increased by a small amount 
and the relaxation applied again until the field derivatives are 
sufficiently small to ensure that the surface tension remains stationary.

The determination of the density profiles enables 
the calculation of the surface tension. 
In the non-relativistic case Eq.~(\ref{eq:surfbulk}) showed that the bulk and 
gradient terms in the Hamiltonian density gave equal contributions to
the surface tension. 
A similar result can be obtained here by multiplying 
Eqs.~(\ref{eq:field1})--(\ref{eq:field3}) by the derivative of the 
appropriate field with respect to $z$, combining the equations and 
integrating. This gives
\begin{equation}
{\cal H}_B - \mu n - \mu_\alpha \alpha =
{\textstyle \frac{1}{2}} 
\left( \sigma_0^{\prime~2} - \omega_0^{\prime~2} - \rho_0^{\prime~2} \right)
\;,\label{eq:relbs}
\end{equation}
where ${\cal H}_B$ denotes the bulk Hamiltonian, namely the
non-derivative terms in Eq. (\ref{eq:rhamz}). Thus we can employ
Eq. (\ref{eq:nrend}) to calculate the surface tension here also.
However, it is not possible to reduce the calculation of the surface
energy to a simple quadrature over the density since here derivatives
of three fields are involved. Note that in the relativistic case the
$Q$'s are the coefficients of the field gradients with
\begin{eqnarray}
Q_{\sigma \sigma} = - Q_{\omega \omega} = - Q_{\rho \rho} = 1\;,
\end{eqnarray}
all other coefficients being zero.  

\subsection{Relationship to the Liquid Droplet Model of Nuclei}

The empirical liquid droplet approach, originally formulated in 
Ref.~\cite{Myers69}, provides a useful framework for the description of
nuclei in the equation of state relevant for astrophysical simulations
of supernovae and neutron stars.  The droplet approach for isolated
nuclei can be easily extended to the case in which nuclei are immersed
in a dense medium comprised of electrons, positrons, protons, neutrons
and alpha particles \cite{Lattimer85}.  Historically, the parameters
of the droplet model have been established from fits to binding energy
data for laboratory nuclei (cf., Ref. \cite{Myers69}).
Although such fits cannot independently establish reliable values for the
volume and surface symmetry coefficients, they can determine a strong
correlation between them.  This correlation, however, depends on the
formulation of the droplet model.

The simplest droplet model consists of an extension of the Bethe-von
Weizs\"acker liquid drop model to incorporate the surface asymmetry.  
As demonstrated by Myers and Swiatecki~\cite{Myers69}, 
the nuclear energy can be written as
\begin{eqnarray}\label{drop}
E(A,Z)&=&-BA+E_sA^{2/3}+S_vA{(1-2Z/A)^2\over1+(S_s^*A^{-1/3}/S_v)}+
 E_C{Z^2\over A^{1/3}}\cr
&&+E_{dif}{Z^2\over A}+E_{ex}{Z^{4/3}\over A^{1/3}}+a\Delta A^{-1/2}.
\end{eqnarray}
In this expression, $B\simeq16$ MeV is the binding energy 
per particle of bulk
symmetric matter at saturation, $E_s, E_C, E_{dif}$ and $E_{ex}$ are
coefficients for the surface energy of symmetric matter, the Coulomb
energy of a uniformly charged sphere, the diffuseness correction and
the exchange correction to the Coulomb energy, respectively.  The last
term represents pairing corrections, where $\Delta$ is a constant and
$a=+1$ for odd-odd nuclei, 0 for odd-even nuclei, and $-1$ for
even-even nuclei.  The effects of curvature and higher-order terms are
neglected.  The quantity $S_S^*$ is related to $S_s$ as described below.

Care must be taken in making the assumption that the surface energy
scales as $A^{2/3}$ and the Coulomb energy as $Z^2A^{-1/3}$.
Physically, the Coulomb energy scales as $Z^2/R_p$.  The surface
energy scales with the surface area, but its definition depends on the
exact radius of the surface and is therefore ambiguous \cite{Lattimer85}.
Nevertheless, the total of the bulk plus surface energies is
unambiguous.  This ambiguous splitting of the bulk and surface
energies can be resolved if the surface thermodynamic potential is
used instead of the surface energy.  
In what follows, we describe two ways in which this can be done. 

\subsubsection{``$\mu_n$'' Approach}

In this approach developed by Lattimer et al. \cite{Lattimer85}, $R_p$
is used as the reference surface.  This choice was made since the
Coulomb energy is naturally expressed in terms of this radius.  Since,
in general, $R_n>R_p$, the bulk energy of an additional $N_s$ neutrons
in the neutron skin has to be included. The diffuseness and Coulomb
exchange corrections are also naturally expressed in terms of $R_p$.
These definitions allow the mass formula to be simply written in terms
of the variables $A$ and $I=1-2Z/A$. However, the connection between
$R_p$ and $A$ depends on the density (assumed to be uniform and equal
to $n_L$) of the bulk interior fluid.  This results in the modified
surface symmetry energy parameter
\begin{equation}
S_s^* = 4 \pi \left(\frac{3}{4 \pi n_0}\right)^{2/3} \sigma_{\delta}
\end{equation}
where the true saturation density $n_0$ is used in place of the
fiducial density of 0.16 fm$^{-3}$.
Different parameterizations of the nuclear force predict 
different values for the saturation density, and consequently for the
interior densities $n_L$ in nuclei.  Using $R$ to refer to $R_p$ in
this approach, the total droplet energy becomes
\begin{eqnarray}\label{eq:liqdrop}
E(A,Z)=(-B&+&S_v\delta_L^2)(A-N_s) + 4\pi R^2\sigma(\mu_n^s)
+\mu_n^s N_s  \nonumber \\
&+&{3Z^2e^2\over5R}-{\pi^2Z^2e^2d^2\over2R^3}-
{3Z^{4/3}e^2\over4R}\left({3\over2\pi}\right)^{2/3}+a\Delta A^{-1/2} \,,
\end{eqnarray}                                    
where explicit expressions for $E_C, E_{dif}$ and $E_{ex}$ have been
 substituted.  The quantity $d \approx 0.55$~fm is the surface 
 diffuseness parameter.  Note that the bulk energy per baryon in the
 interior, expanded to second order in $\delta_L$, is
 $-B+S_v\delta_L^2$, and that the total surface energy is $4\pi
 R^2\sigma(\mu_n^s)+\mu_n^s N_s$, where the surface thermodynamic
 potential per unit area $\sigma$ is a function of the surface neutron
 chemical potential $\mu_n^s$.  Technically, the surface tension is a
 function of both $\mu_n$ and $\mu_p$, but if Coulomb and other
 finite-size effects are neglected, the matter pressure vanishes
 throughout the surface and $\mu_n$ and $\mu_p$ can each be expressed
 as a unique function of $\delta_L$.  Thus, the surface tension in
 this limit can be expressed as a function of only one chemical
 potential.  Since by definition only neutrons exist in the neutron
 skin, it is appropriate to take the surface tension as a function of
 $\mu_n$.  Minimizing the total energy with respect to $\mu_n^s$
 requires that $\mu_n$ be the same in the surface as in the bulk
 interior, and its value is thus given by Eq. (\ref{eq:chempot1}).
 The function $\sigma(\mu_n^s)$ is determined through the
 auxiliary variable $\delta_L$ using
 $\sigma=\sigma_0-\sigma_\delta\delta_L^2$.  The number of surface
 neutrons follows from minimizing the total energy with respect to
 $N_s$ and leads to the thermodynamic identity
\begin{equation}
N_s=-4\pi R^2{\partial\sigma(\mu_n)\over\partial\mu_n}=
4\pi R^2{\sigma_\delta\delta_L\over S_v(1-\delta_L)}\,.
\label{eq:ns1}
\end{equation}
Noting that the local asymmetry in the bulk interior is
\begin{equation}\label{eq:localas}
\delta_L={A-N_s-2Z\over A-N_s},
\end{equation}
one has that $N_s=A(I-\delta_L)/(1-\delta_L)$, where the global asymmetry is
$I=(N-Z)/A$.  Thus, it is possible to establish that
\begin{equation}\label{eq:jdelta}
\delta_L=I\left[1+{4\pi R^2\sigma_\delta\over S_vA}\right]^{-1}\,.
\end{equation}

In order to express the modified liquid droplet energy in terms of
$A$ and $I$, it is necessary to relate the nuclear proton radius $R$ to $A$,
the interior density $n_L$ and the asymmetry $I$.  This is
accomplished with the implicit relation 
\begin{equation}
\frac{4\pi}{3} R^3n_L=A-N_s=A{1-I\over1-\delta_L}=
A(1-I)\left[1-{AI\over A+4\pi R^2\sigma_\delta/S_v}\right]^{-1}.
\label{eq:implicit}
\end{equation}
This is a quintic equation for $R$.  It is convenient to write its 
solution in terms of a function $v(A,I,n_L)$ such that
\begin{equation}
R=r_{0}[A v(A,I,n_L)(0.16{\rm~fm}^{-3})n_L^{-1}]^{1/3}\;,
\end{equation}
where, as before, $(4\pi r_{0}^3/3)(0.16{\rm~fm}^{-3})=1$.  We also define
$S_{s}=4\pi r_{0}^2\sigma_\delta$ to provide a measure of the
surface symmetry parameter that is independent of the interior
density, unlike $S_s^*$ in the simple liquid droplet
model. In the case of symmetric matter, the quantity
$v(A,I,n_L)$ would simply equal unity; for asymmetric matter, it also
contains the effects of the neutron skin.  The implicit equation for
$v$ becomes
\begin{equation}
(v-1)(1+\alpha v^{2/3}-I)=-\alpha I v^{2/3}\,,
\label{eq:implicit1}
\end{equation}
where
\begin{equation}\label{eq:alpha}
\alpha={S_{s}\over S_vA^{1/3}}
\left({0.16{\rm~fm}^{-3}\over n_L}\right)^{2/3}\,.
\end{equation}
In practice, solving the quintic
equation for $v$ can be avoided by employing its expansion
\begin{equation}\label{eq:v}
v=1-{\alpha\over1+\alpha}I-{\alpha\over(1+\alpha)^3}
\left(1+{1\over3}\alpha\right)I^2-{\alpha\over(1+\alpha)^5}
\left(1+{2\over3}\alpha+{2\over3}\alpha^2+{1\over9}\alpha^3\right)I^3+\dots
\end{equation}
in terms of the small quantity $I$.
Eliminating $N_s, \delta_L$ and $R$, and setting $w=v(0.16{\rm~fm}^{-3})/n_L$,
%$E_{s,0.16}=4\pi r_{0.16}^2\sigma_0$, 
the liquid droplet mass formula becomes
\begin{eqnarray}
E(A,Z) = &-&BA + 4\pi r_{0}^2\sigma_0(wA)^{2/3}
+S_vAI^2\left[1+{S_{s}w^{2/3}\over 
S_v A^{1/3}}\right]^{-1} 
+{3e^2Z^2\over5r_{0}}\left({1\over wA}\right)^{1/3}\nonumber \\
&-&{\pi^2Z^2e^2d^2\over2r_{0}^3wA}-
{3Z^{4/3}e^2\over4r_{0}}\left({3\over2\pi}\right)^{2/3}
\left({1\over wA}\right)^{1/3}+a\Delta A^{-1/2}.\label{eq:liqdrop1}
\end{eqnarray}                                      
For a given $A$ and $Z$, therefore, the modified liquid droplet energy
is a function of the six parameters $B, \sigma_0, S_v, S_{s}, n_L$
and $\Delta$.  The neutron skin thickness is given by
\begin{eqnarray}
t &=& \frac{2}{3} \frac{r_{0}w}{v} \frac{S_{s}}{S_v} 
\frac{\delta_L}{1-\delta_L^2}\, \quad {\rm with} \quad
\delta_L=I\left[1+{S_{s}w^{2/3}\over S_v A^{1/3}}\right]^{-1}\,.
\label{eq:rnrp_nuclei} 
\end{eqnarray}
A shortcoming of the droplet model described above is it employs the
surface tension along the coexistence curve for which the pressure 
is zero and Coulomb effects are ignored.  Thus, in a nucleus with
$N=Z$, it predicts that there should be no neutron skin.  However, the
Coulomb repulsion of protons results in a lowering of the proton
density inside the nucleus and the development of a proton skin.  The
following section describes the approach adopted by Danielewicz 
\cite{Danielewicz03} to incorporate this Coulomb effect in the droplet
model.

\subsubsection{``$\mu_\alpha$'' Approach}

In this approach, the surface tension is parameterized in terms of 
$\mu_\alpha\equiv (\mu_n-\mu_p)/2$ instead of $\mu_n$. The quantity 
$\mu_\alpha$
is antisymmetric in neutrons and protons.  In this way, the physical
feature that the proton chemical potentials in the two bulk phases
across a boundary differ is captured.  Since neutrons and protons are
treated symmetrically, the surface term $\mu_n^s N_s$ in the mass 
formula (\ref{eq:liqdrop}) is replaced by
\begin{eqnarray}
(\mu_n^s+\mu_p^s)(N_s+Z_s)/2+(\mu_n^s-\mu_p^s)(N_s-Z_s)/2\,.
\end{eqnarray}
The surface thermodynamic potential is to be treated as
function of $\mu_\alpha^s$ alone.  A result of
these definitions is that minimization of the total energy with respect to
$\mu_n^s-\mu_p^s$ and $\mu_n^s+\mu_p^s$ results in the surface
asymmetry density being proportional to
$N_s-Z_s\propto\partial\sigma/\partial\mu_\alpha^s$, whereas the total
surface density is proportional to
$N_s+Z_s\propto\partial\sigma/\partial\mu^s$ and will vanish.  Thus,
$Z_s=-N_s$ in this approach.  In addition, the nuclear radius now
satisfies $4\pi R^3 n_L/3=A$.

The possibility that $Z_s$ protons exist in the surface must be taken
into account in the Coulomb energy.  
Then the Coulomb energy becomes
\begin{equation}
{3e^2\over5R}\left[Z^2-{ZZ_s\over3}+\cdots\right]\,,
\label{newcoul}
\end{equation}
to which diffuseness and Coulomb exchange corrections must be added.
We note that Danielewicz \cite{Danielewicz03} kept an additional term 
in the expansion in
$Z_s/Z$, but this has only a marginal effect on the results.
Minimizing the total energy 
\begin{eqnarray}\label{eq:ealpha}
E(A,Z)=&(&-B+S_v\delta_L^2)(A-N_s-Z_s) + 4\pi R^2\sigma(\mu_\alpha^s)
+\mu^s (N_s+Z_s)+\mu_\alpha^s (N_s-Z_s)  \nonumber \\
&+&{3Z^2e^2\over5R}\left(1-{Z_s\over3Z}\right)-{\pi^2Z^2e^2d^2\over2R^3}-
{3Z^{4/3}e^2\over4R}\left({3\over2\pi}\right)^{2/3}+a\Delta A^{-1/2} \,,
\end{eqnarray}                                    
with respect to $N_s-Z_s$ results in
\begin{equation}
\mu_\alpha^s=2S_v\delta_L-{e^2Z\over10R}\,.
\end{equation}
Using 
\begin{equation}
\sigma=\sigma_0-\sigma_\delta(\mu_\alpha^s/2S_v)^2\,,
\end{equation}
one finds that
\begin{eqnarray}
N_s-Z_s&=&4\pi R^2\sigma_\delta{\mu_\alpha^s\over2S_v^2}=
\alpha A(\delta_L-\beta)\nonumber\\
&=&A(I-\delta_L)=\alpha A{I-\beta\over1+\alpha}\,,\\
\delta_L&=&{I+\alpha\beta\over1+\alpha}\,,
\end{eqnarray}
where $\alpha=4\pi R^2\sigma_\delta/(S_vA)$ has the same definition as in
Eq. (\ref{eq:alpha}) and $\beta=e^2Z/(20 RS_v)$.
The neutron skin thickness is
\begin{equation}\label{eq:tmua}
t=R_n-R_p={2\over3}R{N_s-Z_s\over A(1-\delta_L^2)}=
{2\over3}{r_0\over u_0}{S_s\over S_v}{\delta_L-\beta\over1-\delta_L^2}\,,
\end{equation}
where $u_0=n_L/0.16{\rm~fm}^{-3}$. This  shows explicitly how the Coulomb 
repulsion reduces the skin thickness.
Finally, the total energy can be written as
\begin{eqnarray}\label{eq:emua}
E(A,Z) &=& -BA + 4\pi r_{0}^2\sigma_0 \left({A\over u_0}\right)^{2/3}
+S_vA{I^2+\alpha\beta(2I-\beta)\over1+\alpha}\nonumber\\
&&+{3e^2Z^2\over5r_{0}}\left({u_0\over A}\right)^{1/3}
\left[ 1 - {5\pi^2d^2\over6r_{0}^2} \left({u_0\over A}\right)^{2/3} - 
{5\over4}\left({3\over2\pi Z}\right)^{2/3}\right] 
 + a\Delta A^{-1/2}\,.
\end{eqnarray}

\section{RESULTS}

\subsection{Selection of Models and Their Parameters} 
\label{section:select}

In this section, we wish to establish some generic trends that emerge
from calculations based on the potential model Hamiltonian in
Eq.~(\ref{SkyH}) and the field theoretical Lagrangian in
Eq.~(\ref{FLTL}). Toward this end, the various coupling strengths that
enter in these two approaches are chosen so as to reproduce the
empirical properties of 
\begin{eqnarray}
{\rm equilibrium~binding~energy}~: B &=& -16\pm 1~{\rm MeV}\,, 
\nonumber \\ 
{\rm equilibrium~density}~: n_0 &=&
0.16\pm 0.01~{\rm fm}^{-3}\,, 
\nonumber \\
{\rm incompressibility}~: K&=&(200-300)~{\rm  MeV}\,,
\nonumber \\
{\rm Landau~effective~mass}~: m_L^* &=& (0.6-1.0)~M \,, \quad {\rm and} 
\nonumber \\ 
{\rm symmetry~energy}~: S_v&=& (25-35)~{\rm MeV}\,.
 \end{eqnarray}

In the case of potential models, calculations of nuclei are performed
using the Hartree-Fock-Bogoliubov approach \cite{Reinhard99} that
includes pairing interactions. Hartree calculations \cite{Horowitz81}
are employed in the field theoretical approach, as a treatment of
the exchange (Fock) terms is considerably more complicated than in the
potential model approach.  In both approaches, we require that the
binding energy and the charge radii of closed-shell nuclei are
reproduced to within $2\%$ of the measured values.  The scalar meson
mass in the field theoretical approach was restricted to lie between
450 and 550 MeV.  Insofar as fits to the valence single particle
energies of closed shell nuclei are known to require a detailed
treatment of correlations beyond the mean field level 
(such as short range and RPA correlations~\cite{Foris87} that are 
not considered here) we have
chosen to tolerate slight deviations from the measured energies.
We note, however, that 
Todd and Piekarewicz~\cite{Todd03} have discussed a correlation
between the neutron radius of lead and the binding energy of the
valence orbitals. They found that 
smaller neutron radii lead to
last occupied neutron orbitals that are more weakly bound,
and therefore smaller neutron drip densities. 

In all cases considered the supranuclear EOS was constrained to
yield a maximum neutron star mass of at least $1.44{\rm M}_\odot$, the
larger of the accurately measured neutron star masses in the double
neutron star binary PSR1913+16 (see Ref.~\cite{Lattimer04} for a
compilation of known masses).  This constraint resulted in
the elimination of several non-relativistic models in which either 
the pressure decreased as a function of increasing density
(e.g. SkP~\cite{Dobaczewski84}) or the proton (and hence the electron)
fraction vanished at a finite supranuclear density
(e.g. SkT3~\cite{Tondeur84}).  In the latter case, the condition of
beta-equilibrium requires positrons and anti-protons which potential
models cannot account for in a natural manner.  Recently, a systematic
study of potential models has been performed by Stone et al. 
\cite{Stone03} who adopt a similar procedure to select among 87
Skyrme parameterizations.  In addition to imposing the constraints in 
Ref. \cite{Stone03} for potential models, we have also removed all models
that do not have reasonable values for the Landau parameters at
saturation density \cite{Margueron02}.

Since few relativistic models meet our criteria, we have generated a new class
of relativistic models, es25, es275, es30, es325 and es35 with the
Lagrangian of Eq. (\ref{FLTL}). In these models only $a_2$ and $b_1$
in Eq. (\ref{eq:ffun}) were allowed to be non-zero. They were designed to 
meet our restrictions while offering some variation in the value of the
symmetry energy. We have also exploited the full freedom of the function 
$f$ in Eq. (\ref{eq:ffun}) to produce models SR1, SR2 and SR3 for which 
the symmetry energy has a substantially weaker density dependence than 
in typical relativistic models. This leads to a smaller skin thickness. 
Note that es25 and es275 also have smaller skin thicknesses than typical.

A list of all of the models used in this work together
with their saturation properties, surface energies, and skin
thicknesses is given in Table \ref{tab:modelprop} in Appendix A. 
In Appendix B the coupling strengths of models SR2, es25, es30, and es35 are 
listed in Tables \ref{SR2prop}, \ref{es25prop}, \ref{es30prop}, 
and \ref{es35prop}, respectively.

\subsubsection*{Results for the EOS of APR}

We present here results for nuclei and semi-infinite matter obtained
with the potential (NRAPR) and field-theoretical (RAPR) model fits to
the EOS of APR. These models satisfy all of the constraints mentioned
above. The values obtained for the binding energy/particle and the
charge radii of closed-shell nuclei are given in Table \ref{Props}.
The results of the two models differ by at most a few percent and
compare quite well with the data.  This agreement is gratifying
insofar as the EOS of APR was obtained from many-body theory without
reference to nuclei. The experimental situation regarding skin
thickness is quite model dependent and we show a few representative
values in Table \ref{Props}. In agreement with the prediction, most of
the recent extracted values for the skin thickness in $^{40}$Ca are
small and negative (however positive values were obtained in earlier
work, see the review of Batty et. al.~\cite{Batty89}). For $^{90}$Zr  
the skin thickness is predicted to be 0.084, on average, which is close
to the central experimental value.  In $^{208}$Pb the theoretical
value $\sim0.2$ is at the upper end of current experimental estimates. 

\begin{table}[hbt]  
\begin{ruledtabular}
\caption{Comparison of results from the potential and field 
theoretical approximations to the APR equation of state with
experimental data for selected closed-shell nuclei.}
\begin{tabular}{|l|r|r|r|r|}
Nucleus&Property & Experiment & Potential & Field-theoretical \\
\hline
$^{208}$Pb&Charge radius (fm) & 5.50~\cite{Fricke95} & 5.41 & 5.41 \\
&Binding energy per particle (MeV) & 7.87~\cite{Audi03} & 7.87 & 7.77 \\
&Skin thickness (fm) & $0.12\pm0.05$, $0.20\pm0.04$ & 0.19 & 0.20 \\ 
&&~\cite{Clark03,Starodubsky94}&&\\ 
\hline
$^{90}$Zr&Charge radius (fm) & 4.27~\cite{Fricke95} & 4.18 & 4.17 \\
&Binding energy per particle (MeV) & 8.71~\cite{Audi03} & 8.88 & 8.65 \\
&Skin thickness (fm) & $0.09\pm 0.07$~\cite{Ray78} 
& 0.075 & 0.093 \\ \hline
$^{40}$Ca&Charge radius (fm) & 3.48~\cite{Fricke95} & 3.40 & 3.34 \\
&Binding energy per particle (MeV) & 8.45~\cite{Audi03} & 8.89 & 8.61 \\
&Skin thickness (fm) & $-0.06\pm0.05$, $-0.05\pm0.04$ & $-0.044$ & $-0.046$ \\
&&~\cite{Clark03,Gibbs92}&&
\end{tabular}
\label{Props}
\end{ruledtabular}
\end{table}              

\begin{figure} 
\begin{center}
\includegraphics[scale=0.7,angle=0]{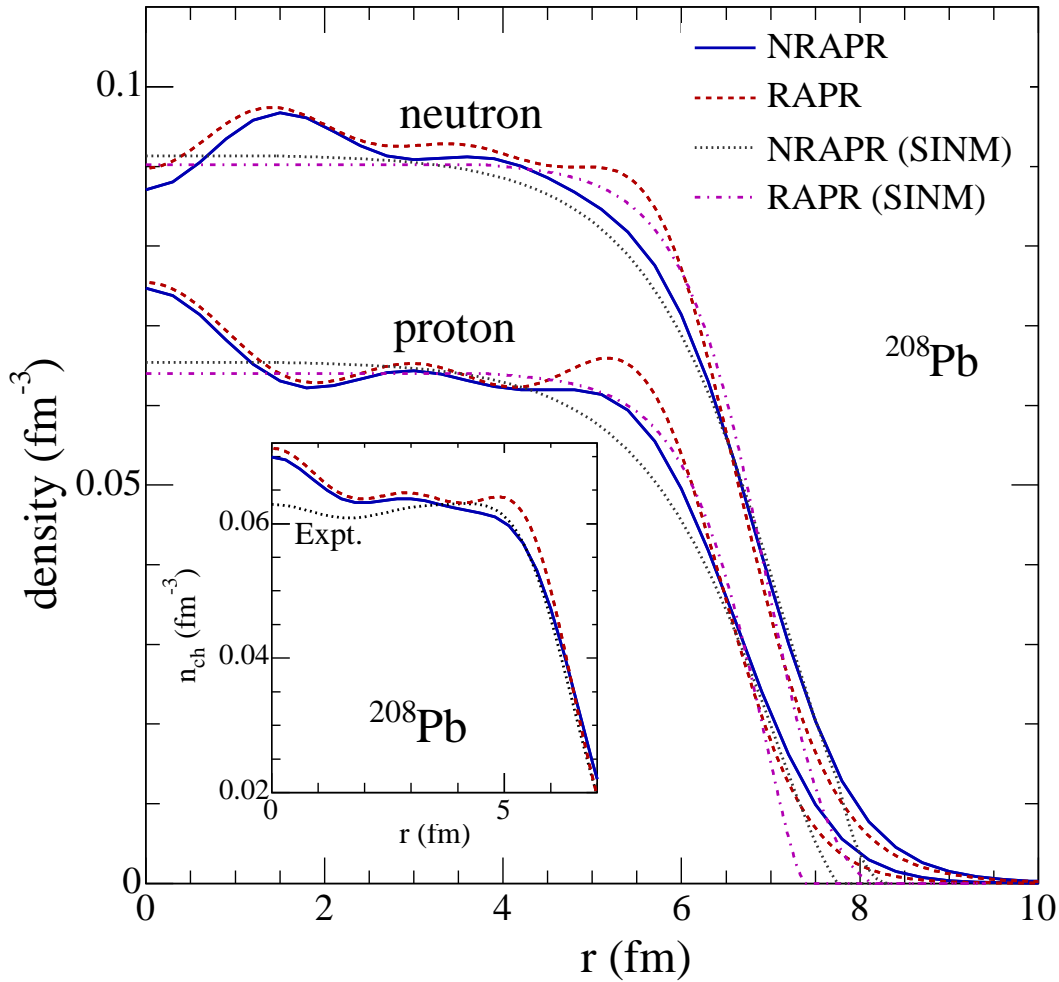}
\end{center}
\caption{Calculated neutron and proton density distributions for
$^{208}$Pb and for semi-infinite matter.  The inset compares the
calculated charge distributions with data \cite{DeVries87}.  }
\label{figs/aprdensity}
\end{figure}

In Fig.~\ref{figs/aprdensity} the neutron and proton density  
distributions for $^{208}$Pb calculated with the NRAPR and RAPR models
are seen to agree reasonably well with each other.
These results are compared with the corresponding results of
semi-infinite matter calculations which also agree reasonably well amongst
themselves. There is, however, a noticeable enhancement of the relativistic 
densities compared to the non-relativistic ones at the onset of the 
surface region ($r\sim5.3$ fm).
In order to specify $\delta_L$ for the semi-infinite
matter calculation, the neutron and proton densities were averaged
from the center of the lead nucleus (for each case) out to 2/3 of its
rms charge radius. This procedure smooths the oscillations in the
calculated density distributions and roughly takes into account the
effects of short range and RPA correlations~\cite{Foris87}. Note that
semi-infinite matter calculations in the Thomas Fermi
approximation are inherently unreliable in the tail of the
distribution. The inset shows that the comparison between the
calculated charge distributions and the data \cite{DeVries87} is
comparable to that of other calculations performed at the
Hartree-Fock-Bogoliubov (for potential models) or Hartree (for  
field-theoretical models) mean field level~\cite{Blaizot80,Serot86}.

\begin{figure} 
\begin{center}
\includegraphics[scale=0.7,angle=0]{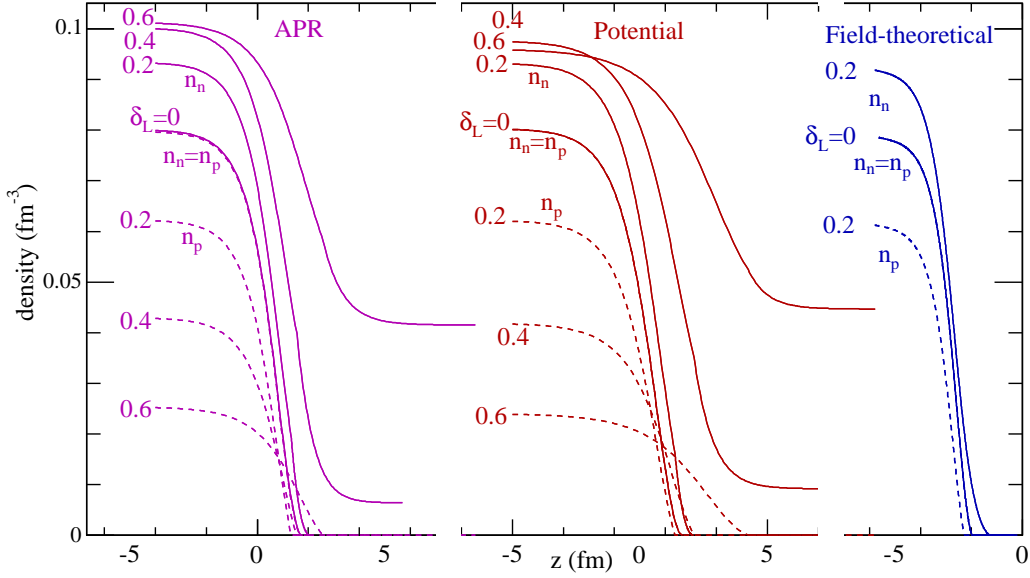}
\end{center}
\caption{The semi-infinite matter neutron (solid curves) and proton 
(dashed curves) density profiles for the indicated values of the 
asymptotic isospin asymmetry, $\delta_L$, as a
function of distance $z$. The profiles are invariant upon translations
along the axis and have been suitably shifted for ease of
display. From left to right, results are for the EOS's of the APR, NRAPR,
and RAPR models.}
\label{figs/siprof}
\end{figure}

Figure~\ref{figs/siprof} shows neutron and proton density profiles from
semi-infinite matter calculations for a range  of
$\delta_L$'s for the NRAPR and RAPR fits to the EOS of APR. The isospin
dependence of the profiles generally matches well with that of the 
semi-infinite matter  profiles calculated directly from the EOS of APR. 
Values of $\delta_L$ larger than about 0.3 (not considered for the  
RAPR case) necessitate the presence of dripped neutrons since the 
neutron chemical potential is positive. Hence the neutron density remains 
finite as $z\rightarrow\infty$.

\begin{figure} 
\begin{center}
\includegraphics[scale=0.4,angle=0]{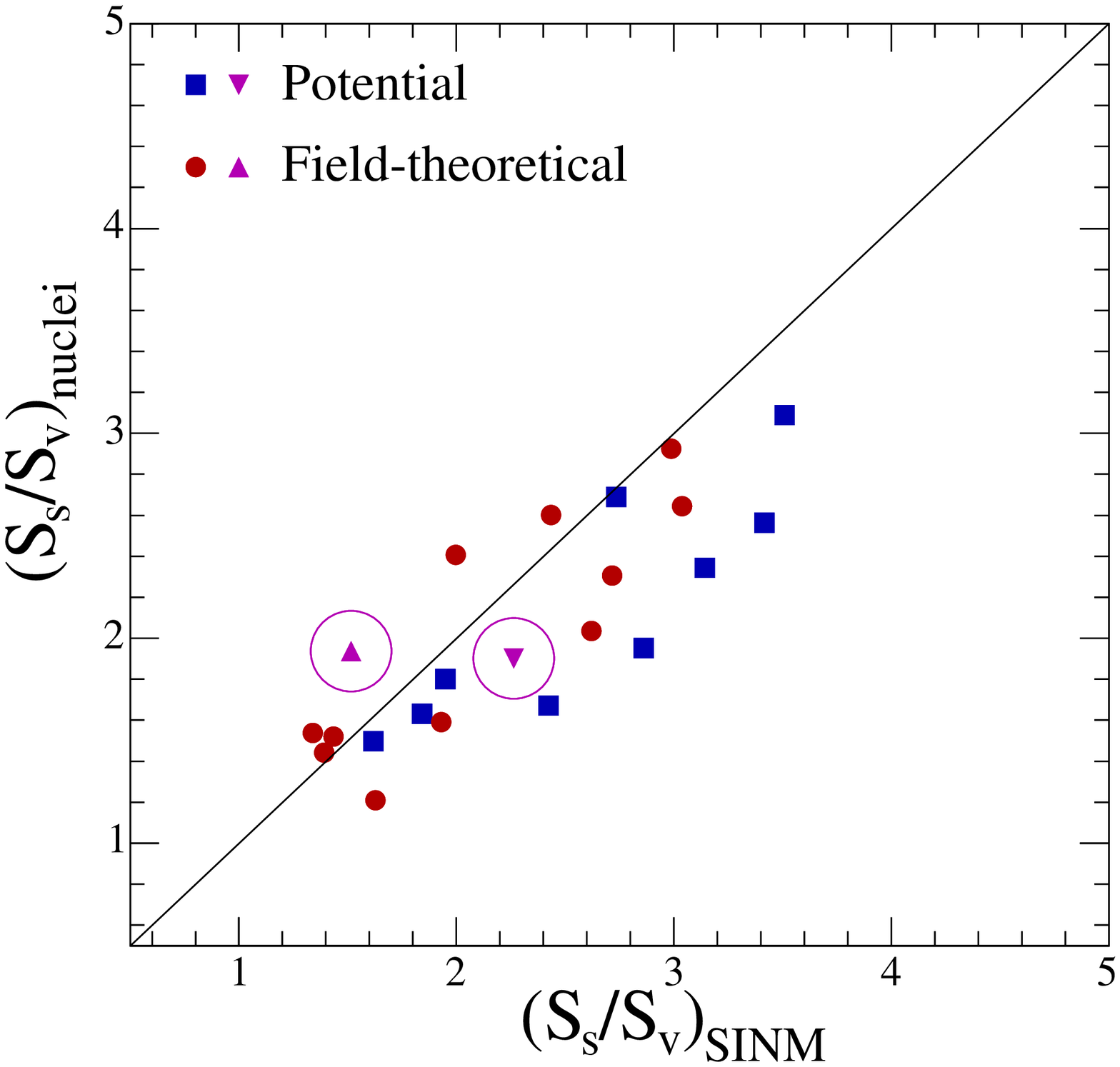}
\includegraphics[scale=0.4,angle=0]{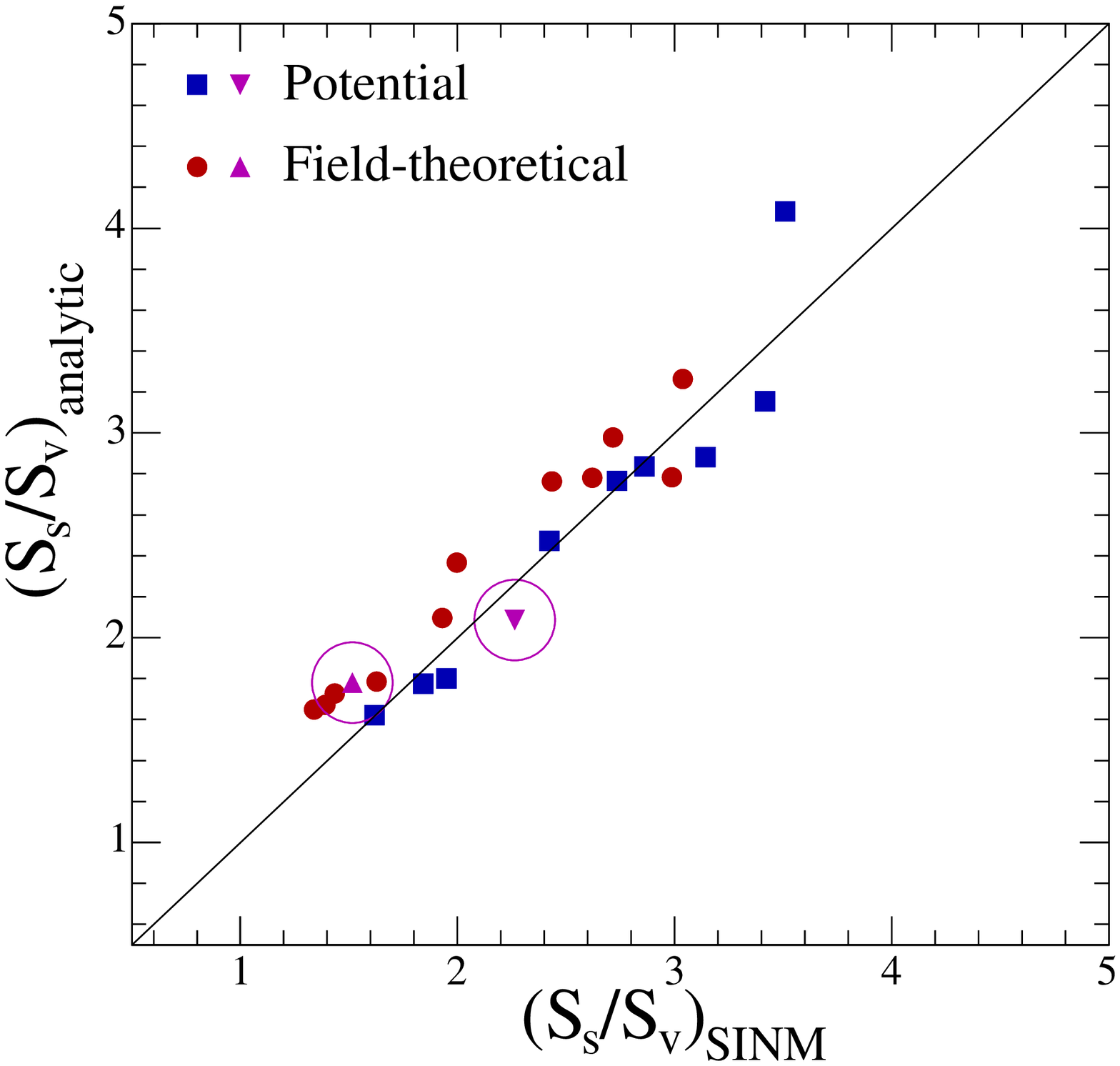}
\end{center}
\caption{Left: The ratio of surface to volume contributions,
$S_s/S_v$, to the total symmetry energy calculated from finite nuclei
versus that calculated from semi-infinite matter.  Right: 
The ratio $S_s/S_v$ calculated from the semi-analytic approximation,
Eq.~(\ref{eq:wd2}), versus that obtained from semi-infinite matter
using Eq.~(\ref{eq:nrend}).  In both panels, the downward
(upward) pointing triangle is the result for the NRAPR (RAPR) model.
}
\label{figs/sssvnuc}
\end{figure}

\subsection{Semi-infinite matter and finite nuclei}

Figure \ref{figs/sssvnuc} compares different methods of calculating the 
ratio of the surface to the volume symmetry energy, $S_s/S_v$. The left 
panel shows the values calculated from nuclei using 
Eq.~(\ref{eq:rnrp_nuclei}) (with $\delta_L$ specified by the averaging 
procedure described earlier) plotted against semi-infinite matter 
results from Eqs.~(\ref{eq:nrend}) and (\ref{eq:sigma_delta}). The 
latter values are expected to be the most reliable. The right panel 
shows the semi-infinite matter results from the approximate relation 
in Eq.~(\ref{eq:wd2}) (using, however, exact expressions for the 
Hamiltonian and the symmetry energy) plotted against the exact 
semi-infinite results described above.  For 
field-theoretical models, the factor $\sqrt{Q_{nn}+Q_{np}}$ needed in 
 Eq.~(\ref{eq:wd2}) is calculated from Eq.~(\ref{eq:quad_profile}). In 
this figure, and in the following figures, potential (field-theoretical) 
model results are denoted by filled squares (circles). 
The 
circled triangles give the APR results from the two fits. 
There is some scatter about the straight line that indicates
perfect agreement between these methods, but the correlation is
sufficiently good that it validates the expressions we have used for
this ratio. 
%(guessing about left panel) 
Notice that, apart from the newly constructed SR1, SR2, and SR3
models, the field-theoretical models tend to have a larger value of
$S_s/S_v$ than the non-relativistic models. 

\begin{figure} \begin{center}
\includegraphics[scale=0.4,angle=0]{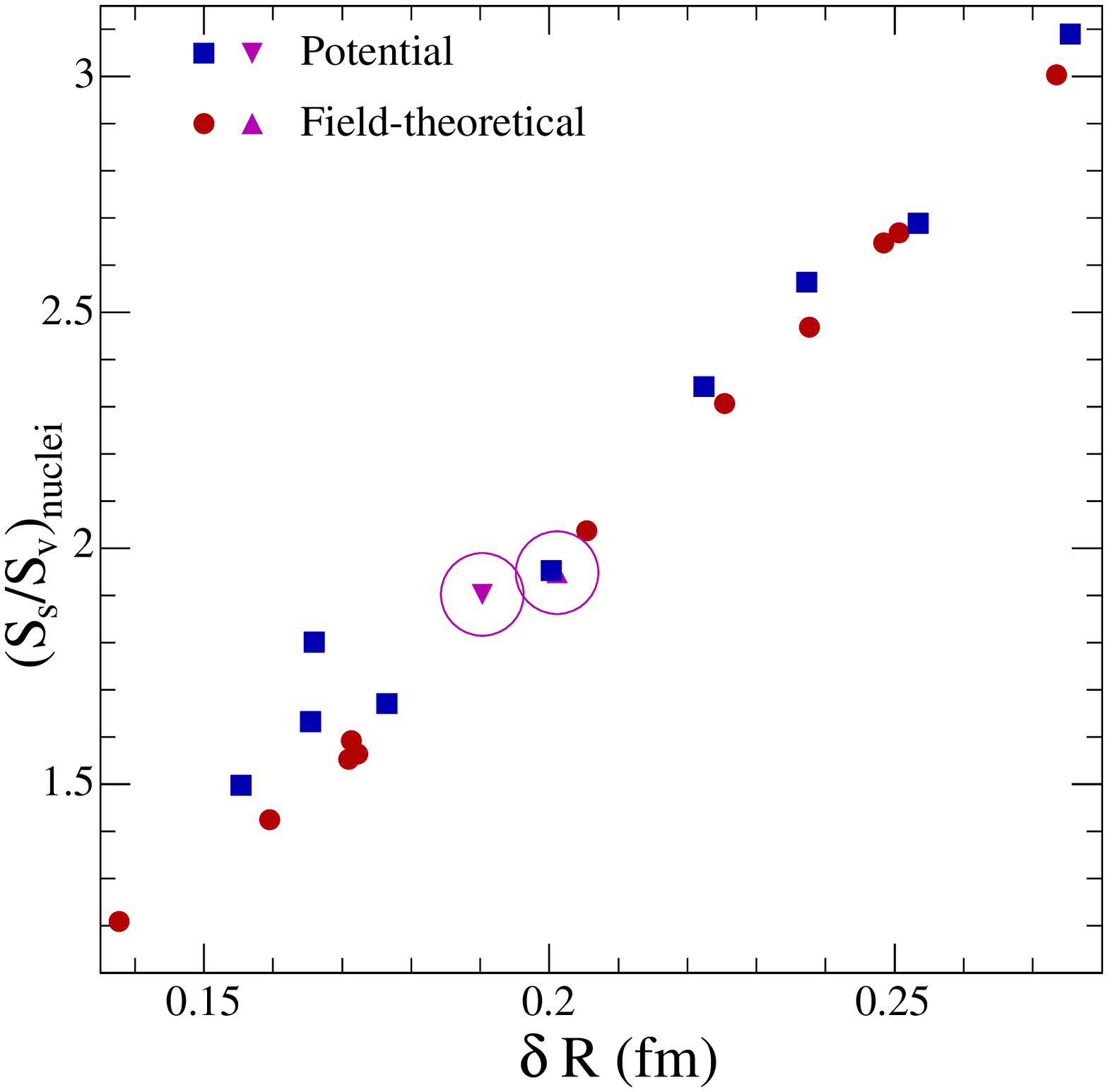}
\includegraphics[scale=0.4,angle=0]{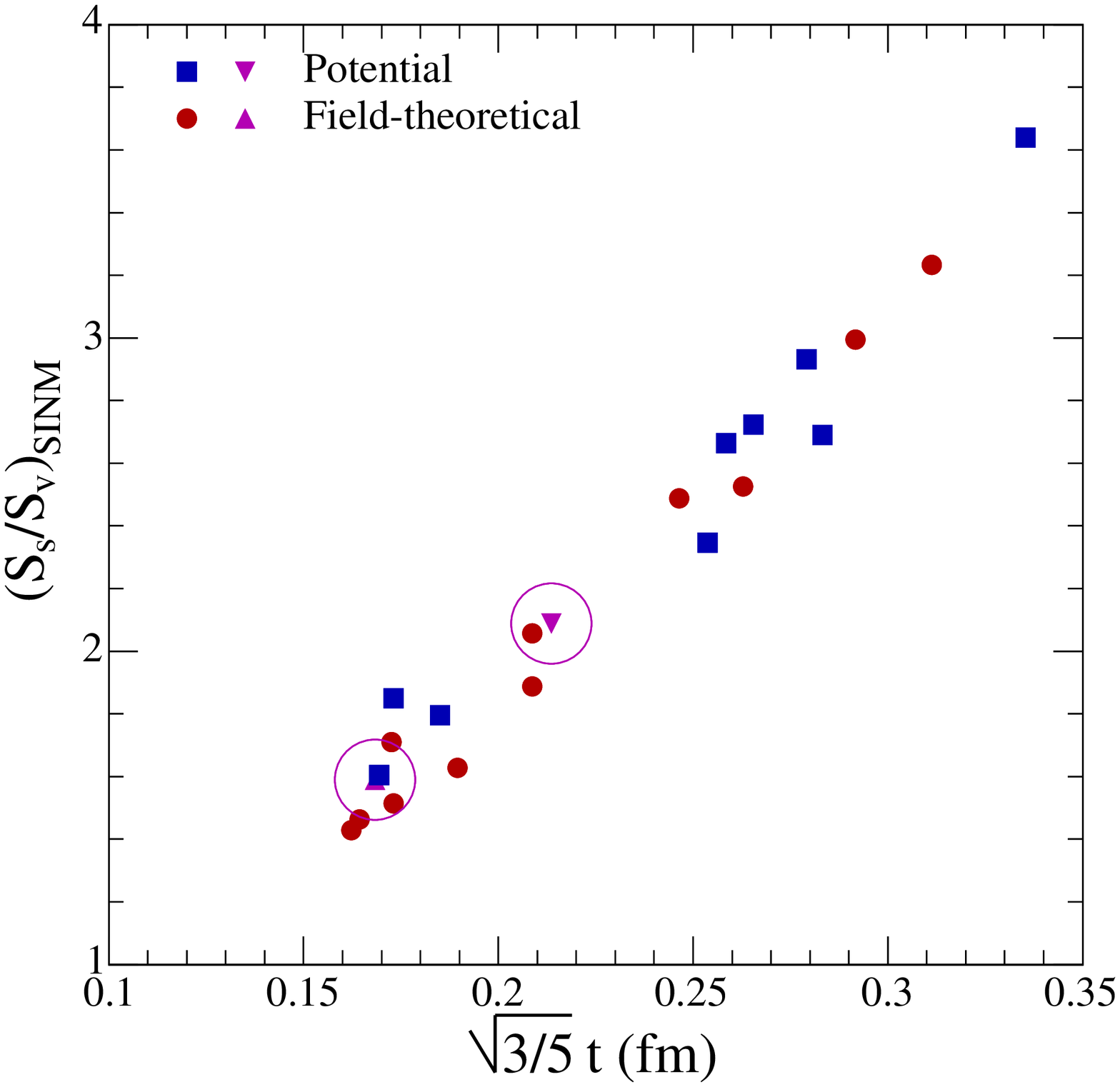}
\end{center}
\caption{Correlation between $S_s/S_v$ and $\delta R$ derived from nuclei 
(left panel) and between $S_s/S_v$ and $\sqrt{3/5}~t$ derived from 
semi-infinite nuclear matter (right panel).}
\label{figs/sssvt}
\end{figure}

In Fig.~\ref{figs/sssvt} we display the correlation between $S_s/S_v$
and the neutron skin thickness $\delta R$, or equivalently
$\sqrt{3/5}~t$.  In the left panel these quantities are derived from
nuclei, whereas for the right panel they are obtained from
semi-infinite matter calculations. For the latter, the definition of
$t$ in Eq. (\ref{eq:tdef}) is employed with $\delta_L$ obtained from
the lead nucleus (see above).  Both panels display a good linear
correlation, as suggested by Eq. (\ref{eq:surften3}),  
and the aforementioned tendency of field-theoretical
models to have larger values of $S_s/S_v$ translates into a tendency
toward larger values for the skin thickness.  

\subsection{Correlations and Their Origins} 
\label{section:correlations}

In this section, we analyze in some detail the correlations that exist
between quantities that are accessible through either laboratory
experiments or astronomical observations that shed light on
isospin-dependent interactions in nucleonic matter. Our emphasis will
be on uncovering the causes of these correlations.

\subsubsection{Nuclear masses and the correlation between $S_s$ and $S_v$}

\begin{figure}[htb]
\begin{center}
\includegraphics[scale=0.6,angle=90]{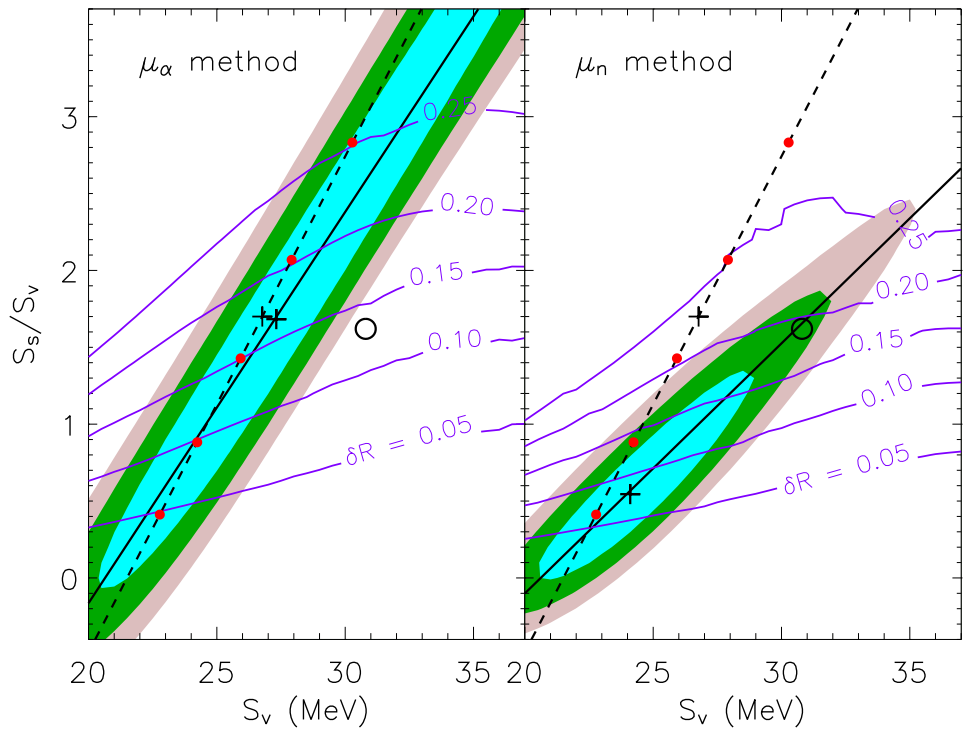}
\end{center}
\caption{Correlations in the symmetry energy parameters from liquid
droplet fits to nuclear binding energies.  The left (right) panel is
for the `$\mu_\alpha$' (`$\mu_n$') method.  Filled-in contours refer to
the droplet models that have average errors 0.01, 0.02 and 0.03
MeV/baryon, respectively, above the minimum values (marked by
plus signs).  The thick solid (dashed) line lies on the minimum $\chi^2$
valley for our (Danielewicz's) droplet models.  Contours of skin thicknesses
($\delta R$) are shown for our droplet models; corresponding values are
indicated on Danielewicz's correlation with solid dots.
The open circles indicate the minimum point found by  
M{\o}ller {\it et al.} \cite{Moller95}.}
\label{chi_droplet}
\end{figure}     

The formalism for the modified liquid droplet models was 
discussed in Subsec. III B. Each model has six parameters and we wish 
to study the correlation between the volume and surface symmetry 
parameters, $S_v$ and $S_{s}$ (see also Lattimer and Swesty 
\cite{Lattimer91}). Therefore as a function of $S_v$ and
$S_{s}$, we have minimized the quantity
\begin{equation}
\chi^2=\sum_i^n{(E_i^D-E_i)^2/n}\,,
\end{equation}
with respect to the other four parameters. Here $E_i^D$ and $E_i$ are
the data \cite{Audi03} and model energies of the $i$th nucleus and
$n=2841$ is the number of nuclei included in the fit (only
experimentally determined masses with $A\ge20$ are included; the
results for a smaller set of nuclei with $A\ge40$ are not
significantly different).  In either case, a nearly linear behavior is
found for the contour of minimum $\chi^2$ between $S_s/S_v$ and $S_v$,
as shown in Fig. \ref{chi_droplet}.  The equations describing the
correlations are given by
\begin{eqnarray}\label{eq:correl}
S_{s}/S_v&=&-5.253+0.254 S_v\qquad {\rm for}\qquad \mu_\alpha{\rm
  ~method}\,,  
\cr
S_s/S_v&=&-3.453+0.163 S_v\qquad {\rm for}\qquad \mu_n{\rm ~method}\,.
\end{eqnarray}    
The best-fit values of $S_v$ and $S_{s}/S_v$ are about 27.3 (24.1) MeV and
1.68 (0.545), respectively, for the `$\mu_\alpha$' (`$\mu_n$') methods for the 
modified droplet models. 

A relevant question is the range over which values of $S_v$ and
$S_{s}$ provide fits that 
differ by a statistically insignificant amount
from the best fit.  Traditional mass formula fits were developed to
assist in interpolating within or to extrapolate to nearby regimes
outside the ranges of measured nuclear masses.  The experimental
errors of measured masses are of order keV's per nucleus, so
interpolations to the same accuracy are desirable.  However, for many
astrophysical applications, which involve extrapolations to ranges of
$A$ and $Z$ well beyond the measured ranges, a model accuracy of a few
hundredths of an MeV/baryon would be acceptable.  Such an error in the
computed energy per particle of a nucleus would translate into a similar
error in the neutron or proton chemical potentials.  Errors in
chemical potentials generated by the use of the ``single-nucleus''
approximation, an essential assumption of liquid droplet high-density
EOS models, are of order $T/A$ \cite{Burrows84}, or several hundredths
of an MeV, for example.

At the best fit for the droplet models, for which the value of $\chi^2$ is
$\chi^2_0\simeq6$, the mean difference, or error, between the
experimental and predicted nuclear energy per baryon is
$\delta\simeq0.028$ MeV, where $\delta$ is defined by
\begin{equation}\label{eq:delta}
\chi^2_0=\sum_i^n (r_i\delta A_i)^2/n\,,   
\end{equation} 
and $r_i$ is a random number between $-1$ and 1. Therefore, increasing
the mean error by a further amount
$\epsilon$ per baryon would increase $\chi^2$ to a value
\begin{equation}\label{eq:epsilon}
\chi^2=\sum_i^n [r_i(\delta+\epsilon)A_i]^2/n\,.
\end{equation}
Contours of mean errors $\epsilon$ of 0.01, 0.02 and
0.03 MeV per baryon above the minimum are shown in Fig. \ref{chi_droplet}.  For
the `$\mu_n$' modified droplet model, and for an additional mean error of 0.01
MeV/baryon, the allowable excursions in $S_v$ are about $(+5,-4)$ MeV,
and in $S_{s}/S_v$, about $(+0.9,-0.5)$, even though the excursions must
be constrained by the obvious correlation.  However, for the `$\mu_\alpha$'
model, the allowed excursions in parameters are much larger.  It is 
apparent that fitting nuclear energies alone cannot constrain these
parameters, even though a relatively tight correlation between them exists.

The origin of the correlations in Fig. \ref{chi_droplet} 
can be most easily understood by examining the
functional form of the total symmetry energy in the simple droplet
model:
\begin{equation}
E_{sym}(A,I) = S_vAI^2~\left[1+{S_s^*\over S_vA^{1/3}}\right]^{-1}\,.
\label{dropsym}
\end{equation}
Assuming that a linear 
correlation $S_s^*/S_v=a+bS_v$ exists between the parameters, minimization of
Eq. (\ref{dropsym}) with respect to $S_v$ yields
\begin{equation}
a=-\left<A^{-1/3}\right>^{-1}=-5.31\;,
\end{equation}
where the average is mass-weighted over the entire range of nuclear masses
used in this study. Noting that
\begin{equation}\label{eq:esym}
{E_{sym}\over A}={I\over4}\left(\mu_n-\mu_p+2E_CZA^{-1/3}+
2E_{dif}ZA^{-1}+{\textstyle{4\over3}}E_{ex}Z^{1/3}A^{-4/3}\right)\,,
\end{equation}
where $\mu_n-\mu_p$ can be taken directly from nuclear masses as
\begin{equation}\label{eq:muhat}
\mu_n-\mu_p\simeq \thalf[E(N+1,Z)-E(N-1,Z)-E(Z+1,N)+E(Z-1,N)]\,,
\end{equation}
one can estimate the average symmetry energy, from which it follows that
\begin{equation}
b=\left<I^2\right>\left<{E_{sym}\over A}\right>^{-1}\left<A^{-1/3}\right>^{-1}
\simeq0.223\,.
\end{equation}
Thus, one has almost precisely the
linear relation shown in Eq. (\ref{eq:correl}) for the `$\mu_\alpha$' 
droplet correlation.

A method similar to the `$\mu_\alpha$' method was suggested by
Danielewicz \cite{Danielewicz03}.  His results resemble ours, but
slight differences in the formulation exist, the most notable being
that we account for variation of the central nuclear density in the
fitting ({\it i.e.}, the ratio $n_L/0.16$ fm$^{-3}\equiv u_0$ in Eqs. 
(\ref{eq:alpha}),
(\ref{eq:tmua}) and (\ref{eq:emua})).  In addition,
Ref. \cite{Danielewicz03} fitted parameters with an absolute energy
deviation minimization, rather than the minimization of $\chi^2$ that
we employ.  Danielewicz determined best-fit values of
$S_s/S_v\simeq1.7$ and $S_v\simeq27$ MeV and found a correlation slope
of 0.32 (compared to 0.25 in our study); Danielewicz's fit is
indicated by the dashed lines in Fig. \ref{chi_droplet}.  We have
reproduced Danielewicz's model fits by employing his stated  
procedures. It is evident that the position of the minimum and the
slope of the correlation are affected by differences in the fitting
procedure and the droplet formulae.  Nevertheless, to within an
average error of 0.005 MeV/baryon, the differences found are not
significant.  Importantly, Ref. \cite{Danielewicz03} found, as do we,
that large excursions in parameter space are not excluded within the
droplet model.  The range of allowed excursions are much
smaller for the `$\mu_n$' modified droplet model.

\begin{figure} 
\begin{center}
\includegraphics[scale=0.5,angle=0]{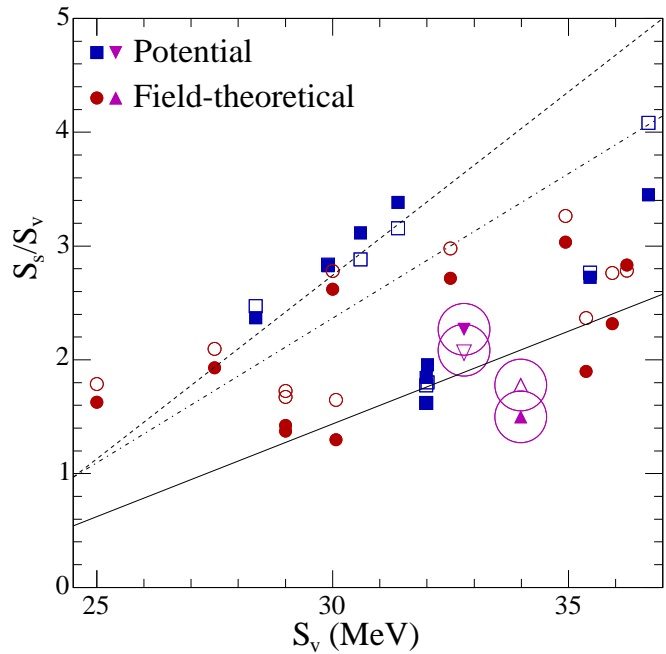}
\end{center}
\caption{The ratio of the surface and volume coefficients of the
symmetry energy versus the volume coefficient for the models used in this
work. The closed symbols show the exact values obtained from a
quadratic fit to the surface tension in semi-infinite nuclear matter,
whereas the open symbols use the value of $S_s/S_v$ obtained via
Eq.~(\ref{eq:wd2}).  The solid (dot-dashed) line is the linear correlation
from the `$\mu_n$' (`$\mu_\alpha$') droplet model Eq.~(\ref{eq:correl}) 
and the short-dashed line is the linear correlation from the droplet model 
of Ref. \cite{Danielewicz03}.}  
\label{figs/sssvsv}
\end{figure}         

It is interesting that the model of M{\o}ller {\it et al.}  
\cite{Moller95} gives $S_v=30.8$ MeV and $S_s/S_v=1.62$. This point is 
indicated by a small open circle in the left and right panels of 
Fig. \ref{chi_droplet}. Clearly it lies much closer to the `$\mu_n$' 
droplet correlation than the `$\mu_\alpha$'  correlation.
Although this point is displaced from the position of the best fit 
for the `$\mu_n$' correlation by over 6 MeV in $S_v$, its average error 
differs from the best fit by less than 0.02 MeV/baryon. In contrast, 
this point differs by about 0.04 MeV/baryon from the best fit 
established with the `$\mu_\alpha$' correlation.
We note that the model of Ref. \cite{Moller95}
contains several additional effects compared to our modified droplet
model, including terms representing compression, curvature,
deformation and shell effects.  It is thus not surprising that the
best-fit parameters differ. However, in spite of the fact that the 
`$\mu_\alpha$' approach contains Coulomb corrections to the surface 
tension while the `$\mu_n$' approach does not, the notable feature 
is that the model of M{\o}ller {\it et al.} seems more consistent with the  
`$\mu_n$' model than the `$\mu_\alpha$' model.

We also compare the volume and surface symmetry coefficients
determined for the potential and field-theoretical models developed in
this paper with those of the published models utilized here and with
the results of mass formula fits to nuclear energies.  The parameters
of the models developed in this paper have been fit to only four
closed shell nuclei, while many of the previously-published forces
have been fit to properties of a wider range of nuclei that span a
larger range of asymmetries and masses.  Thus, we anticipate that a
correlation established from mass fits should be apparent when
comparing previously-published forces, but not necessarily by the
forces developed in this paper.  This is precisely what is found, as
indicated in Fig. \ref{figs/sssvsv}.  It is important to note that
most of the previously-published models follow the correlation of the
`$\mu_n$' modified droplet model as opposed to the steeper `$\mu_\alpha$'
droplet correlation.  This is in spite of the fact that the
`$\mu_\alpha$' model takes into account the polarization of the nuclear
interior due to Coulomb forces which the `$\mu_n$' model ignores.

These results highlight a major difference between the traditional use
of droplet mass formula fits and that of current astrophysical
applications.  Historically, droplet mass formulae were used to
interpolate within known nuclei to establish relatively precise
values.  In astrophysics, however, it is necessary to extrapolate to
extreme conditions of asymmetry.  Extrapolations depend significantly
upon the correlation between $S_s/S_v$ and $S_v$.  Given that the
EOS's in current applications to supernova and neutron star matter are 
based on a modified droplet approach, it is necessary to use
consistent parameter sets obeying the proper correlation.  Otherwise,
connections between quantities such as the neutron skin thickness and
the neutron star radius could be incorrectly interpreted.

Fig. \ref{chi_droplet} also displays the neutron skin thickness
$\delta R$ as a function of $S_s$ and $S_v$ for the two approaches.
In the `$\mu_n$' approach, if the dependence upon $u_0$ were to be ignored,
contours of fixed $\delta R$ would appear as horizontal lines since
$\delta R$ would
be a function of $S_s/S_v$ alone.  This would also be the case in
the `$\mu_\alpha$' approach, if the dependence upon both $u_0$ and
the Coulomb term $\beta$ were ignored.  The dependence of $\delta R$ on
$\beta$ is such that $\partial (\delta R)/\partial S_v<0$ if the $u_0$
dependence is ignored, implying a positive slope for the $\delta R$ contours.
Indeed, we find the slope of the $\delta R$ contours is larger in the
`$\mu_\alpha$' approach than in the `$\mu_n$' approach.  However,
Ref. \cite{Danielewicz03} finds that the $\delta R$ contours have negative 
slope; this discrepancy remains to be resolved. 

The relationship between the skin thickness and the ratio $S_s/S_v$
was studied numerically using a variational approach by
Bodmer~\cite{Bodmer03}. The results of his work can be
understood from our analytical analysis in Section III
C. Ref.~\cite{Bodmer03} also emphasizes the interplay between the 
symmetry and Coulomb energies. Ignoring the latter, as in the `$\mu_n$' 
method, leads to larger values of $\delta R$.

\subsubsection{The neutron skin thickness versus the pressure
of subnuclear neutron-star matter }
\label{section:tb}

Typel and Brown \cite{Brown00,Typel01} have noticed a correlation
between the skin thickness and the pressure of pure neutron matter at
a density of $n=0.1$ fm$^{-3}$.  To the extent that this correlation
can be applied, a measurement of $\delta R$ will help to establish an
empirical calibration point for the pressure of neutron star matter at
subnuclear densities.  Coupled with a neutron star radius measurement
(see the discussion below), that could allow the inference of the
pressure at supranuclear densities, the Typel-Brown correlation would
be valuable in establishing the pressure--density relationship over a
wide range of densities inside neutron stars.

The connection between the neutron skin thickness and the symmetry
energy has been known from Bodmer's work in the
60's~\cite{Bodmer60}. The Typel-Brown correlation between $\delta R$
and $P(5 n_0/8)$, which is closely related to the derivative of the
symmetry energy, demonstrates clearly the new information that could
be obtained by an accurate measurement of the skin thickness in
$^{208}$Pb.  Typel and Brown considered a large set of Skyrme
parameterizations including those that are inadequate to describe
neutron-star matter~\cite{Stone03}.  Such forces lead to small skin
thicknesses. None of the field-theoretical models considered in
Ref.~\cite{Typel01} yields a skin thickness smaller than 0.2 fm.  By
using the function $f$ in Eq.~(\ref{FLTL}), we have constructed
examples of field-theoretical models, SR1, SR2, and SR3 which give
skin thicknesses which are as small as those obtained from some
potential models ($\sim0.15$ fm).  Our models es25 and es275
also have small
skin thicknesses. However, as pointed out by
Furnstahl~\cite{Furnstahl02}, there are other possible ways in which
this could be achieved, including the addition of a $\delta$ meson or
a more complete treatment of pions.  If measurements in $^{208}$Pb
give $\delta R \leq 0.15$ fm, then Skyrme-like parameterizations have 
to be refined at high density to properly describe neutron stars. By
the same token, field-theoretical models will need modifications to
better describe low-density matter.

\begin{figure}
\begin{center}
\includegraphics[scale=0.4,angle=0]{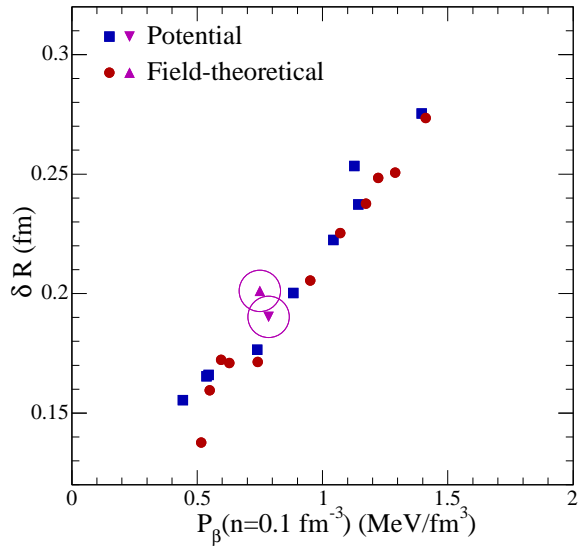}
\end{center}
\caption{The neutron skin thickness $\delta R$ of finite
nuclei versus the pressure of $\beta$-equilibrated matter at a density
of 0.1 fm$^{-3}$. }
%(left panel) and $(5/8)n_0$ (right panel).}
%
\label{fig:tplot}
\end{figure}

We show in Fig.~\ref{fig:tplot} the relation between the skin
thickness $\delta R$ and the pressure of beta equilibrated matter at
0.1 fm$^{-3}$.  We have verified that the correlation is similar also
at a fixed fraction, 5/8, of the saturation density, for a variety of
potential and field-theoretical models.

Some insight into the origin of this correlation can be gained by
examining how the beta-equilibrated pressure $P_\beta$ (the abscissa in
Fig.~\ref{fig:tplot}) and the neutron skin $\delta R$ (the ordinate)
vary with quantities that depend upon isospin.  The pressure of cold,
beta-stable neutron-star matter can be written as 
\begin{eqnarray}
P_\beta(n,x)= n^2\left[E^\prime(n,1/2)+ E_{sym}^\prime(n)
(1-2x)^2\right] + P_e + P_\mu\,,
\label{eq:pbeta}
\end{eqnarray}
where $x=n_p/n$ is the proton fraction, $E(n,1/2)$ is the energy per
particle of symmetric matter, $E_{sym}(n)$ is the bulk symmetry energy
(primes denote derivatives with respect to density), and the last
two terms give leptonic (electron and muon) contributions.
For beta-equilibrated matter, 
\begin{eqnarray}
\mu_n - \mu_p = -\partial E/\partial x &\cong & 4(1-2x)E_{sym}(n) 
\nonumber \\ 
&=& \mu_e = \mu_\mu \,, 
\label{eq:themus} 
\end{eqnarray}
where $\mu_e$ and $\mu_\mu$ are the chemical potentials of the
electron and muon, respectively.  This relation, together with the
charge neutrality condition, $n_p=n_e+n_\mu$, permits the evaluation
of the equilibrium proton fraction.  Since muons begin to appear in
matter for $n \geq n_0$ \cite{Prakash96}, the only leptons at
subnuclear densities are electrons. The electron pressure is
\begin{eqnarray}
P_e = \tquar n_e\mu_e = \tquar nx\mu_e \cong nx(1-2x)E_{sym}(n) \,, 
\end{eqnarray}
where we have used $n_e=n_p$ for $n \leq
n_0$ and Eq.~(\ref{eq:themus}).
Utilizing this relation in Eq.~(\ref{eq:pbeta}), we can write
\begin{eqnarray}
P_\beta(n,x) &=& P(n,1/2) + n(1-2x)E_{sym}(n) 
\left[(1-2x)  \frac {d\ln E_{sym}(n)}{d \ln n} + x \right]\,,
\end{eqnarray} 
where the first term is the pressure of symmetric nuclear matter.
Thus, the total pressure can be written at a particular density in
terms of fundamental nuclear parameters. 
For densities at or below saturation density, the 
equilibrium proton fraction, 
$\tilde x \simeq (4E_{sym}/\hbar c)^3/(3\pi^2n)$, is very small
so that the beta equilibrated pressure and the pure neutron matter 
pressure only differ by a small amount that depends upon
the symmetry properties of the force. Thus, for $n \leq n_0$, $P_\beta(n)
\cong P(n,1/2) + n^2E_{sym}^{\prime}$, where the second term provides
the bulk of the contribution. Thus an approximately linear correlation
would result if in Fig.~\ref{fig:tplot} $\delta R$ were plotted 
against $E_{sym}'(n)$ at densities $n=0.1$ fm$^{-3}$ or $n=5n_0/8$;
such a plot is given in Ref.~\cite{Baldo04}. 

The dependence of $\delta R$ on the isospin asymmetry can be seen
explicitly in Eq.~(\ref{eq:rnrp_nuclei}) for nuclei and in
Eq.~(\ref{eq:surften3}) for the skin thickness of semi-infinite
matter. The latter equation can be manipulated to read 
\begin{equation}
t = \frac{2\sigma_0}{Bn_L}~\frac {\delta_L}{(1-\delta_L^2)} 
~~ \lim_{\xi \to 1} \frac {d}{d\xi} ~~
{ 
\int\limits_0^{1} u^{1/2}
\left[ {h}_B(u,\delta=0) +\xi B \right]^{1/2} 
\left[S_v/E_{sym}(u)-1 \right] d u \over 
\int\limits_0^{1} u^{1/2} \left[ {h}_B(u,\delta=0) +B \right]^{1/2} du
} \;,
\label{eq:newt} 
\end{equation}
where $u=n/n_L$ and $h_B(u,\delta=0)={\cal H}_B(u,\delta=0)/n$.  Since
$t \cong \sqrt{5/3}~\delta R$, this relation illustrates
that the neutron skin of nuclei is proportional to a specific average
of $\left[S_v/E_{sym}(u)-1 \right]$ in the nuclear surface, the averaging
function involving the square root of the energy per baryon of isospin
symmetric matter.

\begin{figure} 
\begin{center}
\includegraphics[scale=0.4,angle=0]{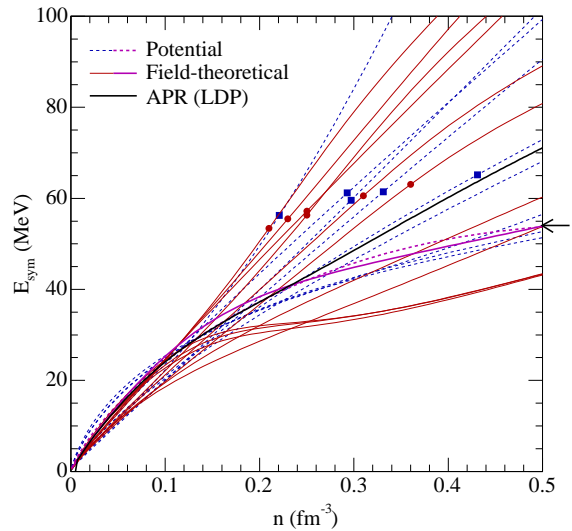}
\end{center} 
\caption{The symmetry energy versus density for
various equations of state.  Solid (dashed) lines are for
field-theoretical (potential) models. The thick solid line shows  
the APR symmetry energy for the low density phase. 
The square (circular) markers
indicate the densities at which potential (field-theoretical) models
allow the direct Urca process to occur.  All models are constrained as
discussed in the text. The arrow identifies the NRAPR (dotted line) and
RAPR (solid line) models which have nearly identical 
symmetry energies at $n=0.5$ fm$^{-3}$. }  
\label{fig:syme}  
\end{figure}

Equations (\ref{eq:pbeta}) and (\ref{eq:newt}) show that the density
dependence of the symmetry energy is the principal cause of the
Typel-Brown correlation. We therefore display the symmetry energies
for the models considered in this work in Fig.~\ref{fig:syme}. The  
thick line indicates the symmetry energy of the APR equation of 
state for the low density phase (this is strictly only applicable 
for $n<0.25$ fm$^{-3}$ for which region the NRAPR and RAPR fits show
good agreement).
Generally the symmetry energies of the field-theoretical models
(solid lines) vary more rapidly with density than those of potential
models (dashed lines). Three exceptions, models SR1, SR2, and SR3,
were constructed in this work by a suitable choice of the function $f$
in Eq.~(\ref{eq:ffun}). These field-theoretical models mimic the density
dependence of potential models near the saturation density, but with
somewhat different behavior at supranuclear densities.  Notice that
the requirements that we have placed on our models strongly constrain
the symmetry energy at an average nuclear density
$n\simeq0.1$ fm$^{-3}$, in agreement with Brown
\cite{Brown00} and Horowitz and Piekarewicz~\cite{Horowitz01} (see
also Ref. \cite{Furnstahl02}). 

For the models considered in Fig.~\ref{fig:syme}, the variation of
$\delta R$ with $P_\beta$  shown in Fig.~\ref{fig:tplot} 
confirms the Typel-Brown correlation.  Additional points that emerge
from our analysis include:

\ni (1) At the densities of interest, the pressure difference between
pure neutron matter and beta-equilibrated    
matter is small enough 
that the correlation is not significantly impacted.

\ni (2) In general, field-theoretical models yield $\delta R >0.2$
fm. The exceptions are models SR1, SR2, SR3, es25, and es275 that were
designed to have special symmetry properties that lead to $\delta R <
0.2$ fm.

\ni (3) If the neutron skin thickness is measured with high accuracy,
the pressure of neutron-star matter at this density could be
determined to within (20--25)\%.  For a fiducial $R_n-R_p=0.2 \pm
0.025$ fm, this correlation predicts that $P_\beta=(0.9\pm 0.3)$
MeV/fm$^3$; for a lower mean value of $R_n-R_p$, the predicted
uncertainty in the pressure is larger. Employing a wider variety of
non-relativistic models than those used by Typel and Brown results in
an enlarged spread in results for models with relatively small values
of $\delta R$.  We emphasize that the linear correlations observed in
Fig.~\ref{fig:tplot} are significantly worsened for slight excursions
from the fiducial densities chosen.

Furnstahl~\cite{Furnstahl02} has presented a correlation between the
neutron form factor and the neutron radius. He also demonstrated
correlations between the skin thickness and, separately, the linear
density dependence of the symmetry energy, the quadratic density
dependence and also $d^2 n_{0}/ d \alpha^2$. These three quantities
are all closely related to $P(5 n_0/8)$ as shown above.

\subsubsection*{Implications for neutrino emission from neutron stars }

The density dependence of the symmetry energy also plays an important
role in determining whether or not the simplest possible $\nu$
emitting processes, the so-called direct Urca processes 
\begin{eqnarray}
p + \ell
\rightarrow n + \nu_\ell\, \qquad {\rm and} \qquad  
n \rightarrow p + \ell + \overline{\nu_\ell}, 
\end{eqnarray} 
where $\ell$ is either an electron or a muon, occur in charge neutral
neutron star matter.  These direct Urca processes can occur whenever
energy and momentum conservation is simultaneously 
satisfied among $n,~p$ and $\ell$ (for
temperatures of interest to long-term cooling, neutrinos would have
left the star).  
In a mixture of neutrons, protons and electrons, the required proton
fraction $x$ for the direct Urca process to occur is 1/9.
When this mixture is in beta equilibrium, $x$
satisfies~\cite{Lattimer91} 
\begin{equation}
x\simeq0.048~(E_{sym}(n)/S_v)^3~(n_0/n)~(1-2x)^3\,,
\label{x}
\end{equation}
which highlights the role of the density dependent symmetry energy.
For example, for a linear symmetry energy, $E_{sym}(n)=S_v(n/n_0)$,
the onset of the direct Urca process occurs for $n_c/n_0 \simeq 2.2$
with $S_v=30$ MeV. In the presence of muons, which begin to appear
around $n_0$, the required proton fraction is about 0.14 and hence the
threshold density is somewhat larger.  In the case that $E_{sym}(n)$
increases less than linearly with density, the direct Urca threshold
density is larger.  If the Urca threshold density is less than the
star's central density, the neutron star will rapidly cool because of
large energy losses due to neutrino emission: the star's interior
temperature $T$ will drop below 10$^9$ K in minutes and reach 10$^7$ K
in about a hundred years.  This is the so-called rapid cooling
paradigm.  If the threshold density is not reached below the central
density, or if the direct Urca process is suppressed due to nucleon
superfluidity (calculated energy gaps are about an MeV or less),
cooling instead proceeds through the significantly less rapid modified
Urca processes  
\begin{equation}
n+(n,p)\rightarrow p+(n,p)+e^-+\bar\nu_e\, \quad {\rm and} \quad
p+(n,p)\rightarrow n+(n,p)
+e^++\nu_e\,,
\end{equation}
in which an additional nucleon enables momentum
conservation.  As the temperature approaches the superfluid critical
temperature, cooling also takes place through neutrino pair emission
from the Cooper-pair breaking and formation processes~\cite{Flowers76}
which are more efficient than the modified Urca process, but
significantly less efficient than the direct Urca process. If the
density dependence of the nuclear symmetry energy is relatively weak,
only the less rapid neutrino cooling processes can occur.

\begin{figure} 
\begin{center}
\includegraphics[scale=0.4,angle=0]{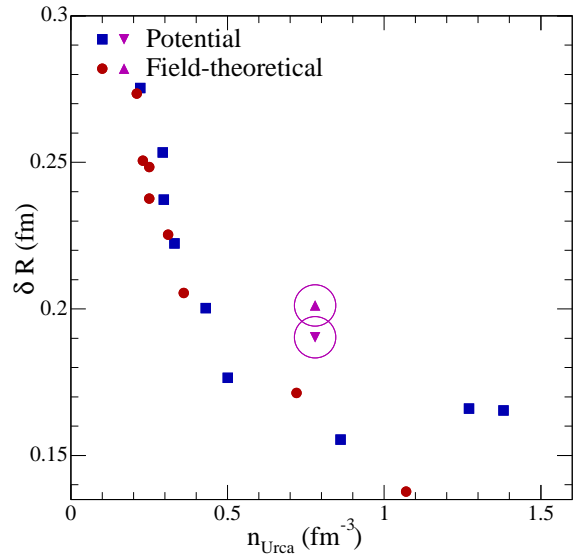}
\end{center} 
\caption{ Skin thickness versus the threshold density for the direct
Urca process to occur in neutron stars.}
\label{fig:durca}  
\end{figure}

The filled squares and circles in Fig.~\ref{fig:syme} 
indicate the densities at which the direct Urca process is allowed by 
energy and momentum conservation conditions. Some of the models whose 
symmetry energy exhibits a moderate density dependence allow the 
direct Urca process to take place, but at densities beyond the 
maximum of 0.5 fm$^{-3}$ plotted.  For some other models the   
direct Urca process does not occur at all since it requires densities 
larger than the central densities of the maximum mass stars.

Figure \ref{fig:durca} shows the skin thickness versus the threshold
density for the direct Urca process (note that the NRAPR and RAPR 
skin thicknesses are plotted against $n_{Urca}$ for the exact APR EOS.).  
The results in this figure
supplement those of Horowitz and Piekarewicz~\cite{Horowitz02} by also
including results of potential models.  A determination that $\delta
R$ has a value greater than 0.2 fm would imply that the
nucleon direct Urca process should exist in 1.4 M$_\odot$ neutron
stars, unless it is suppressed by superfluidity.  For a recent
assessment in this regard, see Ref.~{\cite{Page04}} which studied
neutron star cooling in the absence of enhanced neutrino emission, and
compared the results with data on cooling neutron stars.

\subsubsection*{Implications for the crustal fraction of the moment of
inertia}

Another astrophysical application concerns the thickness of the crust
of a neutron star, which is related to the density dependence of the
symmetry energy by physics similar to that of the thickness of the
neutron skin of nuclei.  The neutron star crust thickness might be
measurable from observations of pulsar glitches, the occasional
disruptions of the otherwise extremely regular pulsations from
magnetized, rotating neutron stars.  The canonical model of Link et
al. \cite{Link99} suggests that
glitches are due to the transfer of angular momentum from superfluid
neutrons to normal matter in the neutron star crust, the region of the
star containing nuclei and nucleons that have dripped out of nuclei.  This
region is bounded by the neutron drip density ($\simeq
1.5\times10^{-3}n_0\simeq4\times10^{11}$ g cm$^{-3}$) and the
transition density $n_t$ ($\approx n_0/2\simeq1.5\times10^{14}$ g
cm$^{-3}$) at which nuclei merge into uniform nucleonic matter.  Link %pje
et al. \cite{Link99} concluded from glitches of the Vela pulsar that
at least 1.4\% of the total moment of inertia resides in the crust of
the Vela pulsar.  The fraction of the star's moment of inertia
contained in the crust is connected to the crust's thickness, and both
are controlled by the pressure of matter, $P_t$, at the transition density
as well as the neutron star's mass $M$ and radius $R$.  
The fractional moment of inertia, $\Delta I/I$, where $\Delta I$ is %pje
the moment of inertia in the star's crust and $I$ is the
star's total moment of inertia, can be expressed as~\cite{Lattimer01}~:
\begin{equation}{\Delta I\over I}\simeq{28\pi P_t R^4\over3
GM^2}{(1-1.67\beta-0.6\beta^2)\over\beta}\left[1+{2P_t(1+5\beta-14\beta^2)
\over n_tm_bc^2\beta^2}\right]^{-1}\,,
\label{dii}
\end{equation}
where $\beta = GM/Rc^2$.

The core-crust transition density is not too far from 0.1  %pje
fm$^{-3}$, and matter at these densities is nearly pure neutron
matter, which suggests that the Typel-Brown correlation between $n_t$ and
$\delta R$ has astrophysical significance.  However, the Typel-Brown
correlation refers to the fiducial density, 0.1 fm$^{-3}$.  Thus, even
if the transition density was always 0.1 fm$^{-3}$, the dependence of
$R$ (for a given $M$) on the symmetry energy   %pje
has to be taken into account also.  The major dependence on the
EOS is therefore through the quantity $P_t R^4/M^2$.

A reasonable approximation to the core-crust boundary can be obtained
by considering the zero temperature phase equilibrium between two
homogenous nucleonic fluids, a dense phase approximating matter inside
nuclei (labelled by $i$) and a dilute phase approximating the drip 
nucleons outside nuclei (labelled $o$).  Finite-size effects, such as
those due to the surface and Coulomb energy can be included as in 
Ref.~\cite{Lamb83}, but are ignored here to obtain a first orientation.  The
electrons can be treated as a free Fermi gas which uniformly fills
space in both phases and we define $Y_e$ to be the ratio of electron %pje
and baryon densities. 

Phase equilibrium follows from minimization of the free energy density
allowing for the two conserved quantitites of baryon number and
charge.  We define $u$ to be the fractional volume occupied by the
dense phase $i$.  The dense phase $i$ has a proton fraction $x_i$; the
dilute phase $o$ has a proton fraction $x_o$.  Phase equilibrium,  %pje
while maintaining beta equilibrium, is then described by
\begin{eqnarray}
n&=&n_iu+n_o(1-u)\,,\qquad nY_e=n_ix_iu+n_ox_o(1-u)\,,\qquad
P_i=P_o\,,
\nonumber \\
\mu_{ni}&=&\mu_{no}\,\qquad \mu_{pi}=\mu_{po},,\quad \mu_{ni}-\mu_{pi}=\mu_e\,.
\end{eqnarray}
The transition density occurs at
the baryon density $n=n_t$ for which $u\rightarrow0$.

\begin{figure} 
\begin{center}
\includegraphics[scale=0.4,angle=0]{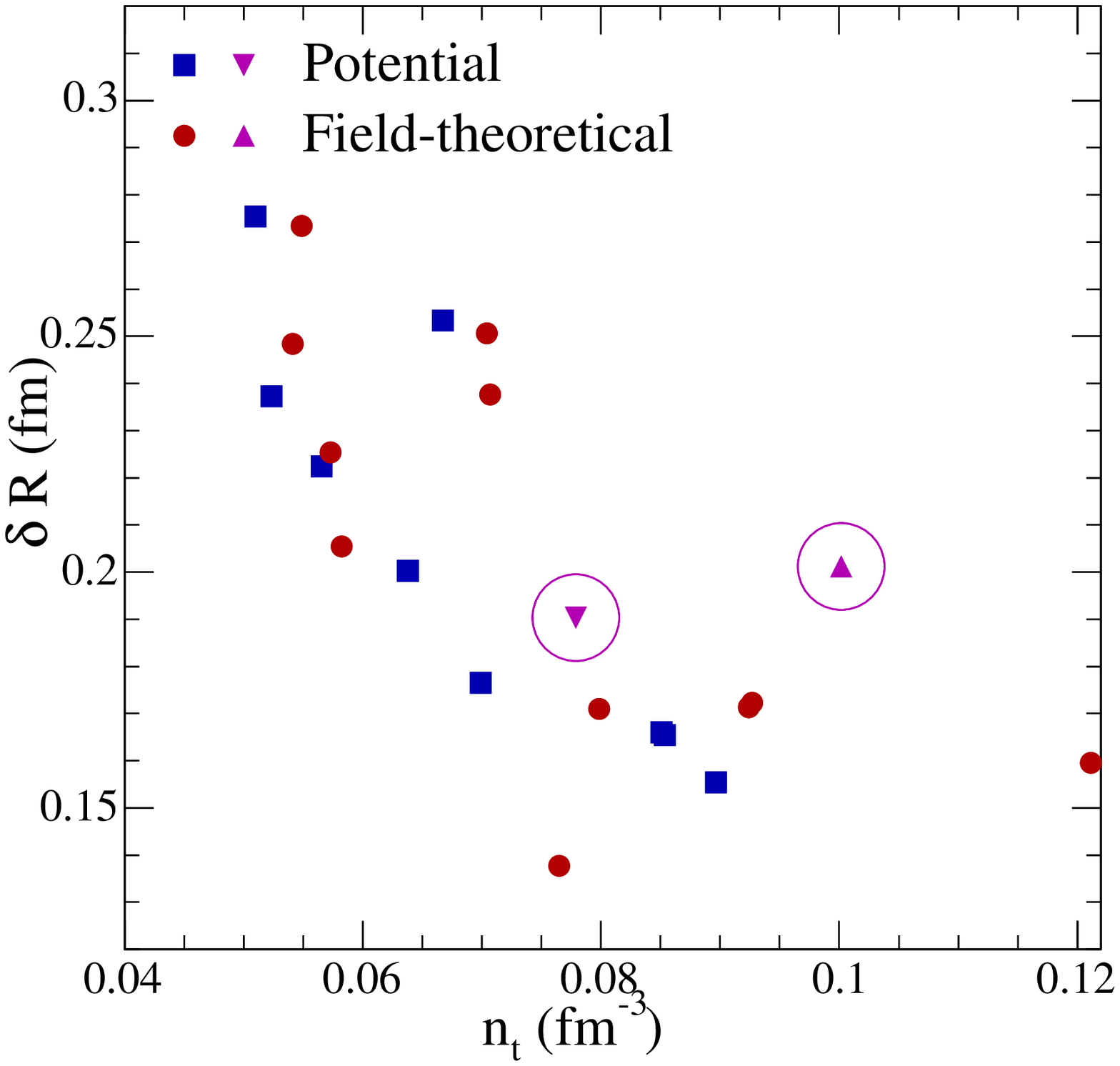}
\includegraphics[scale=0.4,angle=0]{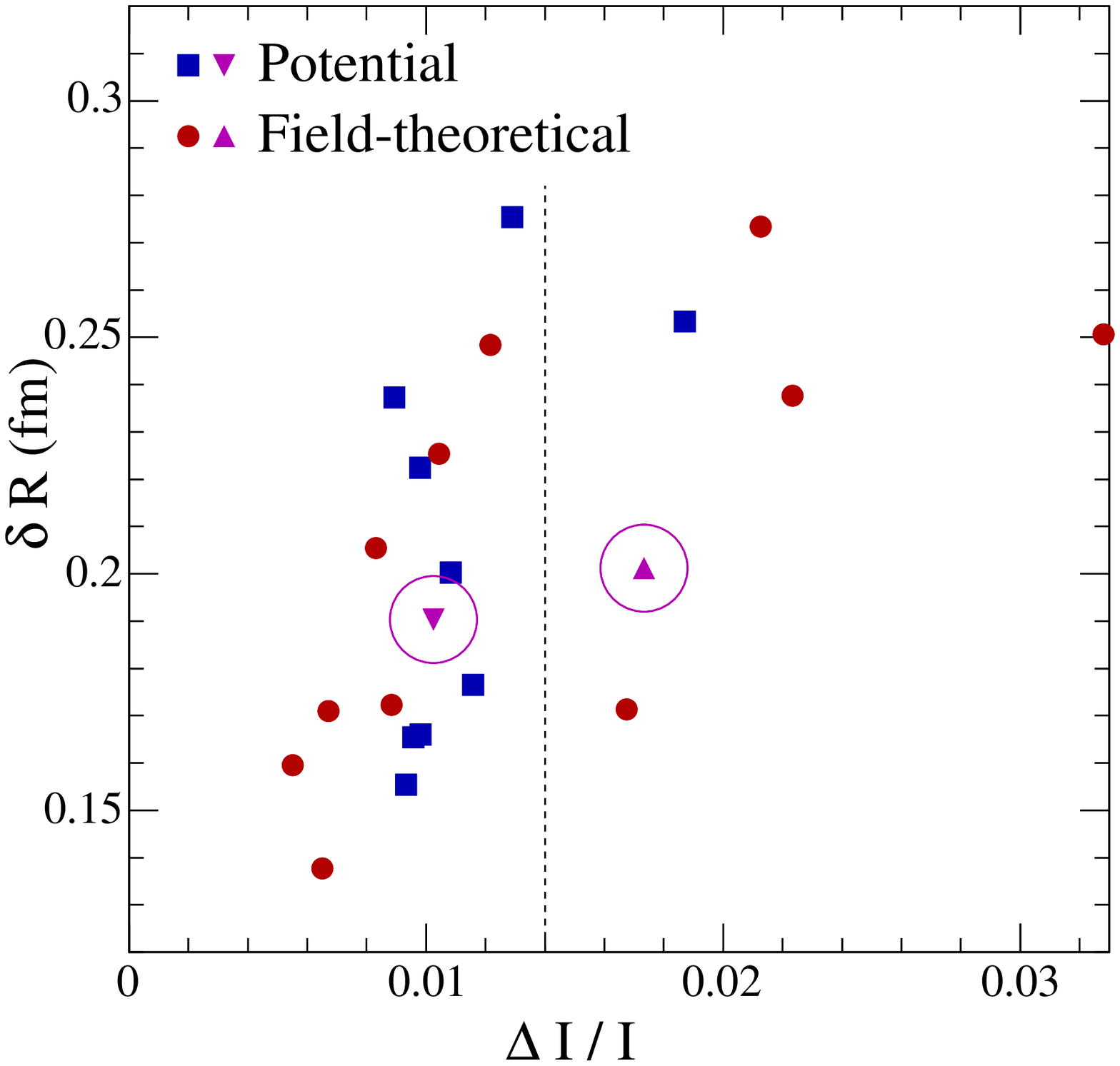}
\end{center} 
\caption{Neutron skin thickness versus the core-crust interface density
$n_t$ (left panel) and  the fractional moment of inertia $\delta I/I$ 
in the crust of a 1.4M$_\odot$ neutron star (right panel).
The dashed line for  $\Delta I/I=0.014$ is deduced for the Vela  
pulsar, assuming a mass for Vela of 1.4${\rm M}_\odot$. }
\label{fig:trans}  
\end{figure}

Figure \ref{fig:trans} displays the relation between $\delta R$ and  
$n_t$ (left panel) and the relation between $\delta R$ and $\Delta
I/I$ (right panel) for a 1.4 M$_\odot$ star.  The transition density
varies over a range of about a factor of 2 in density, from 0.05
fm$^{-3}$ to 0.12 fm$^{-3}$, and there is a negative correlation
between $\delta R$ and $n_t$, but it is not very robust when all the
models are considered. We note that Horowitz and
Piekarewicz~\cite{Horowitz01} computed the transition density as the
density at which an instability to small perturbations occurs in a
uniform liquid of nucleons and electrons (see also
Ref. \cite{Pethick95}).  They obtained a much
tighter correlation between $n_t$ and $\delta R$ by varying the
symmetry energy in a few relativistic models. Using the exact APR 
EOS our method yields $n_t=0.094$ fm$^{-3}$ in close agreement with 
Ref. \cite{Pethick95}; this lies between the RAPR and NRAPR results
plotted in Fig. \ref{fig:trans} being somewhat closer to the former.

Although we do not display it, there is little correlation
between $P_t$ and $\delta R$.  One would expect high values of
$S_v^\prime$ to correspond to high pressures in general. However, high
values of $S_v^\prime$ also tend to have low transition densities, so
the effects somewhat cancel, leaving little net correlation between
$P_t$ and $S_v^\prime$.  However, as discussed above, the
astrophysical observable $\Delta I/I$ is nearly proportional to the
quantity $P_tR^4/M^2$ and not $P_t$ alone.  This is plotted in the right
panel in Fig. \ref{fig:trans}. Although the correlation is not very
strong, it does suggest that, at least in the case of a
1.4M$_\odot$ Vela pulsar, a value of $\Delta I/I \sim 0.014$ implies a
skin thickness $>0.15$ fm for $^{208}$Pb. A larger mass for the
Vela relaxes this lower bound toward lower values of $\delta R$.

\subsubsection{The neutron star radius versus the pressure of
supranuclear neutron-star matter}

\begin{figure}
\begin{center}
\includegraphics[scale=0.4,angle=0]{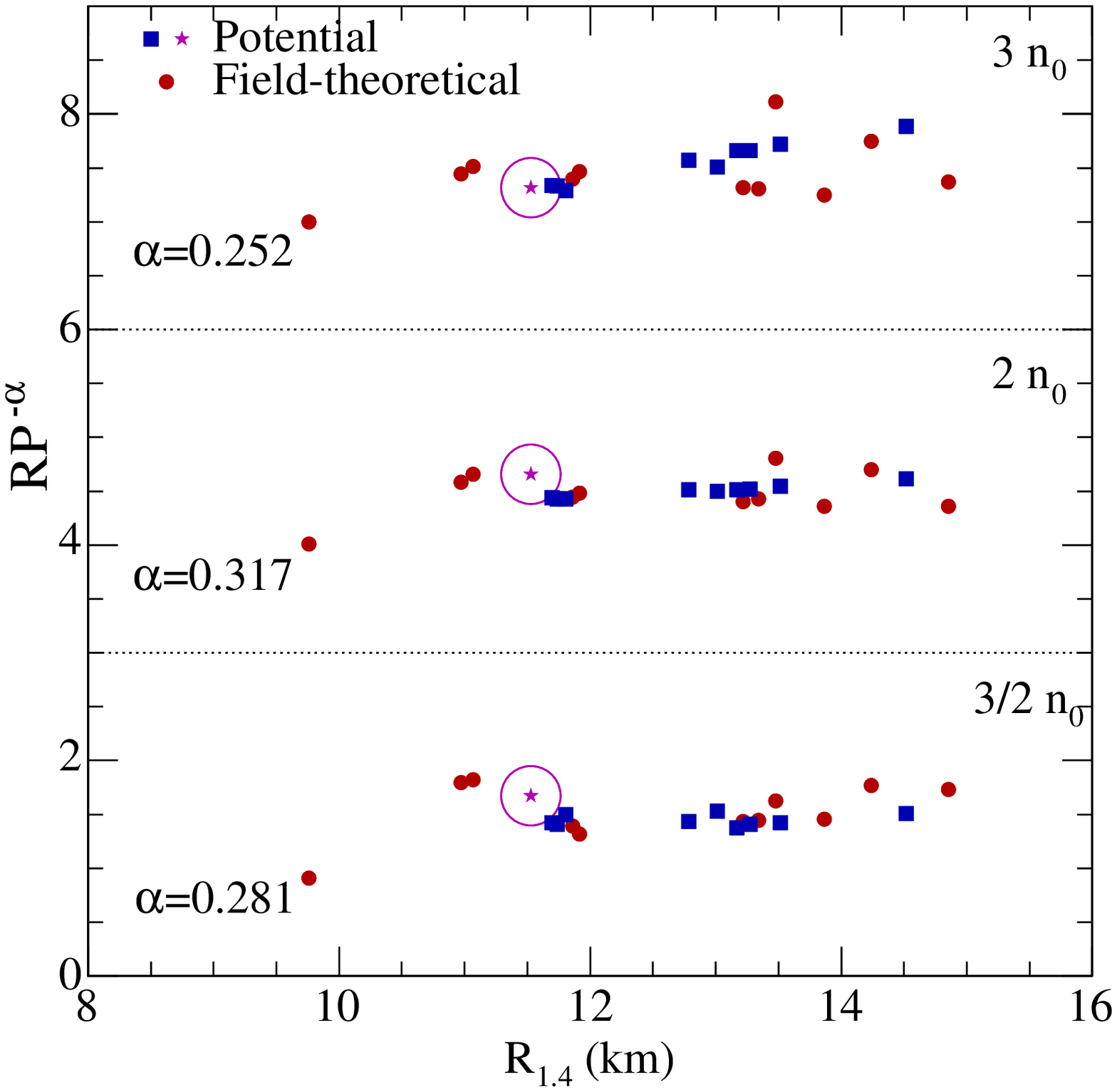}
\includegraphics[scale=0.4,angle=0]{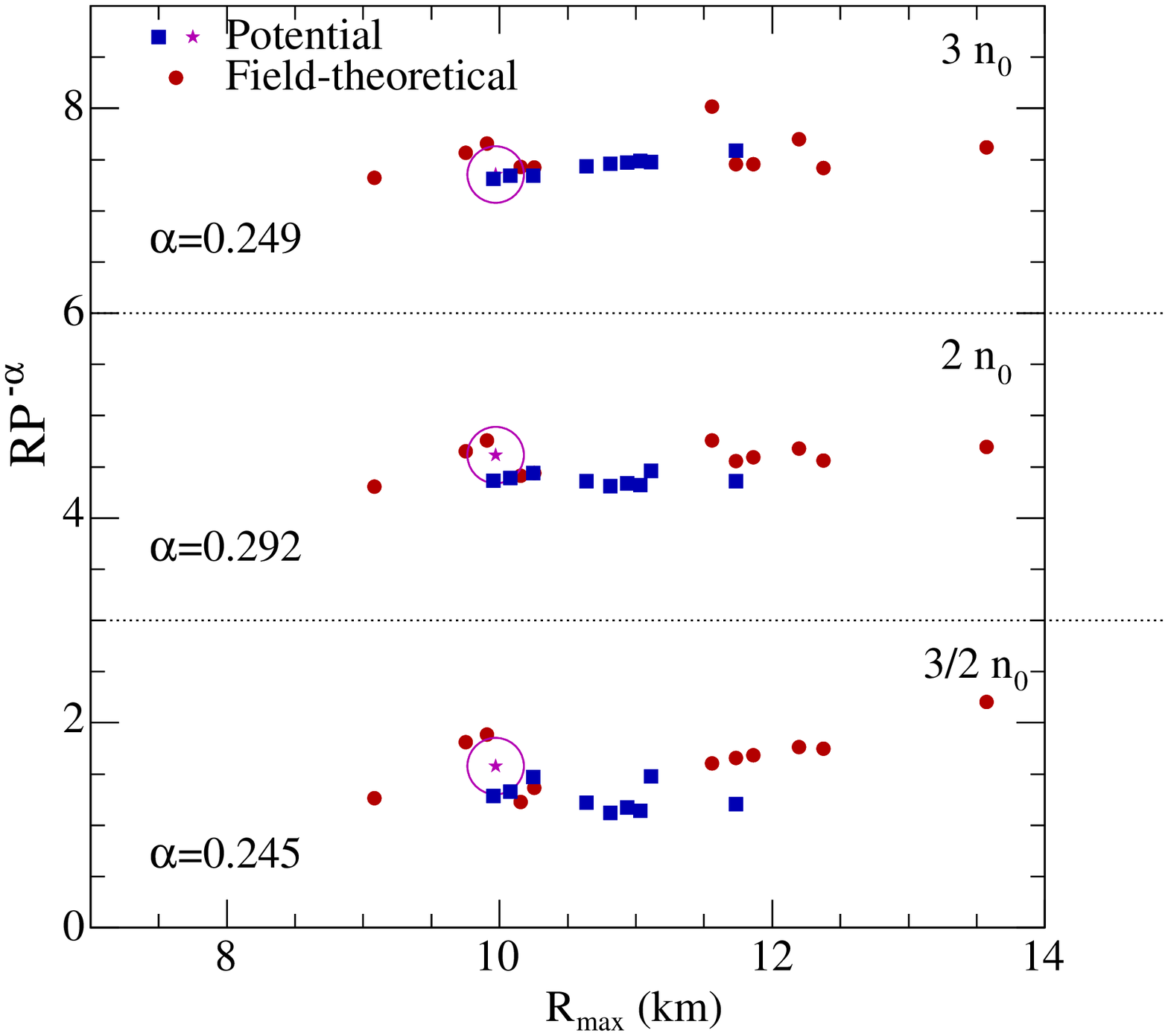}
\end{center} 
\caption{The quantity $RP^{-\alpha}$ as a function of the stellar 
radius $R$, for pressures $P$ determined at 3/2, 2 and 3 times 
equilibrium nuclear matter density.  For each density, 
the best-fit value for the exponent
$\alpha$ is as indicated.  Results for
$1.4M_\odot$  (maximum mass) stars are in the left (right) panel. Circled 
stars indicate the results obtained with the APR equation of state.}
\label{fig:latpi}  
\end{figure}

Lattimer and Prakash \cite{Lattimer01} found that the radius $R$ of a
neutron star exhibits a correlation that has the form of a power law:
\begin{equation}
R \simeq C(n,M)~[P(n)]^{0.23-0.26}\,,
\label{correl}
\end{equation}
where $P(n)$ is the total pressure inclusive of leptonic contributions
evaluated at a density $n$ in the range 1 to 2$n_0$, and    
$C(n,M)$ is a number that depends on the density $n$ at which the
pressure was evaluated and on the stellar mass $M$. In  
Fig.~\ref{fig:latpi} the left and right panels show this 
correlation as $RP^{-\alpha}$ 
versus $R$ for stars of mass $1.4{\rm M}_\odot$ and $M_{max}$,
respectively, for the EOS's considered here and densities $n=1.5-3
n_0$.  In each case, the exponent $\alpha$ was determined by a
least-squares analysis to give the best correlation.  For the optimum
exponent $\alpha = d \ln R/d \ln P \sim 0.25$, the radius increases
very slowly with mass.  As shown in Ref.~\cite{Lattimer01}, general
relativity reduces the value of $\alpha$ from $1/2$, the value
characteristic of an $n=1$ Newtonian polytrope. For the EOS of
Buchdahl \cite{Buchdahl67}, this role of general relativity was
demonstrated analytically \cite{Lattimer01}.

In studying the Lattimer-Prakash radius-pressure
correlation~\cite{Lattimer01}, we have excluded cases in which the
high density EOS was softened because of the presence of non-nucleonic
degrees of freedom.  The correlation, however, is not greatly improved
by this omission, primarily because softening components do not often
affect the equation of state strongly until the density exceeds two to
three times the saturation density.

The discussion in Sec. \ref{section:tb} showed that the pressure at
densities close to equilibrium density is dominated by the derivative
of the symmetry energy. Thus one could alternatively achieve a linear
correlation by plotting $E_{sym}'(3n_0/2)$ versus $R^{1/\alpha}$.

\subsubsection*{Sizes of neutron star radii in potential and
field-theoretical models} 
\label{section:radii}

\begin{figure} 
\begin{center}
\includegraphics[scale=0.4,angle=0]{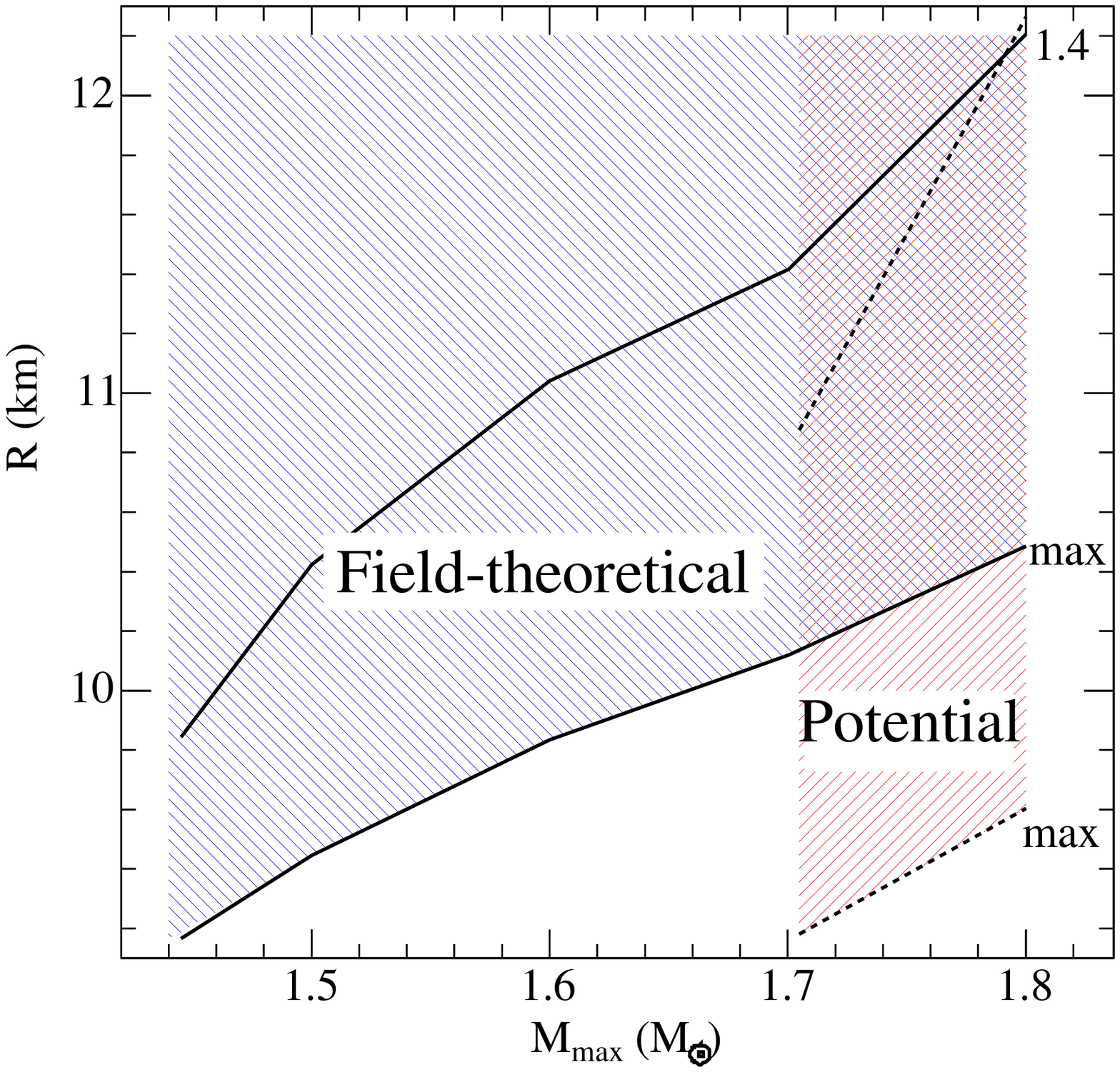}
\includegraphics[scale=0.4,angle=0]{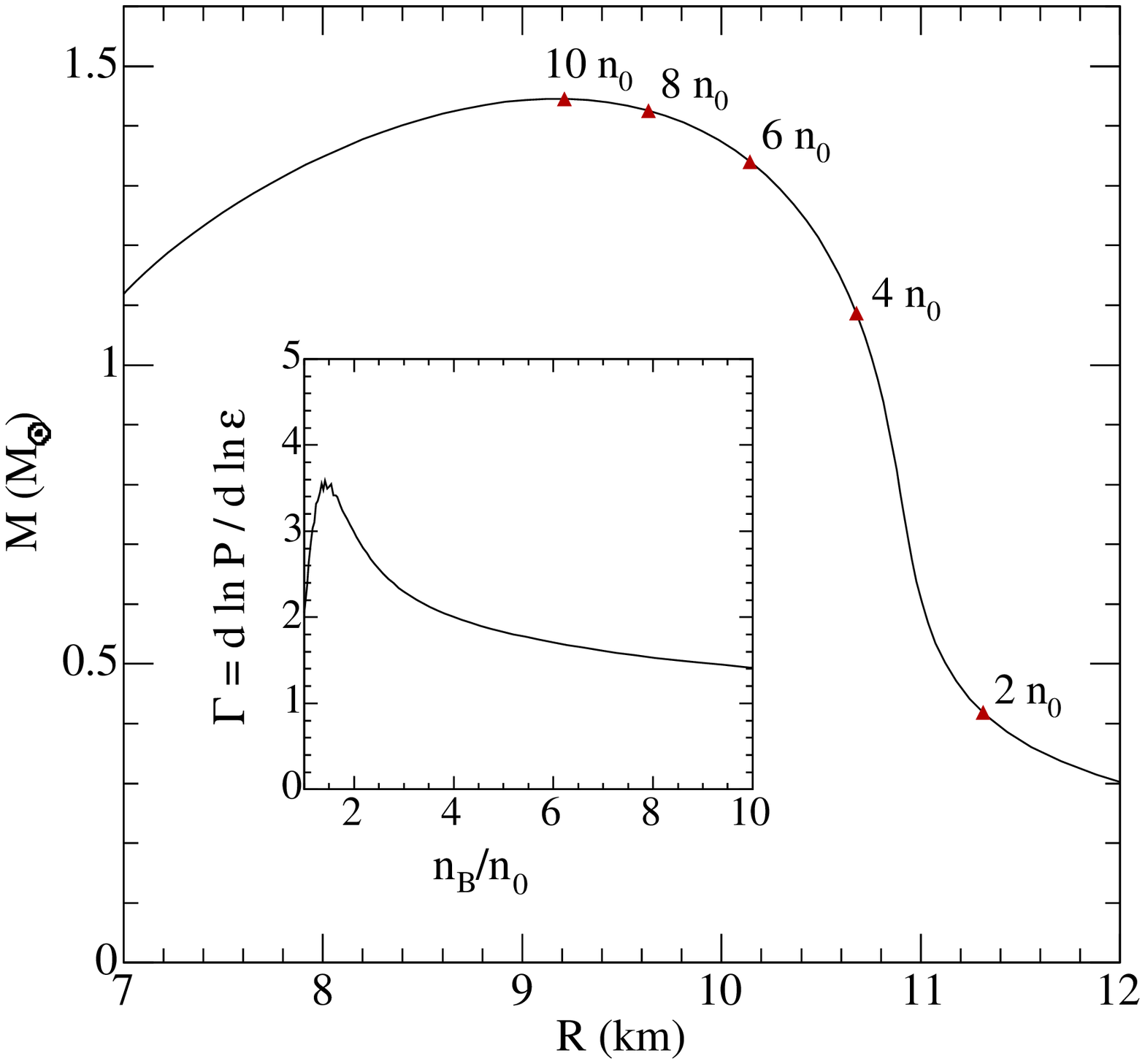}
\end{center} 
\caption{Left panel: The smallest radii of neutron stars given by
our set of EOS's 
as a function of the corresponding maximum masses for field-theoretical 
and Skyrme models (respectively, solid and dashed lines 
labelled ``max''). The radii for $1.4{\rm M}_\odot$ neutron
stars are labelled ``1.4''. The shaded regions indicate radii that are
accessible for a particular class of EOS. Right panel:
The mass--radius relation for the field-theoretical model SR1. The
central densities of stars of different masses are as indicated. The
inset shows the adiabatic index $\Gamma$ versus density for this
model. }
\label{fig:srmass}  
\end{figure}

We have not considered in this work the possible presence of hyperons, 
Bose (pion or kaon) condensates or quarks. Such components generally 
soften the EOS which results in neutron stars with masses and radii 
that are smaller than those with nucleons-only matter. Our
interest here lies in  establishing the minimum radii that 
EOS's with only nucleonic matter yield.  Fig.~\ref{fig:srmass} displays 
this minimum radius as a function of maximum mass for potential and
field-theoretical EOS's that are constrained as described in
Sec. \ref{section:select}.

Each point on the solid line labeled ``max'' is the radius of the
maximum mass configuration for the field-theoretical EOS which yields 
the smallest 
possible radius. The solid line labeled ``1.4'' is the radius of a 
1.4${\rm M}_\odot$ star for the same EOS.  Analogous results for potential 
models are shown by the dashed lines. The hatched regions indicate where
other reasonable potential or field-theoretical models may exist.  For
comparison, the APR equation of state supports a star of maximum mass 
2.2$M_{\odot}$ for which the radius is 10.9 km. The
field-theoretical models SR1, SR2, and SR3 have maximum masses of
1.44, 1.6, and 1.8 $M_{\odot}$, respectively.  
The corresponding radii are 
\begin{eqnarray}
 R_{1.4} &=& 9.84,~ 11.04,~{\rm and}~ 12.21~{\rm km} 
\qquad {\rm for}~M=1.4{\rm M}_\odot \nonumber \\ 
R_{max} &=& 9.17,~ 9.83,~ {\rm and}~ 10.49~{\rm km}~
\qquad {\rm for}~M=M_{max} \,.
\end{eqnarray}
The mass-radius
relation of the model SR1 with a maximum mass of 1.44 $M_{\odot}$
is shown in the right panel of Fig. \ref{fig:srmass}. 
Of the many field-theoretical 
models considered in this work, this model yields the smallest radius.
%Fig.~\ref{fig:smallr}.

\begin{figure}
\begin{center}
\includegraphics[scale=0.4,angle=0]{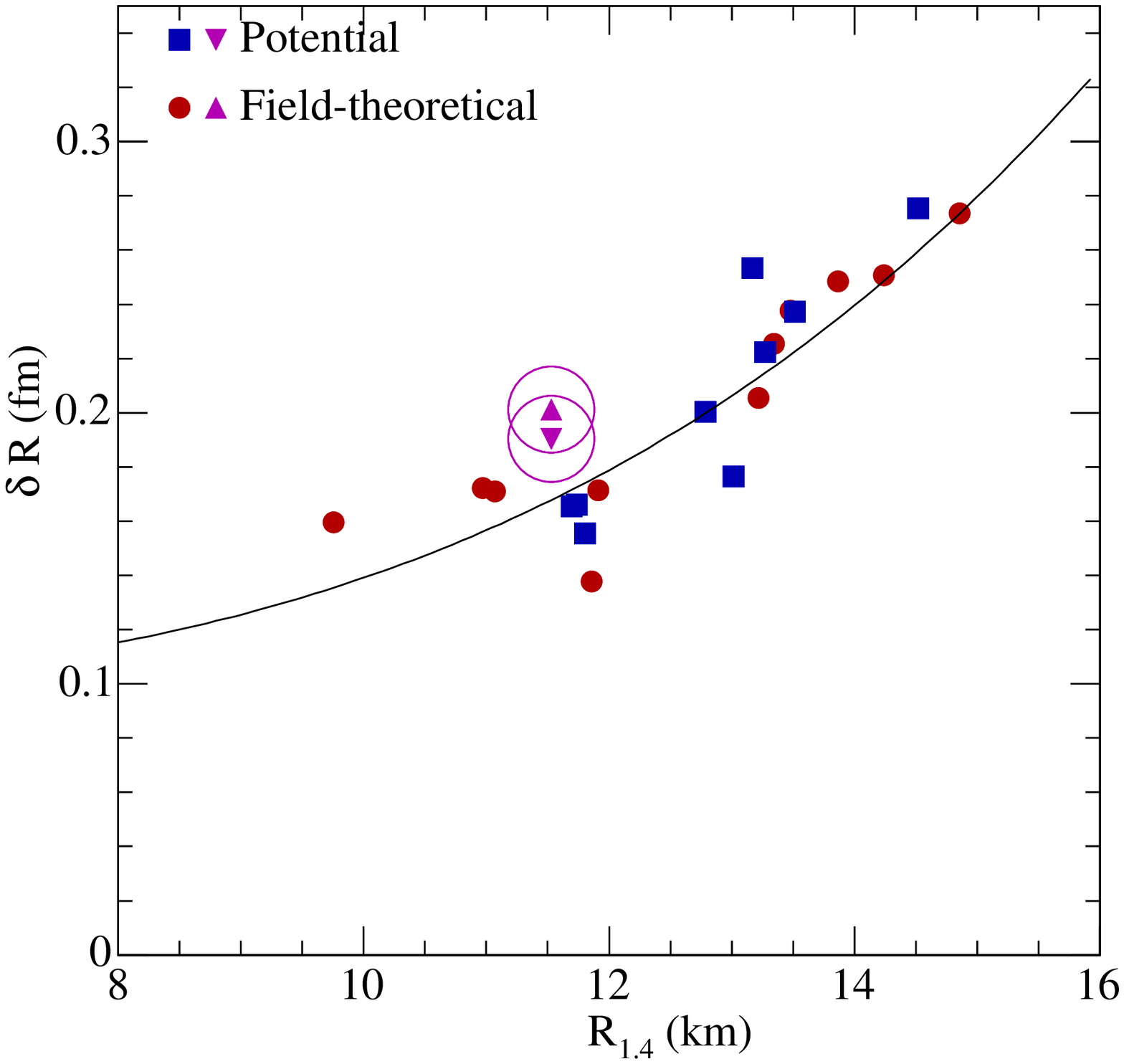}
\includegraphics[scale=0.4,angle=0]{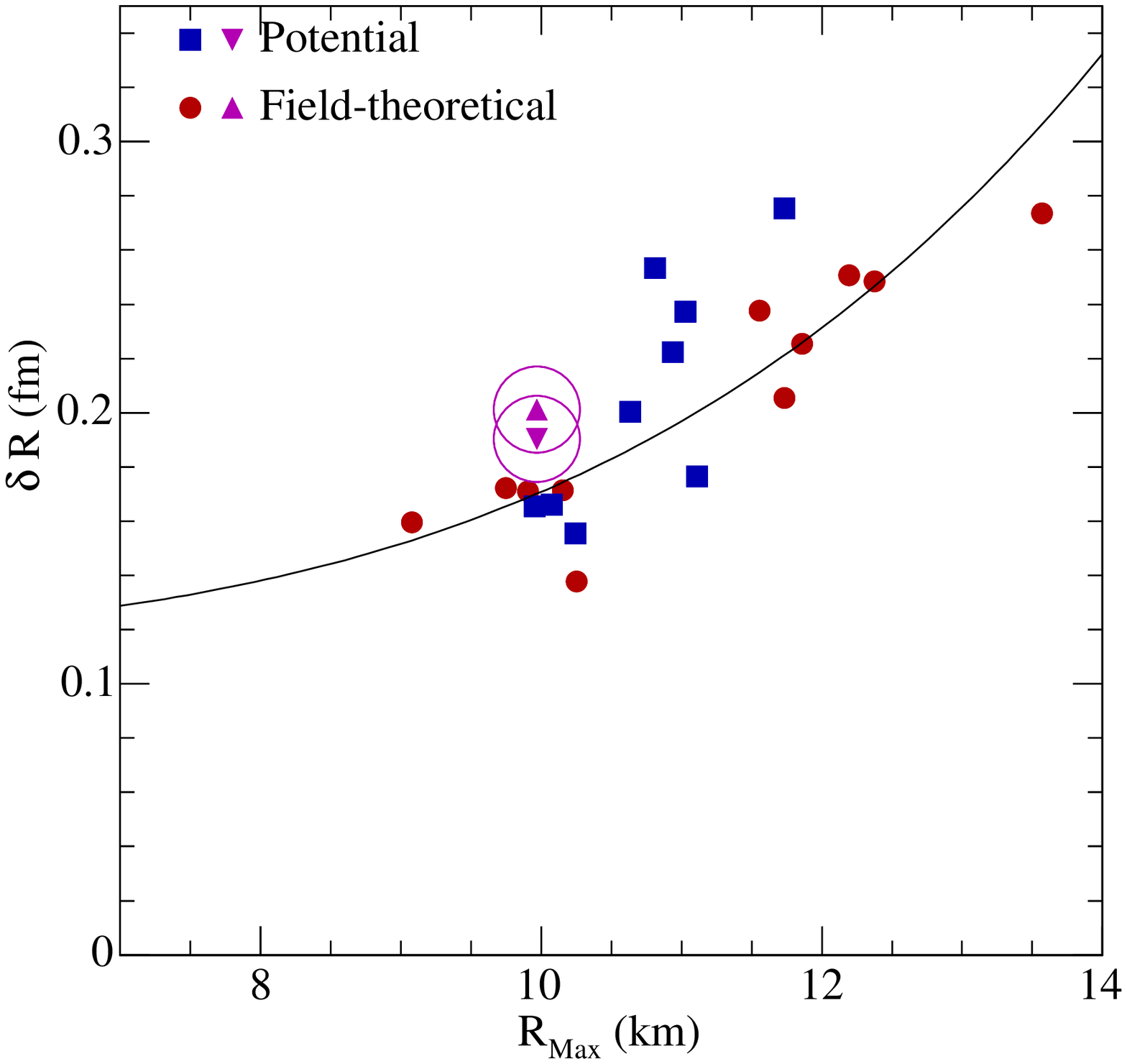}
\end{center}
\caption{Calculated neutron skin thickness $\delta R$ of nuclei versus
the radius of $1.4M_\odot$ stars (left panel) and of maximum mass
stars (right panel). The solid lines are described in the text.}  
\label{figtcorr2}
\end{figure}

\subsubsection{The neutron skin thickness versus the neutron star radius} 

We turn now to the Horowitz-Piekarewicz correlation for the models 
considered in this work.  Our results are shown in
Fig.~\ref{figtcorr2} for a $1.4{\rm M}_\odot$ (left panel) and for the
maximum mass  (right panel) 
configuration for each EOS. Note especially
the results of models SR1, SR2, and SR3 which buck the trend of
field-theoretical models that generally yield large neutron star
radii. As noted in Ref.~\cite{Lattimer01}, large radii are a
consequence of symmetry energies that vary rapidly with density beyond
the nuclear saturation density.

\begin{figure}
\begin{center}
\includegraphics[width=0.5\textwidth]{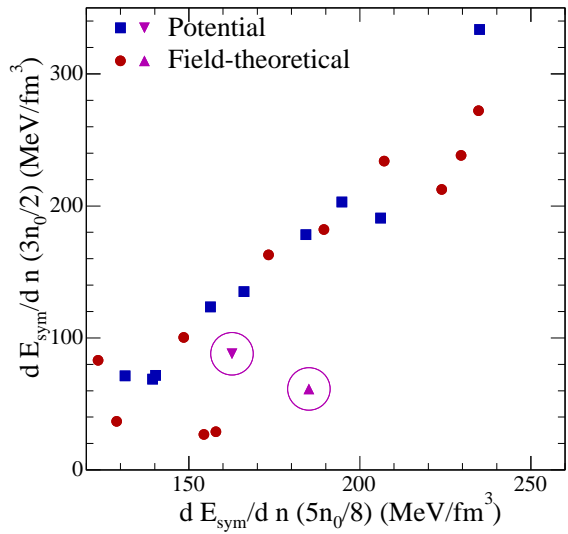}
\end{center}
\caption{The derivative of the symmetry energy at  $3n_0/2$  versus
that at $5n_0/8$.  }
\label{chimchim}
\end{figure}

The Typel-Brown correlation demonstrates that the neutron skin
thickness $\delta R$ is linearly correlated with $E_{sym}'(5n_0/8)$,
whereas the Lattimer-Prakash correlation for the neutron star radius
establishes a linear correlation between $R^{1/\alpha}$ and
$E_{sym}'(3n_0/2)$. These two correlations can be synthesized by
comparing $E_{sym}'$ at the two different densities. This is carried
out in  Fig.~\ref{chimchim} where,
notwithstanding some scatter in the results, there is an overall
linear correlation between $E_{sym}'(5n_0/8)$ and $E_{sym}'(3n_0/2)$.
This suggests a relationship of the form $\delta R=a+bR^{1/\alpha}$.
Determining the coefficients by a least-squares fit to the results 
in  Fig.~\ref{figtcorr2} yields
\begin{eqnarray}
\delta R &=& \left(0.09558  + 1.201\times 10^{-5} ~ 
R^{1/0.281}\right)~{\rm fm}~ \qquad 
{\rm for}~M=1.4{\rm M}_\odot \nonumber \\ 
\delta R &=& \left(0.1160  + 4.540\times 10^{-6} ~ 
R^{1/0.245}\right)~{\rm fm}~ \qquad 
{\rm for}~M=M_{max} \,,
\label{delrral}
\end{eqnarray}
These relationships are depicted by solid lines in Fig.~\ref{figtcorr2}.
For a fiducial $\delta R=0.2\pm0.025$ fm,
$R_{1.4} = (13\pm 0.5)$ km and $R_{max} = (11 \pm 0.5)$ km. In the
latter case, $M_{max}$ is found to vary up to $2.2{\rm
M}_\odot$. These values give a measure of the extent to which the
neutron star radius can be constrained if the neutron skin thickness
is measured to the specified accuracy. For a lower mean value of
$\delta R$ the uncertainty in the radius is larger.

Schramm \cite{Schramm03} further explored the Horowitz-Piekarewicz
correlation~\cite{Horowitz01,Horowitz01c} between the neutron skin 
thickness and the neutron star radius by using covariant
field-theoretical models based on SU(3) symmetry. He found that the
correlation was weak because of the restricted variation allowed in
the fourth-order vector-isovector couplings.  In our work we have
eliminated EOS's which do not satisfy minimal astrophysical and
experimental nuclear constraints, which results in a tight correlation
despite the enlarged freedom in the description of the symmetry
energy. In fact, combining the correlations  as discussed above, 
we find a skin thickness which scales approximately with the stellar 
radius raised to a power $\sim$ 4.

\section {OTHER RELATED CORRELATIONS}

We turn now to address a few other correlations connected with isospin
asymmetry.  Supplements to our brief account here are contained in
several reviews, some of which are alluded to below.

\subsection{Giant Dipole Resonances}

The analyses of giant resonances, particularly dipole resonances, have
long served to delineate the role of volume and surface effects in
nuclei; for a review see Ref.~\cite{Lipparini89}. In medium to heavy
nuclei in which magnetic contributions are small, the 
inverse-energy-squared weighted photoabsorption cross section can be 
related to the
static polarizability $p$ as
\begin{equation}
\sigma_{-2} = \int \frac{\sigma(\omega)}{\omega^2}\,d\omega 
=2\pi^2(e^2/\hbar c)p \,.
\end{equation} 
In nuclei with mass number $A \geq 100$, $\sigma_{-2} =
(2.9\pm0.2)A^{5/3}~\mu{\rm b~MeV}^{-1}$ parameterizes the  
data~\cite{Bergere77,Laszewski79}. The dipole polarizability can be
evaluated as the response of a nucleus to an external dipole field
$\eta D$, where $\eta$ denotes the strength and $D=(1/2)\sum_{i=1}^A
z_i\tau_i^3$ is the dipole operator. Explicitly,
\begin{equation}
p = 2\sum_{n \neq 0} \frac {\left|\langle 0\left| D \right|
  n\rangle\right|^2}{\omega_n - \omega_0} 
%= 2 \left.\frac {d^2E(\eta)}{d\eta^2}\right|_{\eta=0} 
\,, 
\end{equation}
where $|n\rangle$ and $\omega_n$ are the eigenstates and eigenenergies
of the nuclear Hamiltonian responsive to the dipole operator $D$.
Microscopic RPA calculations~\cite{Bohigas81} of $p$ employing
Skyrme-like interactions reproduce the general trends of the data. The
interplay between volume and surface effects are, however, difficult
to extract from RPA calculations. Semiclassical methods, in which the
energy density formalism is coupled with a hydrodynamic approach to
describe collective excitations, have thus been employed to explore
how the volume and surface symmetry energies, $S_v$ and $S_s$,
determine the value of $p$ across the periodic table
\cite{Lipparini82,Krivine82,Lipparini89}.  Using the Hamiltonian
density in Eq.~(\ref{eq:basicham}), and writing the transition density
as $\delta[\rho_n(r)-\rho_p(r)] = \eta \phi(r)\cos\theta$, the
polarizability is calculated from
\begin{equation}
p = \frac {2\pi}{3} \int \phi(r) r^3 \, dr \,,
\end{equation}
where $\phi(r)$ is the solution of the corresponding Euler-Lagrange
(integro-differential) equation.  
It is found that similar results
for $p$ are obtained for correlated values of $S_v$ and $S_s$. 
Numerical results for
$^{40}$Ca, $^{120}$Sn, and $^{208}$Pb show that the choice of $S_v$
in the range 27--42 MeV requires $|S_s/S_v|$ = 1.2--2.2; lower values of
$S_v$ correlate with lower values of $S_s$ for good fits \cite{Krivine82}.
The slope of this correlation can be compared to that found using 
ground state energetics in  Sec. IV C 1 (see Fig. \ref{chi_droplet}).
It is in close agreement with the correlation obtained with the
`$\mu_n$' model.

A qualitative understanding of this correlation can be gained by using
a schematic symmetry energy density functional and 
a leptodermous expansion of the density ~\cite{Lipparini89}, 
whence one obtains the result
\begin{equation}
p = \frac {A}{24} \frac {\langle r^2 \rangle}{S_v}
\left( 1 + \frac 53 \frac {S_s}{S_v} A^{-1/3} + \cdots  \right) \,,
\end{equation}
where 
$\langle r^2\rangle$ is the mean
square radius of the nucleus and 
higher order terms contain corrections from the diffuseness and
the skin thickness. Although values of $S_v=32.5$ MeV and $|S_s/S_v|=2.2$
describe the data adequately~\cite{Lipparini89}, correlated variations
in these numbers are allowed as pointed out in Ref.~\cite{Krivine82}.
                                                                             
Sum rules have been particularly useful to relate experiments and
theory in discussing the mean excitation energies, widths, and the
spreading of the excitation strengths \cite{Lipparini89}. The moments
$m_p$ of the strength function $S(\omega) = \sum_{n > 0} \left|
\langle n \left| F \right| 0 \rangle \right|^2 \delta (\omega -
\omega_n) $ defined by
\begin{equation}
m_p = \int_0^\infty S(\omega) \omega^p\,d\omega = \sum_{n > 0} 
\left| \langle n \left| F \right| 0 \rangle \right|^2 \omega_n^p \,, 
\end{equation}
where $F$ is the physical operator exciting the nucleus from its
ground state $| 0 \rangle$ to its eigenstate $| n \rangle$, are
especially helpful in this regard. For example, a good measure of the
mean excitation energy is provided by $E(D) = \sqrt
{m_1/m_{-1}}$, for which results from RPA and hydrodynamic
calculations have been compared with data \cite{Lipparini89}.  
Using a droplet model coupled with a hydrodynamic approach to excite
the dipole resonance, Ref.~\cite{Lipparini88} obtains
\begin{equation}
E(D) = {\sqrt {
\frac {6\hbar^2(1+K_D)}{M\langle r^2 \rangle} 
\frac {S_v}{1+\frac 53 \frac {S_s}{S_v}A^{-1/3}} }} \,,
\end{equation} 
where $M$ is the nucleon mass.  The quantity $K_D$ is a model
dependent enhancement factor characterizing the relative contribution
of the nuclear interaction to the $m_1$ sum rule and depends
critically on the value of the energy up to which the energy
integration is carried out in analyzing the experimental data. Using
values of $S_v=32.5$ MeV and $S_s/S_v \simeq 2$ from a droplet model
fit to nuclear energies, and $K_D=0.2$ corresponding to $E_{max}=30$
MeV, Ref.~\cite{Lipparini88} finds that a hydrodynamic approach is
able to reproduce the experimental mean excitation energy of nuclei
ranging from $^{40}$Ca to $^{208}$Pb reasonably well.  Several Skyrme
interactions with different values of $S_v$ and $S_s/S_v$ are also able to
account for the data to the same level of accuracy.  However, since
the values of $S_v$ and $S_s$ for these forces are correlated because
they are fit to experimental masses, the additional constraint on the
permitted ranges of $S_v$ and $S_s$ resulting from fitting dipole 
resonances does not seem to be significant.

It must be emphasized that while a hydrodynamic approach is able to
account for the gross features of nuclear mass dependence, several 
detailed features of the data such as strength fractionation,
spreading widths, etc., are not naturally incorporated in its
scope. For such details one must adopt a more microscopic
approach that includes, for example, contributions from 2p--2h
excitations, etc. For a detailed account, see, for example,
Ref.~\cite{Khamerdzhiev97}.

\subsection{Heavy-Ion Collisions}

\subsubsection{Multi-fragmentation}

The breakup of excited nuclei into several smaller fragments during an
intermediate-energy heavy-ion collision probes the phase diagram of
nucleonic matter at sub-saturation density and moderate ($\sim 10-20$
MeV) temperatures. In this region of the phase diagram the system is
mechanically unstable if $(dP/dn)_{T,x}<0$, and/or chemically
unstable if $(d\mu_p/d x)_{T,P}<0$. (A pedagogical account of 
such instabilities can be found in
Ref.~\cite{Chomaz04}). These instabilities, which are directly related
to the symmetry energy at sub-saturation densities~\cite{Li97}, are
believed to trigger the onset  of multifragmentation. Because of the
instabilities, matter separates into coexisting liquid and gas phases,
which each have different proton fractions, {\it i.e.} ``isospin
fractionation''~\cite{Xu00}.  This fractionation is observed in the
isotopic yields which can potentially reveal information about the symmetry
energy.  Also, the scaling behavior of ratios of isotope yields  
measured in separate nuclear reactions, ``isoscaling'', is sensitive
to the symmetry energy~\cite{Tsang01,Ono03}. This scaling is expressed
in the empirically observed ratio of fragment yields from two similar
systems with different neutron-to-proton ratios:
\begin{equation}
Y_2(N,Z)/Y_1(N,Z) \sim \exp^{\alpha N+\beta Z} \,,
\end{equation}
where the constants $\alpha$ and $\beta$ can be related to the
neutron and proton chemical potentials in a canonical ensemble 
description, and thus to the proton fraction of the source.
To date, there are many suggestions of
how the symmetry energy may affect multifragmentation~\cite{Das04}. Ongoing
research is concerned with an extraction of reliable constraints on
the symmetry energy from the presently available experimental
information.

\subsubsection{Collective Flow}

Nuclear collisions in the range $E_{lab}/A = 0.5-2$ GeV offer the
possibility of pinning down the equation of state of matter above
normal nuclear density (up to $\sim 2 ~{\rm to}~ 3n_0$) from a study
of matter, momentum, and energy flow of nucleons \cite{Gutbrod89}.
The observables confronted with theoretical analyses include (i) the
mean transverse momentum per nucleon $\langle p_x \rangle /A$ versus
rapidity $y/y_{proj}$ \cite{Danielewicz85}, (ii) flow angle from a
sphericity analysis \cite{Gustafsson84}, (iii) azimuthal distributions
\cite{Welke88}, and (iv) radial flow \cite{Siemens79}.  Flow data
gathered to date are largely for protons (as detection of neutrons is 
more difficult) and for collisions of laboratory nuclei in which the
isospin asymmetry is not large.  Theoretical calculations have
generally been performed using Boltzmann-type kinetic equations. One
such equation for the time evolution of the phase space distribution
function $f({\vec r},{\vec p},t)$ of a nucleon that incorporates both
the mean field $U$ and a collision term with Pauli blocking of final
states is (see, for example, Ref. \cite{Bertsch88})
\def\sst{\scriptscriptstyle}
\begin{eqnarray}
\frac {\partial f}{\partial t} + {\vec \nabla}_p U \cdot {\vec \nabla}_r f
- {\vec \nabla}_r U \cdot {\vec \nabla}_p f =
&-& \frac {1}{(2\pi)^6} \int d^3p_2\,d^3p_{2^{\prime}}\,d\Omega 
\frac {d\sigma_{\sst NN}}{d\Omega} \, v_{12} 
(2\pi)^3\,\delta^3( {\vec p} + {\vec p}_2 - {\vec p}_{1^{\prime}} 
- {\vec p}_{2^{\prime}} ) \nonumber \\ 
&&  \times               
\left[ ff_2 (1-  f_{1^{\prime}}) (1-  f_{2^{\prime}})
- f_{1^{\prime}} f_{2^{\prime}} (1- f) (1- f_2) \right] \,.
\label{BUU}
\end{eqnarray}
Above, ${d\sigma_{\sst NN}}/{d\Omega}$ is the differential
nucleon--nucleon cross--section and $v_{12}$ is the relative velocity.
In general, the mean field $U$ depends on both the density $n$ and the
momentum ${\vec p}$ of the nucleon.  Equation (\ref{BUU}) contains
effects due to both hard collisions and soft interactions, albeit at a
semiclassical level.  Theoretical studies that confronted data have
thus far used isospin averaged nucleon-nucleon cross sections and mean
fields of symmetric nuclear matter.  It is now well established that
much of the collective behavior observed in experiments stems from
momentum dependent forces at play during the early stages of the
collision \cite{Gale90}. The conclusion that has emerged from several
studies is that as long as momentum dependent forces are employed in
models that analyze the data, a symmetric matter compression modulus
of $\sim 220$ MeV, as suggested by the analysis of the giant monopole
resonance data \cite{Youngblood99}, fits the heavy-ion data as well
\cite{Danielewicz02}.
 
The prospects of rare isotope accelerators (RIA's) that can collide
highly neutron-rich nuclei has spurred further work to study a system
of neutrons and protons at high neutron excess
\cite{Das03,Li04,Li04b}. Generalizing Eq.~(\ref{BUU}) to a mixture,
the kinetic equation for neutrons is
\begin{equation}
\frac {\partial f_n}{\partial t} + {\vec \nabla}_p U \cdot {\vec \nabla}_r f_n
- {\vec \nabla}_r U \cdot {\vec \nabla}_p f_n = J_n = \sum_{i=n,p} J_{ni}\,, 
\end{equation}
where $J_n$ describes collisions of a neutron with all other neutrons
and protons.  A similar equation can be written down for protons with
appropriate modifications.  On the left hand side of each coupled
equation the mean field $U\equiv U(n_n,n_p;{\vec p}\,)$ depends
explicitly on the neutron-proton asymmetry. The connection to the
symmetry energy arises from the fact that $U$ is obtained from a
functional differentiation of a Hamiltonian density, such as that in
Eq.~(\ref{eq:basicham}).  Examples of such mean fields may be found in
Refs.~\cite{Prakash97,Li04,Li04b}. Observables that are expected to
shed light on the influence of isospin asymmetry include
neutron-proton differential flow and the ratio of free neutron to
proton multiplicity as a function of transverse momentum at
midrapidity.  Experimental investigations of these signatures await
the development of RIA's at GeV energies.  In this connection, it will
be important to detect neutrons in addition to protons.

\section{DISCUSSION AND CONCLUSIONS }

In this work, we have investigated the relationship between the symmetry
energy of nucleonic matter and the neutron skin thicknesses of
neutron-rich nuclei as well as the radii of neutron stars.  Precision
measurements of neutron skins and neutron star radii are due to become
available in the near future.  We have studied how these measurements
can constrain the dependence of the symmetry energy on baryon density,
$E_{sym}(n)$, in the vicinity of the nuclear saturation density $n_0$.
This knowledge is crucial to understanding many astrophysical
phenomena, including neutron star evolution, supernova explosions and
nucleosynthesis, and binary mergers involving neutron stars.
Simulations of these phenomena involve extrapolations of the symmetry
energy to supranuclear densities.

A crucial aspect of our presentation is that care was taken to ensure
that the nuclear force parameterizations, whether for potential or 
field-theoretical models, were constrained by fits to closed-shell
nuclei.  In addition, all models were constrained to yield a maximum
neutron star mass of at least 1.44M$_\odot$, the larger of the
accurately measured neutron star masses in the binary pulsar 1913+16.
Without such constraints, much weaker correlations relating the
neutron skin thickness to astrophysical quantities such as neutron
star radii would have resulted.

Analytical representations of the relation between the symmetry energy
and the neutron skin were developed for semi-infinite surfaces
calculated in potential and field-theoretical models.  To lowest
order, the neutron skin thickness $\delta R ={\langle
r_n^2\rangle}^{1/2} - {\langle r_p^2\rangle}^{1/2}$ is proportional to
$\delta_L S_s/S_v$ where $\delta_L$ is the neutron excess in the
nuclear center and $S_v [\equiv E_{sym}(n_0)]$ and $S_s$ are related
to the expansion parameters of the volume and surface symmetry
energies, respectively, in the liquid drop or droplet models of
nuclei.  This representation is validated by comparisons with results
of finite nucleus calculations performed in the
Hartree-Fock-Bogoliubov (for potential models) and Hartree (for
field-theoretical models) approximations.  For the first time, results
for nuclei are presented for the Akmal-Pandharipande-Ravenhall EOS
parameterized according to both potential and field-theoretical models. 

The semi-infinite surface representation predicts that the ratio
$S_s/S_v$, or equivalently, the skin thickness $\delta R$, is a
particular average of the density-dependent symmetry energy in the
surface.  In addition, phenomenological comparisons indicate that
there is a relatively tight correlation between $\delta R$ and the
pressure of neutron star matter at a typical average surface density
of $5n_0/8$.  These results are not independent, but together imply
that the determination of $\delta R$ offers a valuable constraint on
the density dependence of the symmetry energy.  It was explicitly
shown that the $\delta R$ -- pressure correlation is a consequence of
a more general correlation between $\delta R$ and the derivative of
the symmetry energy at the same density.  
In contrast to most, but  
not all, previous results, we have also demonstrated that field
theoretical models can yield a neutron skin thickness in $^{208}$Pb of
less than 0.2 fm.  Since a measurement of the neutron radius to about
1\% accuracy implies an error of about 0.05 fm, the uncertainty in the
neutron skin thickness will probably range from 10\% to 30\%,
depending upon the magnitude of $\delta R$.  On this basis, it is
expected that the pressure at 5/8 $n_0$ could be determined to
20--50\%.

An independent method of constraining $S_s/S_v$ is from a least
squares fit of nuclear models to nuclear binding energies.  However,
this fitting can only reliably establish a correlation between
$S_s/S_v$ and $S_v$.  We have examined this in the liquid droplet
approach using two plausible models.  The surface tension was
parameterized either in terms of $\mu_n$, implying a neutron skin, or
in terms of $\mu_n-\mu_p$ which allowed for the Coulomb interaction of
protons. While these models gave fits to the data of similar accuracy,
the valley in $\chi^2$ differed in the two cases. The behavior of the
`$\mu_n$' model was shown to be more consistent with the systematic
correlation derived from parameter optimization of both potential and
field-theoretical models of laboratory nuclei.

The density dependence of the symmetry energy has a number of
astrophysical consequences.  One of the most important is its role in
determining the composition of matter.  In beta equilibrium, the
proton fraction is proportional to $E_{sym}(n)^3$ for small proton
fractions.  The charge fraction of matter plays an essential role in
establishing the threshold densities of hyperons and the quark-hadron
phase transition.  If any of these threshold densities are exceeded in
a neutron star, the possibility of enhanced neutrino emission by a
direct Urca process exists.  The threshold for the direct Urca process
involving nucleons alone is also determined by the charge fraction
(for example, it is 1/9 in the absence of muons).  We have shown that
the determination of $\delta R$ to have a value greater than
approximately 0.2 fm would imply that at least the nucleon direct Urca
process exists in 1.4 M$_\odot$ neutron stars.

An additional astrophysical application is the relation we established
between $\delta R$ and the radii $R$ of neutron stars.  Although
Lattimer and Prakash established a phenomenological relation between
$R$ for a given mass star and the pressure of neutron star matter
above, but near, the nuclear saturation density, there exists
additional uncertainty in relating this algebraically to the $\delta R
- P$ correlation.  However, as shown in this work, a useful relation
directly relating $\delta R$ and, for example, the radius of a
1.4M$_\odot$ star, $R_{1.4}$, can be established.

The fact that the neutron skin thickness, which measures symmetry
properties below nuclear density, and the astrophysical quantities
involving the neutron star radius and the Urca threshold density, can
be correlated is a consequence of a generic trend in the overall
density dependence of the symmetry energy.  This trend was
specifically demonstrated by displaying the linear relationship which
exists between the density derivatives of the symmetry energy at
$5n_0/8$ and $3n_0/2$.  This relationship is independent of the
parameterization of the nuclear force. 

Observation of neutron star radii may provide qualitative information
on the stellar composition. Using the parametric freedom of the EOS's
considered, while satisfying the aforementioned constraints, we
concluded that the minimum radius achievable for a star composed of
just nucleons and leptons is about 9 km. Observation of a
significantly smaller radius would likely imply that some
softening component (bose condensates, hyperons, quarks, etc.) is
present in dense matter, although it could also indicate that the EOS's
given by potential and field-theoretical models is qualitatively
incorrect. 

We have also explored the influence of the density dependence of the
nuclear symmetry energy on the transition pressure marking the
boundary between the core and the crust in a neutron star.  This
transition pressure is closely approximated by the phase boundary
separating uniform matter from matter in which nuclei exist.  It was
found that the transition pressure was not correlated with the neutron
skin thickness, although the transition density did show some rather
weak correlation. More interesting is the fraction of the star's
moment of inertia residing in the crust which is measurable and which
depends on the transition pressure, mass and radius. Here there was
some correlation with larger values of this fraction tending to imply
larger values of the neutron skin thickness, but the correlation was
not very robust. Probably the neutron skin and the transition pressure
should be viewed as independent quantities whose measurement would
provide separate constraints on the EOS.

\section*{ACKNOWLEDGMENTS} 

Research support of the U.S. Department of Energy under grant numbers
DOE/DE-FG02-87ER-40328 (for AWS and PJE), and DOE/DE-FG02-87ER-40317
(for MP and JML) is acknowledged.  We thank Chuck Horowitz and David
Dean for providing us with computer codes for the calculation of the
properties of finite nuclei.

\clearpage
\appendix
\section{Model properties}
\begin{table}[hbt]
\begin{ruledtabular}
\caption{Saturation and surface properties 
for the models selected in this work.  Symbols are $n_0$:
nuclear matter saturation density in fm$^{-3}$, $B$: the binding
energy per particle in MeV, $K$: compression modulus in MeV, $S_v$:
symmetry energy in MeV, $S_v^\prime$: density derivative of the
symmetry energy in MeV fm$^{3}$, $\sigma_0$: surface tension of
symmetric matter in MeV fm$^{-2}$, $\sigma_\delta$: asymmetry
parameter of the surface tension in MeV fm$^{-2}$, and $\delta R$:
skin thickness of $^{208}$Pb in fm.  The symbol ``TW'' in the second
column refers to new models constructed in this work.}
\begin{tabular}{cccccccccc}
Model & Ref. & $n_0$ & $B$ & $K$ & $S_v$ & $n_0S_v^{\prime}$ & $\sigma_0$ & 
$\sigma_{\delta}$ & $\delta R$ \\
\hline
Potential & & & & & & & & & \\
Gs & \cite{Friedrich86} & 0.15765 & 15.590 & 237.57 & 31.384 &
31.351 & 0.94198 & 6.4724 & 0.23730 \\
Ly5 & [TW] & 0.16059 & 15.986 & 229.94 & 32.010 &
16.048 & 1.1469 & 3.8139 & 0.16600 \\
Rs & \cite{Friedrich86} & 0.15778 & 15.589 & 237.66 & 30.593 &
28.574 & 0.94257 & 5.8063 & 0.22240 \\
SGI & \cite{VanGiai81} & 0.15477 & 15.895 & 263.79 & 28.373 &
21.321 & 1.1156 & 4.0972 & 0.17650 \\
SLy0 & \cite{Chabanat95} & 0.16030 & 15.973 & 229.67 & 31.982 &
15.704 & 1.1447 & 3.5963 & 0.16540 \\
SLy230a & \cite{Chabanat97} & 0.15998 & 15.990 & 229.89 & 31.985 &
14.772 & 1.1472 & 3.1525 & 0.15540 \\
SkMP & \cite{Bennour89} & 0.15714 & 15.562 & 231.36 & 29.899 &
23.447 & 1.1063 & 5.1515 & 0.20030 \\
SkT4 & \cite{Tondeur84} & 0.15902 & 15.957 & 235.56 & 35.457 &
31.380 & 0.98109 & 5.8878 & 0.25340 \\
SkI5 & \cite{Reinhard95} & 0.15596 & 15.848 & 256.95 & 36.697 &
43.185 & 1.1492 & 7.7179 & 0.27540 \\
NRAPR & [TW] & 0.16058 & 15.856 & 225.70 & 32.787 &
19.880 & 1.0391 & 4.5347 & 0.19030 \\
\hline
Field-theoretical & & & & & & & & & \\
NL4 & \cite{Nerlo-Pomorska04} & 0.14761 & 16.158 & 270.35 & 36.239 &
38.307 & 1.1706 & 6.2556 & 0.27346 \\
S271 & \cite{Horowitz01} & 0.14840 & 16.250 & 271.00 & 35.927 &
24.240 & 0.95401 & 5.0735 & 0.25063 \\
Z271 & \cite{Horowitz01} & 0.14840 & 16.250 & 271.00 & 35.369 &
22.334 & 0.89856 & 4.0940 & 0.23765 \\
SR1 & [TW] & 0.16558 & 16.359 & 202.15 & 29.000 &
9.0730 & 1.0593 & 2.5161 & 0.15953 \\
SR2 & [TW] & 0.15245 & 16.376 & 224.64 & 30.071 &
9.8437 & 0.95494 & 2.3783 & 0.17229 \\
SR3 & [TW] & 0.15000 & 16.272 & 222.55 & 29.001 &
11.303 & 0.93076 & 2.4286 & 0.17096 \\
es25 & [TW] & 0.16000 & 16.000 & 211.73 & 25.000 &  
15.107 & 0.94625 & 2.4807 & 0.13769 \\
es275 & [TW] & 0.16000 & 16.000 & 205.33 & 27.500 &
19.253 & 0.94558 & 3.2336 & 0.17135 \\
es30 & [TW] & 0.16000 & 16.000 & 215.36 & 30.000 &
26.853 & 1.0281 & 4.7907 & 0.20542 \\
es325 & [TW] & 0.16000 & 16.000 & 212.45 & 32.500 &
29.808 & 1.0232 & 5.3820 & 0.22536 \\
es35 & [TW] & 0.15972 & 16.000 & 209.97 & 34.937 &
35.386 & 1.0369 & 6.4629 & 0.24843 \\
RAPR & [TW] & 0.15705 & 16.362 & 276.70 & 33.987 &
18.993 & 0.97931 & 3.1029 & 0.20114 \\
\end{tabular}
\label{tab:modelprop}
\end{ruledtabular}
\end{table}

\clearpage
\section{Coupling strengths of models SR2, es25, es30, and es35 }
\vspace*{-20pt}
\begin{table}[hbt!]
\begin{ruledtabular}
\caption{Coupling strengths for the model SR2 with nuclear
matter equilibrium density $n_0=0.15~{\rm fm}^{-3}$, binding energy
$B=16.38$ MeV, compression modulus $K=224.5$ MeV, Dirac effective mass
$M^*(n_0)=0.78M$, and symmetry energy $E_{sym}=30.1$ MeV. The dimensions of
$a_i$ and $b_i$ are such that the Lagrangian in Eq.~(\ref{FLTL}) is in
MeV$^4$.}
\begin{tabular}{|cccccc|}
$m_{\sigma}$ & $g_{\sigma}$ & $g_{\omega}$ &
$g_{\rho}$ & $\kappa$ & $\lambda$ \\
440.93 MeV &
7.1513 &
8.9644 &
11.217 &
14.055 MeV &
0.010091 \\
$\zeta$ & $\xi$ & $a_1$ & $a_2$ & $a_3$ & $a_4$ \\
0.052964 &
3.0351 &
35.470 &
5.0570 $\times 10^{-3}$ &
2.3109 $\times 10^{-6}$ &
1.0686 $\times 10^{-3}$ \\
$a_5$ & $a_6$ & $b_1$ & $b_2$ & $b_3$ & \\
1.0060 $\times 10^{-11}$ &
6.8908 $\times 10^{-7}$ &
0.22299 &
2.2908 $\times 10^{-7}$ &
3.0076 $\times 10^{-13}$ & 
\end{tabular}
\label{SR2prop}
\end{ruledtabular}
\end{table}
\vspace*{-20pt}
\begin{table}[hbt!]
\begin{ruledtabular}
\caption{Coupling strengths for the model es25 with nuclear
matter equilibrium density $n_0=0.16~{\rm fm}^{-3}$, binding energy
$B=16$ MeV, compression modulus $K=211.7$ MeV, Dirac effective mass
$M^*(n_0)=0.76M$, and symmetry energy $E_{sym}=25$ MeV. For the 
couplings $a_i$ and $b_i$, only the non-zero entries are given.}
\begin{tabular}{|cccccc|}
$m_{\sigma}$ & $g_{\sigma}$ & $g_{\omega}$ &
$g_{\rho}$ & $\kappa$ & $\lambda$ \\
459.52 MeV &
7.5425 &
9.1120 &
8.3802 &
16.504 MeV &
-0.052591 \\
$\zeta$ & $\xi$ & $a_2$ & $b_1$ & & \\
1.5174 $\times 10^{-3}$ &
0.12740 &
3.4477 &
1.5931 &
&
\end{tabular}
\label{es25prop}
\end{ruledtabular}
\end{table}
\vspace*{-20pt}
\begin{table}[hbt!]
\begin{ruledtabular}
\caption{Coupling strengths for the model es30 with nuclear
matter equilibrium density $n_0=0.16~{\rm fm}^{-3}$, binding energy
$B=16$ MeV, compression modulus $K=215.4$ MeV, Dirac effective mass
$M^*(n_0)=0.66M$, and symmetry energy $E_{sym}=30$ MeV.  
Only non-zero $a_i$ and $b_i$ are given.}
\begin{tabular}{|cccccc|}
$m_{\sigma}$ & $g_{\sigma}$ & $g_{\omega}$ &
$g_{\rho}$ & $\kappa$ & $\lambda$ \\
503.18 MeV &
9.2236 &
11.173 &
7.5106 &
7.3353 MeV &
-0.028115 \\
$\zeta$ & $\xi$ & $a_2$ & $b_1$ & & \\
2.9571 $\times 10^{-3}$ &
0.20174 &
0.57839 &
0.22445 &
& 
\end{tabular}
\label{es30prop}
\end{ruledtabular}
\end{table}
\vspace*{-25pt}
\begin{table}[hbt!]
\begin{ruledtabular}
\caption{Coupling strengths for the model es35 with nuclear
matter equilibrium density $n_0=0.1597~{\rm fm}^{-3}$, binding energy
$B=16$ MeV, compression modulus $K=210.0$ MeV, Dirac effective mass
$M^*(n_0)=0.64M$, and symmetry energy $E_{sym}=34.93$ MeV. 
Only non-zero $a_i$ and $b_i$ are given.}
\begin{tabular}{|cccccc|}
$m_{\sigma}$ & $g_{\sigma}$ & $g_{\omega}$ & 
$g_{\rho}$ & $\kappa$ & $\lambda$ \\
508.88 MeV &
9.5598 &
11.576 &
8.1475 &
6.6428 MeV &
-0.028032 \\
$\zeta$ & $\xi$ & $a_2$ & $b_1$ & & \\
-5.8216 $\times 10^{-5}$ &
0.067440 &
0.024880 &
0.014600 &
&
\end{tabular}
\label{es35prop}
\end{ruledtabular}
\end{table}

\clearpage

\bibliography{paper}
\bibliographystyle{apsrev}
\end{document}